\documentclass[notitlepage,twocolumn,prl,nobalancelastpage,amssymb,superscriptaddress,showpacs]{revtex4-1}

\usepackage{Symbols}
\usepackage{Shortcuts}
\usepackage{Text}

\usepackage{epsfig}
\usepackage{subfigure}
\usepackage[colorlinks=true,citecolor=blue,linkcolor=blue]{hyperref}
\usepackage{verbatim}
\usepackage{xcolor}

\def \va{\boldsymbol{a}}
\def \vh{\boldsymbol{h}}
\def \vd{\boldsymbol{d}}
\def \vc{\tb{c}}
\def \vk{{\boldsymbol{k}}}
\def \vq{{\boldsymbol{q}}}
\def \vr{{\boldsymbol{r}}}

\def \vx{\boldsymbol{x}}
\def \vy{\boldsymbol{y}}
\def \vz{\boldsymbol{z}}
\def \ble{\boldsymbol{e}}

\def \vs{\boldsymbol{\s}}

\begin{document}
\title{ Electronic Floquet Liquid Crystals}


\author{Iliya Esin}
\affiliation{\mbox{Physics Department, Technion, 3200003 Haifa, Israel}}
\author{Gaurav Kumar Gupta}
\affiliation{\mbox{Physics Department, Technion, 3200003 Haifa, Israel}}
\author{Erez Berg}
\affiliation{\mbox{Department of Condensed Matter Physics, Weizmann Institute of Science, Rehovot, 76100, Israel}}
\author{Mark S. Rudner}
\affiliation{\mbox{Center for Quantum Devices and Niels Bohr International Academy,}
\mbox{Niels Bohr Institute, University of Copenhagen, 2100 Copenhagen, Denmark}}
\author{Netanel H. Lindner}
\affiliation{\mbox{Physics Department, Technion, 3200003 Haifa, Israel}}

\begin{abstract}
``Floquet engineering'' -- designing band structures ``on-demand'' through the application of coherent time-periodic drives -- has recently emerged as a powerful tool for creating new topological and anomalous phases of matter. In this manuscript, we show that the same principle can be applied to create non-equilibrium correlated states with spontaneously broken symmetry in a lightly doped semiconductor. The periodic drive provides means for obtaining large electronic densities of states necessary for the broken symmetry phase. The phase transition occurs in the steady-state of the system achieved due to interplay between the coherent external drive, electron-electron interactions, and dissipative processes arising from the coupling to phonons and the electromagnetic environment. We obtain the phase diagram of the system using numerical calculations that match predictions obtained from a phenomenological treatment and discuss the conditions on the system and the external drive under which spontaneous symmetry breaking occurs. Our results imply that Floquet engineering of the density of states provides a new route for generating and controlling correlated states of electrons with external fields.

\end{abstract}

\maketitle

\section{Introduction}



Strongly-correlated phases of electronic systems emerge from the competition between the potential energy savings and kinetic energy costs of developing correlations that allow electrons to avoid each other.
In materials with band structures that feature large densities of states (DOSs), the kinetic energy costs that oppose the formation of correlations are small.
Such materials therefore provide a rich platform for realizing exotic phases of matter where interparticle interactions crucially alter the ground state properties of the system. 

In two-dimensional systems, a prominent route to achieving high DOS bands 
is through the application of strong out-of-plane magnetic fields, which gives rise to flat Landau levels. At certain rational filling fractions, the resulting macroscopic degeneracy is lifted by the formation of strongly-correlated fractional quantum Hall states~\cite{Tsui1982,Laughlin1983}. 
Recently, a rich phase diagram of correlated states arising from flat band formation has also been uncovered  for twisted bilayer graphene, when the twist angle between layers is tuned close to the ``magic angle''~\cite{Bistritzer2011,Cao2018a,Cao2018,Yankowitz2019,Sharpe2019,Zondiner2020}. 

Two-dimensional systems in which the minimum of the single-particle dispersion occurs along a ring in momentum space (rather than at a single point, as for a standard parabolic dispersion), provide an alternative route for achieving large densities of states and novel correlated phases~\cite{Yang2006,Wang2010,Wu2011,Gopalakrishnan2011,Jian2011,Barnett2012,Sedrakyan2012,Berg2012,Zhou2013,Ruhman2014,Silvestrov2014,Sedrakyan2014,Sedrakyan2015}.
This occurs, for example, in two-dimensional materials with strong Rashba-type spin-orbit coupling~\cite{Rashba1959,Bychkov1984}.
The ring-minimum in such systems leads to a large degeneracy and a divergent DOS at energies approaching the bottom of the band. 
At low densities, inter-particle interactions may lead to a plethora of possible symmetry-broken phases. In particular, for short-ranged interactions, novel electronic liquid-crystalline ground states were predicted in  Ref.~\onlinecite{Berg2012}. These phases exhibit spontaneously broken rotational symmetry, with extremely anisotropic Fermi surfaces and related susceptibilities.

\begin{figure}
	\centering
	\includegraphics[width=8.6cm]{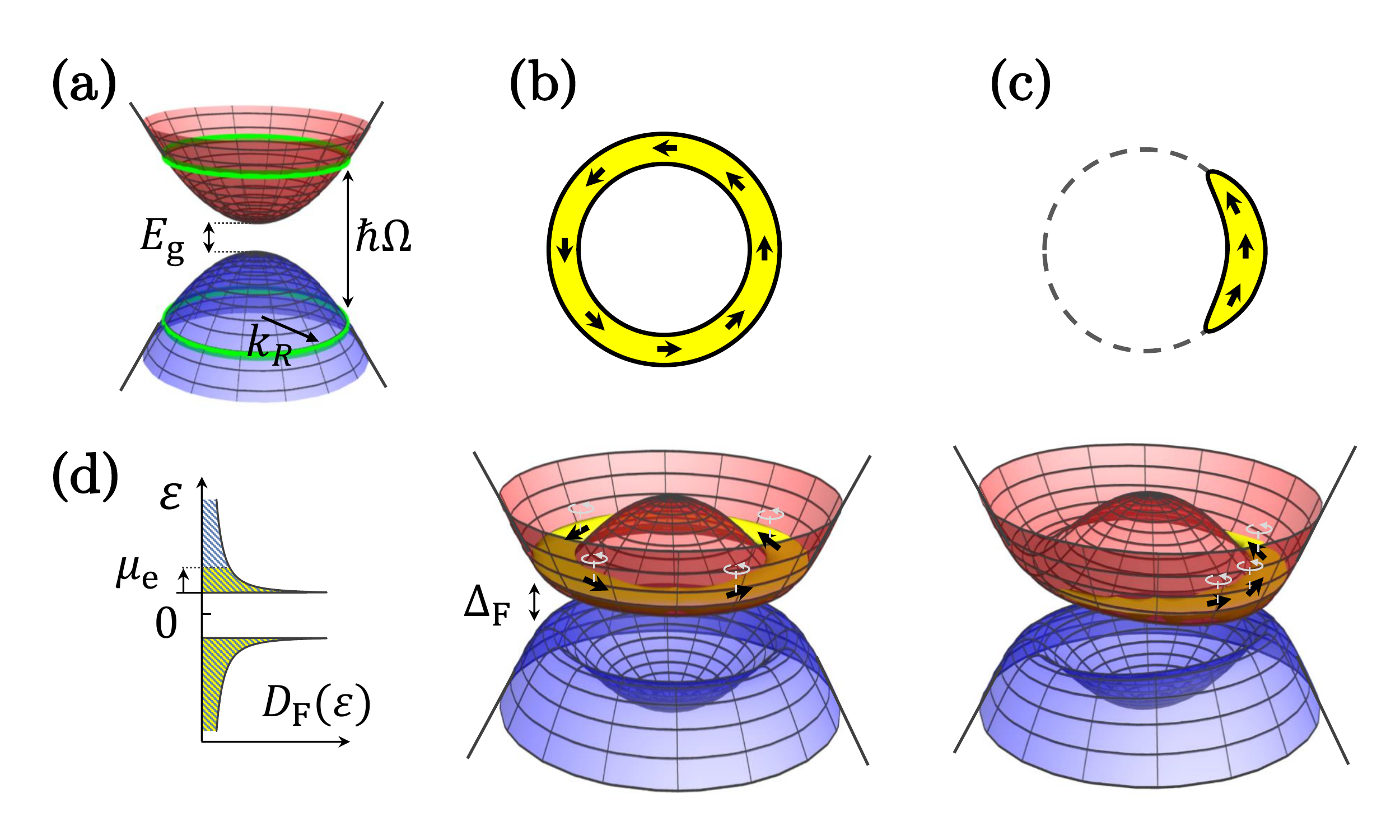}\\
	\caption{Single-particle band structure near the $\G$-point. (a) The band structure of the non-driven semiconductor. The resonance rings of the external drive are indicated by the green curves. (b,c) Floquet quasienergy bands arising from the semiconductor's band structure and the resonant drive around $\ve=0$. The yellow area represents the occupation of the upper Floquet band in the ``ideal'' distribution scenario, analogous to the zero-temperature Gibbs state for the quasi-energy spectrum. Black arrows represent the pseudospin direction of the Floquet states near the resonance ring. The texture of the pseudospins arises from the pseudospin-momentum locking induced by the semiconductor. In addition, each pseudospin rotates in the $x$-$y$ plane with the frequency of the periodic drive as is indicated on the figure by the light-gray thin arrows attached to each pseudospin. In the symmetric phase, (b), due to rotational symmetry the Floquet states near the resonance ring are uniformly occupied, as is indicated below panel (b). Panel (c) demonstrates the single-particle Floquet bands in the broken symmetry phase. In this case, the resonance ring is tilted towards 
	a spontaneously chosen direction. The occupation of the bands is then biased toward this direction, 
	signalling a ferromagnetic alignment of the pseudospins. 
	(d) The density of Floquet states as a function of the quasi-energy around $\ve=0$ in the paramagnetic phase. The density of states features square-root Van Hove singularities in each Floquet band, i.e.,  $D_\yF(\dl\ve)\sim \dl\ve^{-1/2}$ in the upper Floquet band, where $\dl \ve\eqv \ve-\D_\yF/2$. A similar relation holds for the lower Floquet band. \label{fig:Bands}}
\end{figure}

Here we present a ``Floquet engineering'' approach for inducing analogous liquid crystalline phases in electronic systems using time-periodic driving. Floquet engineering \cite{Bukov2015,Eckardt2017,Oka2019,Cooper2019, Rudner2020a} has recently emerged as a powerful technique  for creating topological bands \cite{Oka2009,Inoue2010,Kitagawa2010,Lindner2011,Kitagawa2011, Lindner2013, Gomez-Leon2013, Cayssol2013,Usaj2014,Klinovaja2016,Roy2017, Lubatsch2019, Harper2020} and for inducing novel non-equilibrium phases \cite{Rudner2013,Grushin2014,Sacha2015, Khemani2016, Else2016,Potter2016,Titum2016,VonKeyserlingk2016a,VonKeyserlingk2016b,Po2016,Else2016b,Choi2017,Zhang2017,Rudner2019}. Along with extensive theoretical progress, Floquet engineering has been experimentally realized in solid-state, as well as atomic, molecular and optical (AMO) systems \cite{Wang2013,Rechtsman2013,Jotzu2014,Aidelsburger2015,Lohse2016,Mahmood2016,Nakajima2016,Maczewsky2017,Flaschner2018,McIver2020}.

The non-equilibrium phase transition that we describe results from an interplay between coherent driving, electron-electron interactions, and dissipative dynamics due to the system's coupling to its environment \cite{Iadecola2013,Dehghani2014,Iadecola2015,Iadecola2015a,Seetharam2015,Liu2015,Dehghani2015,Genske2015,Shirai2015,Dehghani2016,Shirai2016,Iwahori2016, Esin2018,Esin2019,Seetharam2019}. The coherent drive is used to produce a Floquet band structure that features a ring-like minimum analogous to that 
of the Rashba system described above. In turn, the interactions and dissipative dynamics determine the steady state of the system and the symmetry breaking that it exhibits.

A ring-like dispersion minimum is natural to obtain in a direct band gap system subjected to a coherent drive, where the drive frequency $\W$ is larger than the system's band gap (see Fig.~\ref{fig:Bands}a). The structure of the modified (Floquet) bands is most easily visualized in a rotating frame. Starting from the original bands as depicted in Fig.~\ref{fig:Bands}a, we transform to a rotating frame in which the energies of all states in the valence band are rigidly shifted upwards by $\hb\W$~\cite{Lindner2011}.
In the rotating frame, the (shifted) valence and conduction bands cross along a continuous ``resonance ring'' of points in momentum space where the original conduction and valence bands were separated by $\hbar\Omega$ (see green curves in Fig.~\ref{fig:Bands}a).
After transforming to the rotating frame, the driving field obtains a static (co-rotating) part, and a component that oscillates with integer multiples of the drive frequency $\W$. Within the rotating wave approximation we keep only the static part of the drive in the rotating frame, and discard the oscillating components.
As we show in detail below, under appropriate conditions on the material's band structure and the form of the drive, the co-rotating part of the drive opens a ``Floquet gap''  all the way around the resonance ring. The minima and maxima of the resulting upper and lower Floquet bands correspondingly occur along a ring in momentum space (see Fig.~\ref{fig:Bands}b), yielding a DOS for the Floquet bands, $D_\yF(\varepsilon)$, with square-root divergences near the two band extrema (Fig.~\ref{fig:Bands}d). Along these ring-extrema, the Floquet-Bloch states may be characterized by a pseudo-spin with a non-trivial winding, see Fig~\ref{fig:Bands}b, in close analogy to the spin winding that occurs around the ring minimum of a Rashba-type band structure.

Our goal is to find the conditions under which the divergence in the DOS promotes spontaneous symmetry breaking in the electronic steady-state of the system. Throughout this paper we study a system lightly doped above half filling. To preview the considerations involved, consider first an ideal situation where the steady-state is a zero-temperature Gibbs state of electrons in the band structure obtained in the rotating wave approximation \cite{Galitskii1970}.  In the absence of electron-electron interactions this zero temperature state corresponds to a full lower Floquet band and an annulus-shaped Fermi sea at the bottom of the upper band, as indicated in Fig~\ref{fig:Bands}b. At low doping density, the DOS at the Fermi energy is strongly enhanced due to the divergence at the band bottom. Sufficiently strong electron-electron interactions make it energetically favorable for electrons to change the topology of the Fermi sea from an annulus to a single pocket centered around a spontaneously chosen point on the resonance ring. The symmetry broken state is ferromagnetic as the pseudospins of the electrons are predominantly aligned along one direction, see Fig.~\ref{fig:Bands}c. This leads to a reduction in the potential energy (as the electronic wavefunction overlaps are suppressed for parallel pseudospins due to Pauli exclusion). As a consequence the mean-field band structure in the symmetry broken phase acquires a tilt along the spontaneously chosen momentum direction. Interestingly, due to the periodic time-dependence of the Floquet states, the emergent pseudospin magnetization vector rotates in time at the frequency of the drive (in the lab frame).

The discussion above, based on the rotating wave approximation and a zero-temperature Gibbs-type steady state, captures the essence of the symmetry breaking transition. However, the non-equilibrium nature of the setup implies that the steady-state cannot be described by a simple Gibbs state. Even when the electronic system is coupled to a zero-temperature heat bath, unavoidable scattering processes  create electron-hole excitations in the Floquet bands. These excitations may suppress the tendency towards ordering. In this work, we introduce a self-consistent treatment of coupled kinetic and Floquet-Hartree-Fock equations that capture the interplay between the steady state of the system and its renormalized Floquet band structure, with the possibility of spontaneously broken symmetry. Using this treatment, we will obtain the non-equilibrium phase diagram for different doping densities, interaction strengths, and properties of the external heat baths.

\section{Model System and problem setup}
To study the phase transition in the steady-state of a periodically driven direct bandgap semiconductor, we introduce an effective 
model that describes the single-particle electronic states near the semiconductor's $\G$-point. We consider a two-band, two-dimensional (2D) system, with topologically trivial bands. (This model, which lacks time reversal symmetry, may be taken to represent half of the degrees of freedom of a time-reversal symmetric semiconductor \cite{Bernevig2006, Lindner2011}.) We 
assign a common effective mass $m_\ast$ for the electrons in the conduction band and holes in the valence band, and denote the gap separating these bands by $E_\yg$, see Fig.~\ref{fig:Bands}a. The Hamiltonian describing the 
electronic system and the time-periodic drive, near the $\G$-point, reads
\Eq{
\hat\cH(t)=\sum_{\vk} \hat \vc_\vk\dg [H_0(\vk)+H_\yd(t)]\hat \vc_\vk+\sum_\vq \cV_\vq \hat \ro_{\vq}\hat \ro_{-\vq}.
\label{eq:HamiltonianBHZ}
}
Here $H_0(\vk)=E_0+(\hb^2 |\vk|^2/2m_\ast+E_\yg/2)\s^z+\lm_{0}\, \vk\cdot\vs$, where $\vk = (k_x, k_y)$ is the two-dimensional momentum, $\hat \vc_\vk\dg=(\hat c_{\vk\aup}\dg,\hat c_{\vk\adn}\dg)$ is the two-component spinor for the pseudospin degree of freedom, $\cV_{\vq}$ describes an effective short-ranged electron-electron interaction, 
$\hat \ro_{\vq}=\sum_{\vk} \hat \vc\dg_{{\vk+\vq}}\hat \vc_{\vk}$, and $E_0$ is an energy offset. We denote the pseudospin-orbit coupling by $\lm_0$ and use $\vs=(\s^x,\s^y)$, where $\s^\a$, $\a=x,y,z$, is a Pauli matrix in the pseudospin space. The bandstructure of the system in the absence of the drive is given by the spectrum of $H_0$. We denote the energies of the valence and conduction band by $E_v(\vk)$ and $E_c(\vk)$, respectively.

We consider a uniform driving field that couples to the electrons through $\s^z$, $H_\yd (t)=V\cos(\W t)\s^z$ \footnote[10]{More realistic time-dependent electromagnetic fields can be incorporated in this model, see \cite{Lindner2011}.}, with an above-gap frequency $\hb\W=E_\yg+\dl E$, where $\dl E$ is much smaller than semiconductor's full bandwidth. The Floquet state solutions of $\hat\cH(t)$ for $\cV_\vk=0$ satisfy $\bS{i\hb\frac{\dpa }{\dpa t}-H_0(\vk)-H_\yd(t)}\ket{\y_{\vk\n}(t)}=0$, with $\ket{\y_{\vk\n}(t)}=e^{-i\ve_{\vk \n} t/\hb}\ket{\phi_{\vk\n}(t)}$. Here $\ket{\f_{\vk\n}(t)} = \ket{\f_{\vk\n}(t + \cT)}$ is periodic with period $\cT = 2\pi/\Omega$ and $\ve$ is the quasienergy (which is periodic in $\hb \W$). Throughout, we use the convention $-\hb\Omega/2\leq\ve<\hb\Omega/2$. For convenience we take $E_0=\hbar\Omega/2$ such that $\ve=0$ at the center of the Floquet gap (see Fig.~\ref{fig:Bands}d).

The drive resonantly couples valence and conduction band states along a ring in momentum space for which
$\hbar \Omega= E_c(\vk)-E_v(\vk)$. We denote the radius of this ring by $k_\yR$. At the resonance ring, a gap of magnitude $\D_\yF=2\lm_0k_\yR V/\hb\W$ opens in the Floquet quasienergy spectrum. This gap separates the ``upper Floquet'' ($\n=+$) and ``lower Floquet'' ($\n=-$) bands, corresponding respectively to $\ve>0$ and $\ve<0$. Here, we will focus on the parameter regime $\D_\yF\ll\dl E$, where the ring minimum is well developed. Each of the bands has a ring of degenerate states associated with square-root van Hove singularities in the density of Floquet states: near the bottom of the upper Floquet band, the density of states takes the form $D_\yF(\dl \ve)\eqa\frac{m_\ast}{2\p\hb^2}\sqrt{\frac{\D_\yF}{\dl \ve}}$,  where $\dl \ve=\ve-\D_\yF/2$, see Fig.~\ref{fig:Bands}. A similar expression holds for quasienergies near the top of the lower Floquet band. Below we show how these van Hove singularities promote spontaneous symmetry breaking in the driven system.

\section{Order Parameter and Floquet self-consistent mean field approach}

In this work we will look for spontaneous symmetry breaking that emerges in the steady state of the driven system. The steady state arises from an interplay between the time-periodic drive, electron-electron interactions, and  the coupling of the electrons to the electromagnetic and phononic modes of their environment. In this interplay, the electron-electron interactions play a dual role, as they lead to formation of order parameters as well as to incoherent scattering which may suppress the tendency towards order.

In order to capture the coherent part of the electron-electron interaction, which leads to order parameter formation, we use a mean-field approximation \footnote[20]{We assume here only the ferromagnetic phase, which we can easily treat within the Hartree-Fock analysis. We leave the analysis of the nematic paramagnetic phase, which can potentially emerge in our system \cite{Berg2012,Ruhman2014}, for future studies.} in which we assume that the steady state is Gaussian (i.e., obeys Wick’s theorem). We assume translation invariance is maintained, and consider a mean-field decoupling of the Hamiltonian Eq.~\eqref{eq:HamiltonianBHZ} with  ferromagnetic nematic order parameter
\Eq{
\vh_\vk(t)=-\sum_{\vk'}\cV_{\vk-\vk'} \av{\hat \vc_{\vk'}\dg\vs\hat \vc_{\vk'}}_{\rm MF}.
\label{eq:MeanField}
}
The expectation value in Eq.~\eqref{eq:MeanField} is taken with respect to the time-periodic steady-state of the system. The corresponding mean-field Hamiltonian is given by $\hat \cH_{\rm{MF}}(t)= \sum_\vk \hat \vc_{\vk}\dg H_{\rm MF}(\vk,t)\hat\vc_\vk$, where
\Eq{
H_{\rm MF}(\vk,t)=H_0(\vk)+H_\yd(t)+\vh_\vk(t)\cdot\vs.
\label{eq:MeanFieldHamiltonian}
}
Note that if $\vh_\vk(t)$ has the same time-period as the drive, $\hat{\mathcal{H}}_{\rm MF}(t)$ is also time-periodic and therefore defines a new Floquet problem.

The time-periodic steady state used in Eqs.~(\ref{eq:MeanField}) and~(\ref{eq:MeanFieldHamiltonian}) is determined self-consistently by solving the kinetic equation for the populations of electrons in the Floquet bands of $\hat{\mathcal{H}}_{\rm MF}(t)$.
These populations are defined as $f_{\vk\n}(t)\eqv\av{\hat\f_{\vk\n}\dg(t)\hat\f_{\vk\n}(t)}$,
where $\hat\f_{\vk\n}\dg(t)$ is a creation operator corresponding to the Floquet state $\ket{\f_{\vk\n}(t)}$. Note that the meaning of the index $\nu$ and the values of populations $f_{\vk\n}$ depend on the order parameter, $\vh_\vk(t)$, as it determines the Floquet bandstructure of $\hat{H}_{\rm MF}(t)$.
The kinetic equation includes scattering rates due to electron-phonon interactions, $I_{\vk\n}^{\ys}$, radiative recombination, $I_{\vk\n}^{\ell}$, and electron-electron collisions, $I_{\vk\n}^{\rm{ee}}$, and is given by
\Eq{
\dot f_{\vk\n}=I^{\ys}_{\vk\n}(\bC{f})+I^{\ell}_{\vk\n}(\bC{f})+I^{\rm ee}_{\vk\n}(\bC{f}),
\label{eq:RateEquation}
}
where the steady state is determined by $\dot f_{\vk\n}=0$. The notation $\{f\}$ refers to the full set of populations over all momenta and band indices.

In writing the kinetic equation in terms of the populations $f_{\vk\nu}$ we have assumed that the Gaussian steady state is approximately described by a single particle density matrix which is diagonal in the Floquet basis. This condition is satisfied when the scattering rates in the steady state are small, $\hb/(\ta_{\rm scat}\D_\yF)\ll1$ \cite{Seetharam2015}. Here $1/\ta_{\rm scat}$ is the total scattering rate of the electrons.

The scattering rates $I_{\vk\n}^{\ys}$ and $I_{\vk\n}^{\ell}$ describe scattering processes in which a boson (phonon, ${\rm s}$, or photon, $\ell$) is emitted or absorbed by the electronic system.  The corresponding rates are determined by the dispersions of these bosons, and the form of the electron-boson coupling.


We denote by $\hat b_{p\vq}\dg$ the operator creating an acoustic phonon (for $p=\ys$) or a photon (for $p=\ell$) with the three-dimensional (3D) momentum $\vq=(\vq_\parallel,q_z)$ and frequency $\w_\vq=v_p |\vq|$.   Here $\vq_\parallel$ is the component of $\vq$ within the plane of the 2D electronic system and $v_\ys(v_\ell)$ is the speed of sound (light). Note that the phonons propagate in the 3D substrate of the 2D electronic system.

The electron-boson coupling is described by the Hamiltonian \cite{Mahan2000}
\Eq{\hat\cH_{\rm HB}=\sum_{\vk,p,\vq}\hat \vc_{\vk}\dg \cM_{p}(\vq_\parallel,\w_\vq)\hat \vc_{\vk+\vq_\parallel}(\hat b\dg_{p,\vq}+\hat b_{p,-\vq})+\rm{h.c.},
\label{eq:ElectronBathCoupling}
}
where $\cM_p(\vq_\parallel,\w_\vq)$ is the coupling matrix in pseudospin space. We consider a diagonal electron-phonon coupling matrix in the $\{\uparrow,\downarrow \}$ basis, which captures the conservation of the pseudospin in small-momentum-transfer electron-phonon interactions. In contrast, photon emission requires changing the electronic angular momentum. We account for this by taking an electron-photon coupling matrix that is strictly off-diagonal in the  $\{\uparrow,\downarrow \}$ basis, as these two basis states have opposite parity. Throughout the manuscript, we will assume that the phonons and photons are in thermodynamic equilibrium at zero-temperature.

The rates  $I_{\vk\nu}^{\ys}$ and $I_{\vk\nu}^{\rm{\ell}}$ in Eq.~(\ref{eq:RateEquation}) can be computed through Floquet-Fermi's golden rule \cite{Rudner2020} using the electron-boson coupling in Eq.~(\ref{eq:ElectronBathCoupling}). Similarly, $I_{\vk\nu}^{\rm ee}$ is computed using Floquet-Fermi's golden rule and the  electron-electron interactions appearing in Eq.~(\ref{eq:HamiltonianBHZ}). Explicit expressions for these rates appear in the supplementary material (SM).

\section{Ferromagnetic-nematic steady states}
Before presenting the full steady-state solution to Eqs.~\eqref{eq:MeanField},  \eqref{eq:MeanFieldHamiltonian}, and \eqref{eq:RateEquation}, we introduce a phenomenological model which we will use to characterize the phase diagram of the system. The model includes the key processes required for obtaining the steady-state distribution for the electrons. Our goal is to identify the conditions on the electronic system and its environment under which spontaneous symmetry breaking may occur. A key quantity for describing the steady state is the density of electrons in the upper Floquet band, defined as $n_\ye=\int\frac{d^2\vk}{(2\p)^2}f_{\vk+}$. Likewise, the density of holes in the lower band, $n_\yh$, is computed by integration over $1-f_{\vk-}$. In what follows, we discuss generation and annihilation rates of electron-hole pairs (in the Floquet basis) resulting from collision processes [see Eq.~\eqref{eq:RateEquation}]. We refer to these as heating and cooling processes, respectively.
Of particular importance are Floquet-Umklapp processes, in which the energies of the electrons and bosonic modes in the initial and final states differ by $\hbar \Omega$. At zero bath temperature, these processes provide the only mechanism for heating.



We will be interested in the situation in which the system is doped slightly above half filling. In the absence of Floquet-Umklapp processes and at zero bath temperature, the steady-state is a zero-temperature Gibbs distribution of electrons in the (mean-field) Floquet bands~\cite{Galitskii1970}.
Specifically, in this situation, the steady state features a completely filled lower Floquet band, and a low density Fermi sea of electrons in the upper Floquet band. In the presence of Floquet-Umklapp processes, this ideal distribution is perturbed by the creation of (inter-Floquet-band) electron-hole pairs.
We will focus on the regime where the  densities of electrons and holes in the upper and lower Floquet bands are low: $n_\ye,n_\yh\ll\cA_{\yR}$, where $\cA_{\yR}\eqv \p k_\yR^2$ is the area in reciprocal space 
enclosed by the resonance ring.

The pair creation rate in almost empty upper and almost full lower Floquet bands is approximately independent of the densities of electrons and holes in the respective bands. We denote the total pair creation rate due to collisions with both phonons and photons by $\dot n_{\ye}|_{\rm ph}=\G_{\rm ph}$. Similarly, the pair creation rate due to electron-electron collisions 
is denoted by $\dot n_{\ye}|_{\rm ee}= \G_{\rm ee}$. The parameter $\G_{\rm ee}$ depends on $\cV_{\vq}^2$ at $\vq$ corresponding to the inverse interparticle distance in the nearly filled band. The processes contributing to $\dot n_{\ye}|_{\rm ee}$ are of the Floquet-Umklapp type, and are suppressed by $(V/\hb\W)^2$. In addition, electron-electron scattering gives rise to quasienergy conserving processes, causing thermalization of the populations within each band without changing the electron and hole population densities. These processes therefore do not contribute to $\dot n_{\ye}|_{\rm ee}$. Moreover, as in equilibrium, these elastic scattering processes all together preserve the form of the distribution when the electrons are 
distributed according to the Fermi-function over the quasienergy spectrum.


Once excited, the electrons (holes) rapidly 
relax to the  bottom (top) of the Floquet band through multiple low-energy phonon emissions. The electron-hole pairs then annihilate through inter-Floquet-band scattering processes mediated by phonons \footnote[30]{The electron-hole pair annihilation processes predominantly occur near the resonance ring, where the electrons and holes are concentrated. Note that for these momenta the Floquet states are equal superpositions of the conduction and valence bands [see Eq.~\eqref{eq:SpinMomentumLocking}], and these states are efficiently coupled by acoustic phonons.}. The rate of the pair annihilation processes, $\dot n_\ye |_{\rm cool}$, is proportional to the product of the densities of electrons and holes. Therefore, we estimate $\dot n_\ye |_{\rm cool}=-\Lm_{\rm inter} n_\ye n_\yh$, where $\Lm_{\rm inter}$ is independent of the populations. Note that for this essential cooling process to occur, the Debye frequency of the phonons needs to be larger than the Floquet gap $\Delta_{\rm F}$.

Summing up the cooling and heating rates we obtain a rate equation for the density of 
electrons in the upper Floquet band,
\Eq{
\dot n_\ye=\G_{\rm ph}+\G_{\rm ee}-\Lm_{\rm inter}n_\ye n_\yh.
\label{eq:PhonomenologicalRates}
}
In the steady-state ($\dot n_\ye=0$), Eq.~\eqref{eq:PhonomenologicalRates} leads to $n_\ye n_\yh=\ka$, where we define the ``heating parameter''
$\ka\eqv\ka_{\rm ph}+\ka_{\rm ee}$, with $\ka_{\rm ph}\eqv \G_{\rm ph}/\Lm_{\rm inter}$, $\ka_{\rm ee}\eqv \G_{\rm ee}/\Lm_{\rm inter}$.
Furthermore, the difference between electron and hole excitation densities is fixed by the electron doping, $\D n$, measured relative to half-filling, $n_\ye-n_\yh=\D n$.
Using this relation, together with the steady-state solution to Eq.~\eqref{eq:PhonomenologicalRates} we obtain
\Eq{
n_{\ye/\yh}=\sqrt{\bR{\D n/2}^2+\ka}\pm \D n/2,
\label{eq:Densities}
}
where the plus (minus) sign on the right hand side corresponds to the density of electrons (holes). Note that in the absence of drive-induced heating processes ($\ka = 0$), the ideal steady-state with no holes in the lower band and density $\D n$ in the upper band is obtained. In what follows, we focus on the electron-doped regime, $\D n \ge 0$ (similar considerations apply in the hole-doped regime).



%

Having established the steady-state densities of electrons and holes (concentrated near the Floquet band extrema at the resonance ring), 
Eq.~(\ref{eq:Densities}), 
we are well-positioned to address the conditions for spontaneous breaking of rotational symmetry in the system. In the following, we assume contact interactions 
described by a constant in $\vq$ interaction strength, $\cV_{\vq} =U/\vp$, and $\vk$-independent magnetization $\vh(t)=\vh_\vk(t)$ [see Eq.~\eqref{eq:MeanField}], where $\vp$ is the area of the system. In the steady state, $\vh(t)$ is time periodic with the same time-period as the drive. Therefore, we expand $\vh(t)$ in terms of its Fourier harmonics,
\Eq{
\vh(t)={\rm Re}\bS{\vh_0 + \vh_1 e^{i\W t}+\cdots}.
\label{eq:Magnetization}
}
Here $\vh_0$ and $\vh_1$ are vectors of complex magnitudes, representing the constant and the first harmonic components of the mean-field, respectively, and ``$\cdots$'' represents higher harmonics. The values of the coefficients $\{\vh_i\}$ are determined self-consistently via Eqs.~(\ref{eq:MeanField}), (\ref{eq:MeanFieldHamiltonian}), and \eqref{eq:RateEquation}.

Crucially, a nonvanishing magnitude of the ``in-plane'' ($x$-$y$) component 
of the magnetization $\vh(t)$, which we denote by $\vh^{(xy)}(t)$, does not respect the  rotational symmetry of the microscopic Hamiltonian $\hat{\mathcal{H}}(t)$, see Eq.~\eqref{eq:HamiltonianBHZ}. Therefore, $|\vh^{(xy)}(t)|$ serves as the order parameter for the ferromagnetic-nematic phase that we study. In contrast, a non-vanishing $z$ component of $\vh(t)$ respects the symmetry. Generically,  we  expect  a  non-vanishing $z$ component of $\vh(t)$  in both the symmetry broken and unbroken phases. In particular, we expect a large static $z$ component of $\vh_0$ (with magnitude on the order of $U$) even in the absence of the drive. This static field simply renormalizes the parameters of $H_0$ in Eq.~\eqref{eq:HamiltonianBHZ}, and therefore we do not treat it self-consistently in our analysis.

For simplicity, in the analytical treatment below we take  $\vh_n=0$ for $n\geq2$ since these harmonics are suppressed by powers of $V/(\hb\W)$ for $V/(\hbar\Omega)\ll1$. Furthermore, we note that when the in-plane ($x$-$y$) components of $\vh_0$ are small, $|\vh^{(xy)}_0|\ll E_\mathrm{g}$, their effect on the Floquet band structure via Eq.~\eqref{eq:MeanFieldHamiltonian} is negligible. To facilitate the analysis we thus also take $\vh_0=0$, thereby focusing our attention on the behavior of $\vh_1$, which describes the component of the magnetization that oscillates at the same frequency as the drive. In the next section we will present numerical results in which all harmonics are allowed to freely develop.


In order to understand the expected form of $\vh_1$, it is helpful to examine the Floquet states near the resonance ring.   These states are created by the operators
\Eq{
\hat\f_{\vk\pm}\dg(t)=(e^{-i\W t}\hat c\dg_{\vk\aup}\mp e^{i\q_\vk}\hat c\dg_{\vk\adn})/\sqrt{2}+\cO\bR{V/\hb \W},
\label{eq:SpinMomentumLocking}
}
where $|\vk|=k_\yR$ and $\q_\vk\eqv\arctan(k_y/k_x)$. The pseudospins 
of these states form a rotating-in-time ``vortex'' in the $x$-$y$ plane, see Fig.~\ref{fig:Bands}b. In the low doping limit ($\D n\to 0$) and in the regime where cooling dominates over heating processes ($\ka\ll \D n^2$), the upper Floquet band has a significant population only near the band's bottom.
Above a critical interaction strength, we expect the self-consistent solution to converge to a ferromagnetic nematic steady-state where the electrons localize around a single spontaneously chosen momentum on the ring (see Fig.~\ref{fig:Bands}c). Subsequently, due to the time-dependent pseudospin-momentum locking in Eq.~\eqref{eq:SpinMomentumLocking}, the pseudospins of the electrons will be synchronized. This implies that the ``in-plane'' ($x$-$y$) components of $\vh(t)$ 
should take the form of a rotating (circularly polarized) field, with its dominant harmonic given by $\vh_1^{(xy)} \approx h_1 (\hat \vx-i\hat \vy)/\sqrt{2}$. In our analysis we use $|\vh_1^{(xy)}|$ as the diagnostic for spontaneous symmetry breaking.

We note that in both the symmetry broken and un-broken phases, the system exhibits an oscillating $z$-component of the magnetization $\vh(t)$.
The $z$-component of the harmonic $\vh_1$ 
 renormalizes the amplitude and phase of the drive [see text below Eq.~(\ref{eq:HamiltonianBHZ})]. As we will show below, throughout the parameter regime of interest this renormalization remains weak. Therefore, in estimating the critical interaction strength below, we neglect this component and keep only $\vh_1^{(xy)}$.

We now seek the minimal interaction strength, $U_\yc$, required to 
 achieve spontaneous symmetry breaking for finite values of $\ka$ and $\D n$. To make progress, we approximate the distribution of electrons in the upper Floquet band by a Fermi-Dirac distribution with an effective chemical potential, $\m_{\ye}$, measured from the bottom of the upper band, and temperature (measured in energy units), $T_{\ye}$. Analogously, we parametrize the hole distribution in the lower Floquet band by an effective chemical potential $\m_{\yh}$, measured from the top of the lower band, and temperature $T_{\yh}$. Such a fit well-approximates the distributions in the limit of low density (see Ref.~\onlinecite{Esin2019} and numerical results in SM). Note that the electron and hole populations are generically described by finite effective temperatures, even when the baths are at zero temperature.

In Eq.~\eqref{eq:Densities} above, we found the total densities of electrons and holes in the upper and lower Floquet bands, $n_\ye$ and $n_\yh$, respectively.
However, a given pair of values for $n_\ye$ and $n_\yh$ can be obtained for a continuous family of choices of $\mu_{\rm e/h}$ and $T_{\rm e/h}$.
Below we first derive a general result for the critical interaction strength $U_\yc$, parametrized by the chemical potentials and temperatures that are realized. Later, we will discuss how to determine the values of $\mu_{\rm e/h}$ and $T_{\rm e/h}$ in the steady-state.

To find $U_\yc$, assuming the transition is continuous, we solve Eq.~\eqref{eq:MeanField} by expanding the expectation value on its RHS to linear order in the amplitude of the in-plane ($x$-$y$) component of the magnetization, $|\vh^{(xy)}_1|$, which we take to be circularly polarized. Note that the RHS of Eq.~\eqref{eq:MeanField} depends on $\vh(t)$ through the steady-state distribution, $f_{\vk\n}$, defined in the basis of the eigenstates of $H_{\rm MF}$ [which also depend on $\vh(t)$], see Eq.~\eqref{eq:MeanFieldHamiltonian}.
%
Given that the effective temperature and chemical potential weakly depend on $\vh^{(xy)}_1$, the dominant dependence of $f_{\vk\n}$ on $\vh^{(xy)}_1$  arises from the eigenstates and eigenvalues of $H_{\rm MF}$.

Expanding the RHS of Eq.~\eqref{eq:MeanField} to linear order in $h_1$ yields three terms:
(i) a contribution corresponding to a full lower Floquet band, and the contributions of (ii) the electrons and (iii) the holes in the upper and lower Floquet bands, respectively.
We use the assumed Fermi-Dirac distribution functions for electrons and holes to evaluate each of the terms analytically (for the full derivation see SM), yielding an expression for the critical interaction strength:
\Eq{
\tilde U_\yc\inv=\tilde U_{\rm fb}\inv+\tilde U_{\rm ex}\inv\bR{\frac{\tilde\Q(\m_\ye/T_\ye)}{ \tilde n_\ye}+\frac{\tilde\Q(\m_\yh/T_\yh)}{ \tilde n_\yh}}.
\label{eq:CriticalU}
}
Here $\tilde U_\yc=\cA_\yR U_\yc/\dl E$, $\tilde n_\ye=n_\ye/\cA_{\yR}$, and $\tilde n_\yh=n_\yh/\cA_\yR$ are the normalized interaction strength and population densities, respectively. (Recall that $\cA_\yR$ is the area in reciprocal space enclosed by the resonance ring and $\delta E=\hbar\Omega-E_g$.) The dimensionless function $\tilde{\Theta}$ will be defined below. The contribution to the inverse of $\tilde{U}_\yc$ of type (i) above is given by $\tilde U_{\rm fb}\inv$.
For a hypothetical state with a full lower Floquet band and an empty upper Floquet band, the critical interaction strength would be equal to $\tilde U_{\rm fb}$. The contributions to $\tilde{U}_\yc\inv$  of types (ii) and (iii) are captured by the terms proportional to $\tilde U_{\rm ex}\inv$ in Eq.~\eqref{eq:CriticalU}. At finite doping, and/or with 
a finite density of electron-hole excitations, these terms reduce the critical interaction strength. In the derivation of Eq.~\eqref{eq:CriticalU} we obtain explicit expressions for these coefficients, $\tilde U_{\rm ex}=4\p^4 \dl E/\D_\yF$ and $\tilde  U_{\rm fb}=2\p^2[\log\bR{\frac{8 E_{\rm BW}\dl E}{\D^2_\yF}}-1]\inv$, where $E_{\rm BW}$ is a high-energy cutoff representing the bandwidth of the semiconductor, see SM.


The enhancement of the density of states at the ring extrema of the Floquet bandstructure affects $\tilde{U}_c$ through the terms of type (ii) and (iii) in Eq.~\eqref{eq:CriticalU}.  The unitless function $\tilde\Q(x)$ that appears in this term has the form of a ``smeared'' step function that drops to zero when its argument is negative, and saturates to $1$ in the opposite limit, with a smooth cross-over whose width is $\cO(1)$. Therefore, the contribution of type (ii) is governed by a competition between two effects: on the one hand, for this term to be significant, a small density of electrons is required. On the other hand, to achieve $\tilde\Q(\m_\ye/ T_\ye)\approx 1$ the distribution of the electrons in the upper Floquet band is required to have a sharp Fermi surface  (which is realized for $\m_\ye/ T_\ye\gg1$). 
When these conditions are met, the critical interaction strength is suppressed due to the divergence of the DOS at the ring minimum. Similar considerations hold for the contribution of type (iii) arising from holes in the lower Floquet band.




Equation~\eqref{eq:CriticalU} is a non-equilibrium analogue of the Stoner criterion \cite{Stoner1938,Stoner1939}, which gives the critical interaction strength for spontaneous symmetry breaking in the steady-state of the system.
The criterion crucially depends on the effective chemical potentials and temperatures of electrons and holes in the steady state, which are controlled by the interactions both within the system and between the system and its environment. As discussed above, when the electrons in the upper band form a low-density population with a sharp Fermi surface (such that $\tilde\Q \eqa 1$), the critical interaction strength $U_\yc$ may be reduced. Such a suppression of $U_\yc$ is particularly important for ensuring the possibility that a low-temperature symmetry-broken steady-state can arise in the non-equilibrium system, as the heating rate due to electron-electron scattering 
scales as $U^2$ [see Eq.~\eqref{eq:PhonomenologicalRates}].
In the next section, we will analyze the phase diagram of the system using both numerical simulations and further analysis based on the rate equation approach.

\begin{figure}
	\centering
	\includegraphics[width=8.6cm]{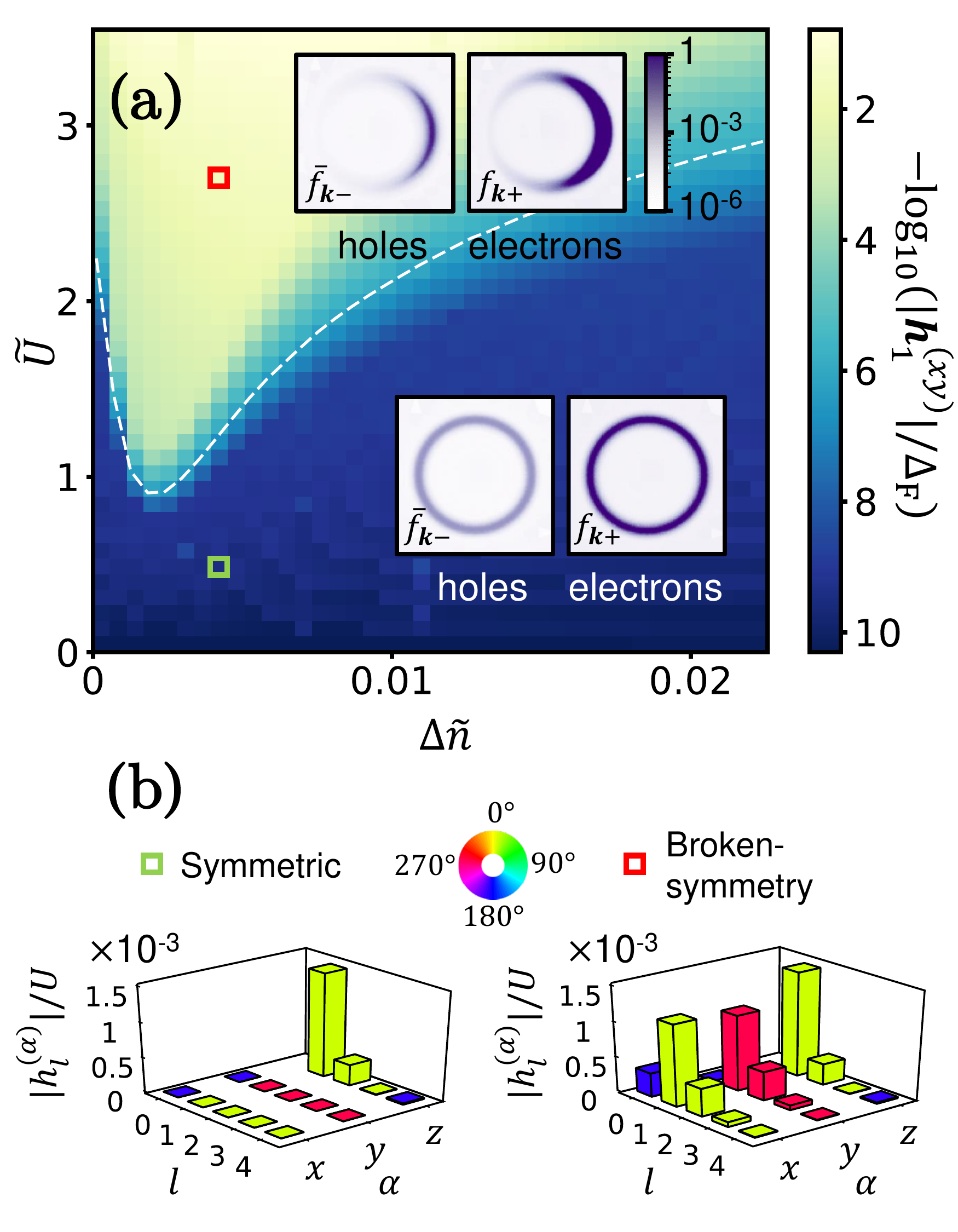}\\
	\caption{(a) Spontaneous magnetization strength, $|\vh_1^{(xy)}|$, obtained from the self-consistent mean-field calculation, as a function of a normalized electron doping, $\D \tilde n\eqv \D n/\cA_{\yR}$ and normalized interaction strength, $\tilde U\eqv \cA_\yR U/\dl E$. The dashed white line represents the phase boundary, corresponding to the critical interaction strength $\tilde U_\yc$, extracted using the same method as in Fig.~\ref{fig:CriticalU} below. The insets show the electron and hole steady-state distributions (respectively $f_{\vk+}$ and $\bar f_{\vk-}\eqv 1-f_{\vk-}$)  in the momentum domain near the resonance ring, for $\D \tilde n=0.004$, $\tilde U=0.44$, indicated by a green square, in the symmetric phase, and $\D \tilde n=0.004$, $\tilde U=2.66$, indicated by a red square, in the symmetry-broken phase. (b) Harmonics of the self-consistent magnetization $\vh(t)={\rm Re}\bS{\sum_{l,\a}  \hat \ble_\a h_l^{(\a)}  e^{il\W t}}$, where $\hat \ble_\a=\hat \vx, \hat \vy,\hat \vz$. We plot $|h_l^{(\a)}|/U$ corresponding to the first five harmonics ($l=0,1,2,3,4)$ at the two points on the phase diagram indicated by the red and green squares in panel (a). The heights and the colors of the bars respectively indicate the amplitudes and phases of the harmonics. The color scale for the phase is shown at the top of the panel. Note that we omit $|h_0^{(z)}|$, which is responsible for the bandgap renormalization of the system in the absence of the drive, see main text. \label{fig:PhaseDiagram}}
\end{figure}


\section{Phase diagram and numerical simulations}
In this section, we introduce a lattice model whose effective description for momenta near the $\G$-point is given by Eq.~\eqref{eq:HamiltonianBHZ}. Our motivation is to demonstrate symmetry breaking from a full self-consistent solution of the coupled kinetic and Floquet mean-field equations in Eqs.~(\ref{eq:MeanField})--(\ref{eq:RateEquation}). In addition, we seek to validate the suppression of  $U_\yc$ due to the enhanced density of states near the resonance ring [exhibited by the term proprotional to $\tilde{U}_{\rm ex}^{-1}$ in Eq.~\eqref{eq:CriticalU}]. To this end, we extend the Hamiltonian in Eq.~\eqref{eq:HamiltonianBHZ} to the entire Brillouin zone of a square lattice with primitive lattice vectors $\va_1=(a,0)$ and $\va_2=(0,a)$. We consider  nearest and next nearest neighbor hopping, described by the modified Hamiltonian $H_0(\vk)=\vd(\vk)\cdot \vs$, where $\vd=(d_x,d_y,d_z)$, with $d_{x(y)}(\vk)=A\sin(ak_{x(y)})+A'[\sin(\vr_1\cdot\vk)\pm\sin(\vr_2\cdot\vk)]$,
$d_z(\vk)=E_\yg/2-B[\cos(ak_x)+\cos(ak_y)-2]-B'[\cos(\vr_1\cdot\vk)+\cos(\vr_2\cdot\vk)-2]$, and Hubbard interaction $\cV_{\vq}= U/\vp$. The coefficients $A'$ and $B'$ denote the next-nearest neighbor hopping along the vectors $\vr_{1,2}=a(\hat \vx\pm\hat \vy)$. In the numerical simulations we set 
$A' = A/4$ and $B' = B/4$ \footnote[40]{Such a fine-tuning of the next-nearest neighbor hopping parameters helps to reduce lattice modulations of the resonance ring. The latter lift the degeneracy at the resonance ring, thus removing the divergence of the DOS essential for satisfying the Stoner criterion.}.
Note that the form of $H_0(\vk)$ used for the numerical simulation agrees with the Hamiltonian $H_0(\vk)$  in Eq.~\eqref{eq:HamiltonianBHZ} for momenta near the $\G$-point, with $A=2\lm_{0}/3a$, and $B=2\hb^2/3m_\ast a^2$.
We consider the case $E_\yg,B>0$ \footnote[50]{Such a choice provides non-inverted bands of the semiconductor.}, and restrict $\hb\Omega>E_\yg/2+4 B$ to ensure that $2\hbar\Omega$ is larger than the total bandwidth, such that there are no second and higher order resonances in the numerical simulation.

Using the lattice model within the mean-field approximation, we numerically solved the rate and the mean-field equations for the steady state in a self-consistent manner according to the procedure described between Eqs.~(\ref{eq:MeanField})-(\ref{eq:RateEquation}).
To this end, we 
computed the occupation function $f_{\vk \n}$ and the scattering rates $I_{\vk\n}^\ys$, $I_{\vk\n}^\ell$, and $I_{\vk\n}^{\rm ee}$ using a non-uniform grid of $8008$ points in momentum space, with enhanced resolution in the vicinity of the resonance ring. We evaluated the scattering rates using Fermi's golden rule with the electron-phonon coupling matrix $\cM_\ys(\vq_\parallel,\w)=g_\ys |\vq_{\parallel}|/\sqrt{\w}$, and electron-photon coupling matrices for two orthogonal photon polarizations, $\cM^{(1)}_\ell=g_\ell\s^{x}$ and $\cM^{(2)}_\ell=g_\ell\s^{y}$.
The densities of states for spontaneous emission of acoustic phonons ($p = s$) and photons ($p = \ell$), traced over the out-of-plane momentum, are given by 
$\ro_{p}(\w,\vq_\parallel)=\ro^0_p\, \w/\sqrt{\w^2-|v_p\vq_{\parallel}|^2}$ when $\w>v_p|\vq_{\parallel}|$ and $\ro_{p}=0$ otherwise, where $\omega$ is the frequency of the emitted phonon or photon and $\vq_\parallel$ is the in-plane component of its momentum. 
The constants $g_{\ys(\ell)}$ and $\ro^0_{\ys(\ell)}$ are material-dependent parameters. In the simulations we tune $g_{\ys(\ell)}$ and $\ro^0_{\ys(\ell)}$ to explore their roles in determining the steady states, and to effectively tune the heating parameter $\ka$ for comparison with our analytical results. In the numerical results presented in the main text, we focus on the regime $\kappa_{\rm ee}\ll \kappa_{\rm ph}$, where 
Floquet-Umklapp electron-electron scattering processes do not significantly contribute to the heating rate. 
We obtain qualitatively similar results in the regime of $\kappa_{\rm ee} \gtrsim \kappa_{\rm ph}$, see SM.


In each iteration of the algorithm, we numerically compute the 
magnetization $\vh(t)$ via Eq.~\eqref{eq:MeanField}. To improve the precision of the momentum integral, we first fit the electron and hole distributions to Fermi functions, then perform the integration using the fits interpolated to a finer grid. In the simulations, we allow for the magnetization to develop components up to the fifth harmonic of the driving frequency. As discussed below Eq.~(\ref{eq:MeanField}), we discard the constant in-time component in the $z$ direction, which simply renormalizes the parameters of the underlying band structure. 

In Fig.~\ref{fig:PhaseDiagram}a we 
show the non-equilibrium phase diagram of the system 
in the plane of doping, $\D n$, and interaction strength, $U$.
The bath parameters are fixed with values that yield $\ka_0 a^4\eqa 10^{-9}$, where the bare heating parameter, $\ka_0$, denotes the value of the heating parameter $\ka$ at $U=0$ and half-filling (see SM for details).
The color scale in Fig.~\ref{fig:PhaseDiagram}a indicates the magnitude of spontaneous magnetization, $|\vh_1^{(xy)}|$, for a lightly electron-doped system. The figure shows two distinct phases: a symmetric phase (blue), $|\vh_1^{(xy)}| = 0$, and a broken-symmetry phase (yellow), $|\vh_1^{(xy)}|>0$.

We present characteristic particle distributions well-inside of each phase in the insets to Fig.~\ref{fig:PhaseDiagram}a.
In the paramagnetic (symmetry-preserving) phase, the electron and hole populations exhibit uniform occupation of states around the resonance ring.
In the ferromagnetic nematic (broken symmetry) phase, the electron and hole populations are concentrated on one side of the resonance ring. 
The magnitudes of the harmonics of $\vh(t)$ for the same representative states in the two phases are shown in Fig.~\ref{fig:PhaseDiagram}b. Here it is evident that in the broken symmetry phase the first harmonic $\vh_1$ gives the dominant contribution, yet the DC component and second harmonic are substantial. Although present, as discussed below Eq.~\eqref{eq:Magnetization}, these harmonics do not significantly affect the Floquet mean-field band structure.

The boundary between the phases occurs at a critical interaction strength $U_\yc$. The dependence of $U_\yc$ on the doping $\Delta n$ can be explained using Eq.~\eqref{eq:CriticalU}. However, to use Eq.~\eqref{eq:CriticalU} we first need to know how $\m_{\ye(\yh)}/T_{\ye(\yh)}$ and $n_{\ye(\yh)}$ depend on $\D n$ and other parameters of the model. The electron and hole densities $n_{\ye(\yh)}$ 
found from the phenomenological rate equation treatment are given in Eq.~\eqref{eq:Densities}.
We now seek two additional equations to fix the ratios $\mu_\ye/T_\ye$ and $\mu_\yh/T_\yh$ 
for these electron and hole populations, respectively. (Recall that the same values of $n_\ye$ and $n_\yh$ can be obtained from a continuous family of values of $\mu_\ye/T_\ye$ and $\mu_\yh/T_\yh$.) Note that $\m_{\ye(\yh)}/T_{\ye(\yh)}$ and $n_{\ye(\yh)}$ depend on $U$ through their dependence on $\kappa$. Therefore, the RHS of Eq.~\eqref{eq:CriticalU} implicitly depends on $U_\yc$.  For simplicity, as in the numerical simulations that lead to Fig.~\ref{fig:PhaseDiagram}, here we focus on the case of $\kappa_{\rm ee}\ll\kappa_{\rm ph}$, where the heating parameter $\kappa$ can be treated as a $U$ independent parameter (see text below Eq.~\eqref{eq:PhonomenologicalRates} for definitions). 


Connecting back to $U_\yc$ given by Eq.~\eqref{eq:CriticalU},  recall that $\tilde{\Theta}(\m_\ye/T_\ye) = \cO(1)$ 
when $\m_\ye/T_\ye> 1$, leading to a suppression of $U_\yc$. In this situation, the population of electrons in the upper band exhibits a sharp Fermi surface. We refer to such a state as a degenerate electronic Floquet metal (EFM). Alternatively, if $\m_\ye/T_\ye<0$, the effective chemical potential lies in the Floquet gap and the electronic distribution corresponds to a non-degenerate Fermi gas. We refer to such a state as an electronic Floquet insulator (EFI). In this state,  $\tilde{\Theta}(\m_\ye/T_\ye)$ is small.

We now discuss the factors that determine the value of $\m_\ye/T_\ye$ and which phase (EFM or EFI) is achieved in the steady state.  The EFM phase is established when the intraband cooling of excited electrons is more efficient than the relaxation of electrons from the upper to the lower Floquet band. In this case, electrons excited from the lower to the upper Floquet band via Floquet-Umklapp processes quickly relax to the bottom of the upper Floquet band, where a Fermi sea is formed.  The flow of electrons into the Fermi sea in the upper Floquet band is balanced by phonon-assisted annihilation of electrons in the Fermi sea with holes in the lower Floquet band. The balance between these interband and intraband rates can be analyzed by extending the rate equation treatment, expressed in Eq.~\eqref{eq:PhonomenologicalRates}, to include an energy-resolved treatment of the electron and hole populations, see Ref.~\onlinecite{Esin2019} and SM.

Deep in the EFM phase,  and for $U\lesssim U_\yc$, the extended rate equation treatment yields $\m_\ye/T_\ye \approx x_\ye^{1/4}$, where $x_\ye\eqv \zeta \, n_\ye^{6}/ (v_\ys^{3} \ka)$, and we estimate $\zeta\approx C\hb^5/(\D_\yF m_\ast^4k_\yR^3)$, where $C$ is a constant of $\cO(10^{-3})$
(see SM for the full details). Since the EFM phase corresponds to large $\m_\ye/T_\ye$ and therefore large $x_\ye$, this phase is favored at large electron density (large doping), small sound velocity $v_s$, and low values of the heating parameter 
$\kappa$.  In particular, lower sound velocities facilitate intraband cooling, as this leads to an enhancement of the density of states for low frequency phonons.

The EFI phase is obtained in the opposite limit, where interband relaxation is more efficient than intraband cooling. In this case, the extended rate equation treatment yields $e^{\m_\ye/T_\ye} \approx {x_\ye^{1/5}}$ (see SM). The extended rate equation treatment can also be used to characterize the hole population in the lower Floquet band. We find that for electron doping, $\Delta n > 0$, the holes form a non-degenerate Fermi gas for all parameter values within our model. Fast, quasienergy conserving electron-hole scattering processes tend to equalize the electron and hole temperatures, $T_\ye$ and $T_\yh$.


\begin{figure}
	\centering
	\includegraphics[width=8.6cm]{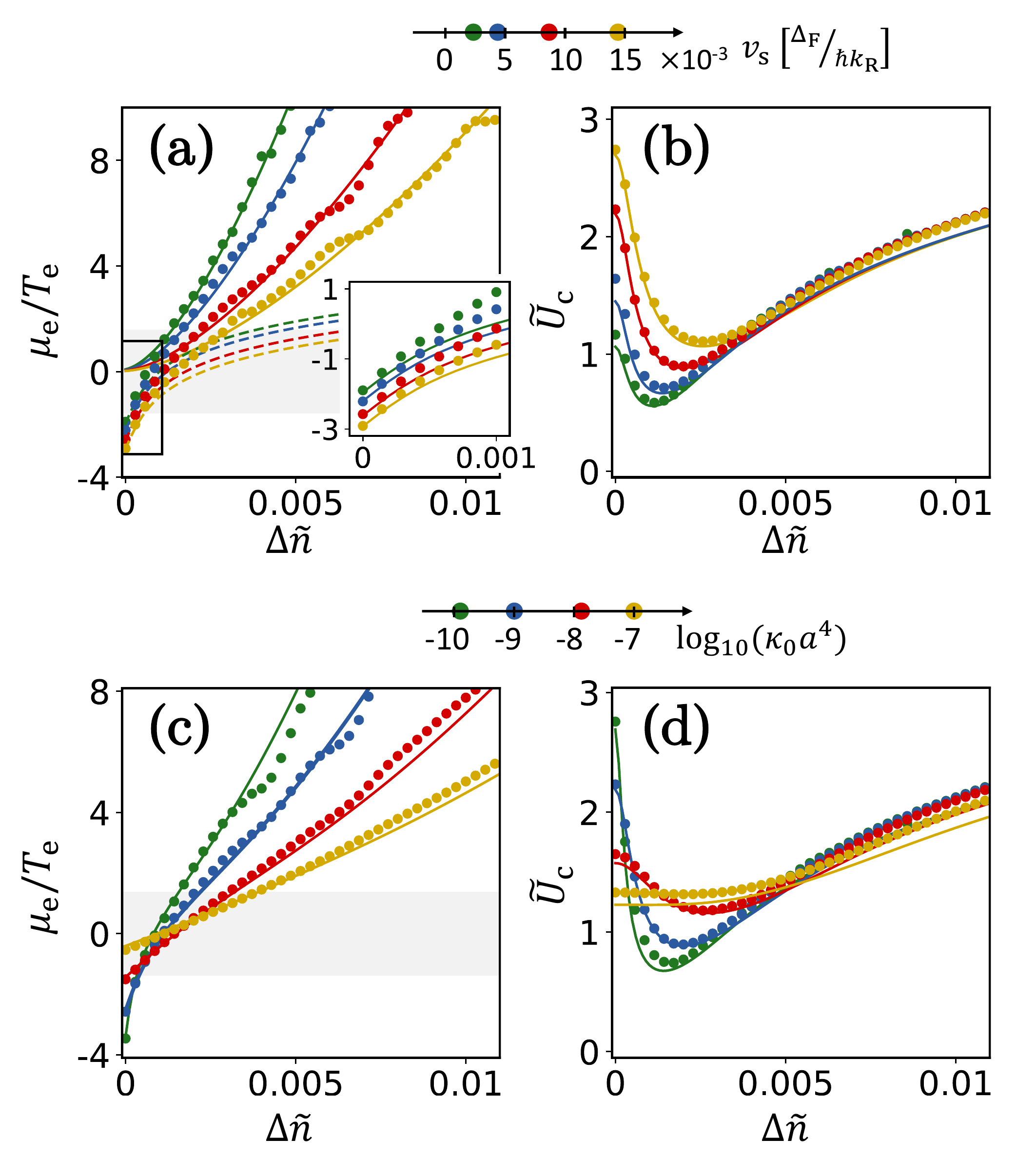}\\
	\caption{(a) 
	Numerically obtained ratio of effective chemical potential to temperature of the electronic population in the upper Floquet band, as a function of the normalized doping (data points). The data are extracted from the  steady-state solution to Eq.~\eqref{eq:RateEquation} for $\ka_0a^4\eqa 10^{-9}$ and four values of speed of sound, $v_\ys$.
	Solid and dashed lines represent the results of the extended rate equation treatment for the EFM and the EFI phases, respectively. The full set of curves is generated using the same value of the single fit parameter $\z$ (see text for definition). The shaded area indicates the EFI-to-EFM crossover range where the $\tilde \Q$-function in Eq.~\eqref{eq:CriticalU} rises from $0$ to $1$. Inset: Zoom-in on the low-doping regime (enclosed by a black frame in the main panel). Solid lines correspond to the analytical curves for the EFI phase. (b) Critical interaction strength $U_\yc$ extracted for the same data set as in panel a (data points). Solid lines represent $U_\yc$ calculated from Eq.~\eqref{eq:CriticalU}, where for the values of $\m_\ye/T_\ye$ we used a function which interpolates between the analytical results deep in the EFM and EFI phases. We use the same value of $\z$ as in panel a, and two additional fitting parameters $\tilde U_{\rm ex}$ and $\tilde U_{\rm fb}$. (c) and (d) Results for $\mu_\ye/T_\ye$ and $U_\yc$, extracted in the same manner as in panels a and b (data points), for $v_\ys=0.0086\,\D_\yF/\hb k_\yR$, and four values of $\ka_0$. Solid lines in the two panels show the interpolated values of $\m_\ye/T_\ye$ and the resulting $U_\yc$. All fitting parameters are the same as in panels a and b.
   \label{fig:CriticalU}}
\end{figure}

 Figure~\ref{fig:CriticalU} shows a comparison of $\m_\ye/T_\ye$ and $U_\yc$ extracted from numerical simulations (data points), to the analytical estimates obtained from the extended rate equations discussed above. We obtain $U_\yc$ using a numerical analogue of the procedure leading to Eq.~\eqref{eq:CriticalU}. Specifically, we compute the expectation value on the RHS of  Eq.~\eqref{eq:MeanField}, using the steady state obtained from Eq.~\eqref{eq:RateEquation} for a system whose electronic Hamiltonian corresponds to  $\hat{\mathcal{H}}_{\rm MF}(t)$ in Eq.~\eqref{eq:MeanFieldHamiltonian}. In this procedure for obtaining $U_\yc$, we use a prescribed form of $\vh(t)$ with a single non vanishing harmonic $\vh^{(xy)}_1$ of small magnitude in $\hat{\mathcal{H}}_{\rm MF}(t)$, see Eq.~\eqref{eq:Magnetization}.
The dashed white line in Fig.~\ref{fig:PhaseDiagram}a shows $U_\yc$ extracted using the above procedure on top of the phase diagram obtained from the full self-consistent 
numerical simulations for the same parameters.

In Fig.~\ref{fig:CriticalU}a  we show $\mu_\ye/T_\ye$ as a function of doping $\Delta n$ for several values of phonon sound velocity $v_\ys$. We extract $\m_\ye$ and $T_\ye$ from the numerical simulations described in the previous paragraph by fitting the electron steady state distribution in the upper Floquet band to a Fermi function with respect to the quasienergies of the mean-field Hamiltonian [Eq.~\eqref{eq:MeanFieldHamiltonian}].
The  fit lines correspond to the analytical forms for $\mu_\ye/T_\ye$ obtained from the extended rate equation treatment. The solid and dashed lines correspond to the forms for $\mu_\ye/T_\ye$ in the EFM and EFI regimes,
respectively. The only freedom in these fits is the parameter $\zeta$ in the definition of $x_\ye$, which was given the same value across all of the curves shown. The extracted value of $\zeta$ is of the same order of magnitude as the analytical estimate given above.

In Fig.~\ref{fig:CriticalU}c we again show $\m_\ye/T_\ye$ as a function of $\Delta n$, this time highlighting the dependence on the value of the bare heating 
parameter $\kappa_0$.
In this plot, the fit lines are given by a function that interpolates between the analytical results for the asymptotic behavior in the EFM and EFI regimes: $\cF_j\left[\lambda(\m_\ye/T_\ye)\right]=  x_\ye^\h$, where $\cF_j$ is the complete Fermi-Dirac integral, $\h=0.174$,  $j=-0.3$, and $\lambda=0.871$. These parameter values are fixed by demanding that $\m_\ye/T_\ye$ displays the correct dependence on $x_\ye$ deep in the EFI $(e^{\m_\ye/T_\ye}\sim x_\ye^{1/5})$ and EFM $(\m_\ye/T_\ye\sim x_\ye^{1/4})$ phases. The value of $\z$ used in $x_\ye$ is the same as used in Fig.~\ref{fig:CriticalU}a.


In Figs.~\ref{fig:CriticalU}b and d we show $U_\yc$ as a function of $\Delta n$ for different values of $v_\ys$ and $\kappa_0$. The data points are obtained from the  numerical procedure discussed above. To obtain the fit lines, we use the interpolated values of $(\m_\ye/T_\ye)$ in Eq.~\eqref{eq:CriticalU}.  We additionally use $\tilde U_{\rm ex}$ and $\tilde U_{\rm fb}$ as fitting parameters. The same values of these parameters were used in all curves shown. The values used for the fits are close to those obtained from the formulas given below Eq.~\eqref{eq:CriticalU}.  For the contribution of the holes, we used the same interpolating function, with $x_\yh$ replacing $x_\ye$. Here $x_\yh$ is defined in the same manner as $x_\ye$, but with $n_\yh$ replacing $n_\ye$ (with the same value of the parameter $\z$). Note that holes are in the analogue of the EFI phase for any $\Delta n>0$, and hence the value of $\tilde\Theta(\m_\yh/T_\yh)$ is small throughout the regime studied.

As is evident in the phase diagram in Fig.~\ref{fig:PhaseDiagram}a, $U_\yc$ obtains a minimal value at an optimal value of the doping, which we denote $\D n_\ast$. Using the extended rate equation treatment, we estimate $\D n_\ast= C_*(v_\ys^3 \ka/\z)^{1/6}$, where $C_*$ is a constant of $\cO(1)$, see SM. The corresponding minimal interaction strength is given by $\tilde U_\yc^{\rm min}\approx \tilde U_{\rm ex}\D  n_\ast/\cA_\yR$ for $\ka\ll k_{\rm R}^4$. For $\D n>\D n_\ast$, electrons in the upper band are in the EFM phase, and exhibit a sharp Fermi surface (note the corresponding values of $\mu_\ye/T_\ye$ in Fig.~\ref{fig:CriticalU}a). As explained below Eq.~\eqref{eq:CriticalU}, the existence of a Fermi surface tends to reduce the critical interaction strength. However, as the doping increases, the phase transition requires stronger interactions as the density of states at the Fermi surface decreases. Below the optimal doping, $\D n<\D  n_\ast$, the electrons in the upper Floquet band are in the EFI regime, which has no Fermi surface. Thus, the suppression of $U_\yc$ is lost for $\D n<\D n_\ast$.

\section{Discussion}
In this work, we demonstrated a mechanism for realizing  electronic liquids crystals in two-dimensional electronic systems through time-periodic driving. The phase that we find exhibits ferromagnetic-nematic order associated with spontaneous breaking of U(1) symmetry in both pseudospin and orbital degrees of freedom. Above the critical interaction strength, the Fermi sea becomes highly anisotropic and occupies a limited sector of the ring minimum of the Floquet bands, see Figs.~\ref{fig:Bands} and \ref{fig:PhaseDiagram}. Due to pseudospin-momentum locking, the Fermi sea in the symmetry broken phase exhibits a
finite magnetization which rotates with the frequency of the drive [cf. Eq.~\eqref{eq:Magnetization}].

Our analysis has been carried out
on a model system with a two-component psuedospin degree of freedom, whose bandstructure is described by $H_0(\vk)$, see Eq.~\eqref{eq:HamiltonianBHZ}. For $E_\yg>0$, the model $H_0(\vk)$ lacks time reversal symmetry. The model can also be taken to describe half of the degrees of freedom of a time-reversal invariant system.  Our analysis can be straightforwardly extended to include the relevant time-reversal partner degrees of freedom.  In this situation, there are many more possibilities for how the system may order. As one example, in the SM we describe a mean-field treatment which shows an instability towards an order in which the magnetizations of the two time-reversal partners are aligned, which yields a breaking of the time-reversal symmetry. We leave a more elaborate study of this interplay for the future investigation.

For simplicity, throughout the paper we considered a driving whose form is described below Eq.~\eqref{eq:HamiltonianBHZ}. It is interesting to consider driving with circularly polarized light, which, like the drive we studied, preserves the $\yU(1)$ symmetry of the system and uniformly opens a gap all the way around the resonance ring.  Depending on the handedness of the drive, the pseudospin may wind twice or zero times around the resonance ring \cite{Lindner2011}. In the case of double winding, each direction of the pseudospin in the $x$-$y$ plane corresponds to two momentum points on the resonance ring. Therefore, in this case, an analogous ferromagnetic-nematic phase would exhibit two electron pockets occupying opposite sectors of the ring minimum.

To put our results in an experimentally relevant context, we estimate the value of the heating parameter $\ka$, employing the definition of $\ka$ appearing below Eq.~\eqref{eq:PhonomenologicalRates}. We base our estimate on typical scattering rates measured in semiconductors \cite{Sundaram2002}. We start with the phonon assisted interband scattering rate, which we estimate by $\Lm_{\rm inter}\eqa (\ta_{\ys}\mathcal{A}_{\rm BZ})^{-1}$, where $\ta_\ys$  is a typical hot-electron  scattering lifetime due to phonons, $\sim 0.1\ \rm ps$, and $\mathcal{A}_{\rm BZ}$ is the reciprocal-space area of the Brillouin zone. In addition, electron-phonon scattering also contributes to heating through  Floquet-Umklapp scattering processes.  We estimate the phonon-mediated excitation rate by $\G_{\ys}\eqa (V/\hb\W)^2 \cA_\yR/\ta_{\ys}$. The area inside the resonance ring $\cA_\yR$ appears due to the form of electron-acoustic phonon coupling and the band inversion 
of the Floquet bands.
Radiative recombination predominantly occurs between states with inverted band indices, i.e., inside the resonance ring. We estimate this rate by $\G_\ell\eqa\cA_{\yR}/\ta_{\ell}$, where $\ta_{\ell}$ is a typical time for the radiative recombination evaluated as $\sim 1 \ \rm ns$. Assuming $V/\hb\W\eqa 10^{-2}$ and $\cA_\yR /\mathcal{A}_{\rm BZ}\eqa 10^{-3}$, we estimate $\ka_{\rm ph}=(\G_\ys+\G_\ell)/\Lm_{\rm inter}\eqa 10^{-7}\mathcal{A}_{\rm BZ}^2$.



Next, we estimate the electron-hole pair generation (heating) rate $\G_{\rm ee}$ due to photon-assisted electron-electron scattering. To lowest order in $(V/\hb\W)^2$, these  processes predominantly excite a pair of electrons from the lower to the upper Floquet band, accompanied by the absorption of one photon from the driving field.
Due to the pseudospin structure of the Floquet states the dominant scattering processes involve one electron that is scattered from the interior of the resonance ring to the exterior, and vice versa for the other electron. Therefore, we expect the scattering rate to be proportional to the squared area of the resonance ring. After averaging over initial and final momenta, we estimate  $\G_{\rm ee}(U)=\frac{\cA_\yR^2 m_\ast U^2}{2\p^4\hb^3} \bR{\frac{V}{\hb\W}}^2$. The critical interaction strength is not significantly changed when $\ka_{\rm ee}(U_{\rm fb})\ll\ka_{\rm ph}$, where $\ka_{\rm ee}(U)=\G_{\rm ee}(U)/\Lm_{\rm inter}$ (see SM for supporting numerical simulations). When the above condition applies, and for $U\lesssim U_{\rm fb}$, the effective temperatures of the electron and hole distributions are dominated by the electron-phonon and electron-photon interactions. The estimate for $\G_{\rm ee}$, when $\cA_\yR$ is small, shows that for realistic parameter choices, the above conditions can be indeed satisfied.
We note that short-range interactions corresponding to this regime can be obtained using screening gates placed near the 2D electronic system.

The phenomenon we discussed can be realized in 2D Dirac systems such as transition metal dichalcogenides and semiconductor quantum wells. To ensure that Floquet-Umklapp processes are suppressed, it is beneficial to use large bandgap materials. In gapless Dirac system such as graphene, driving may induce similar ring-like Floquet-band extrema \cite{Usaj2014}. We leave the exploration of particle dynamics and symmetry breaking in such systems to future studies. The use of periodic driving to create ring extrema in Floquet bands may be utilized to study exotic  phases of fermions and bosons in cold atom systems \cite{Kitagawa2010,Sedrakyan2015a,Wintersperger2020}. In particular, it would be interesting to investigate the possibility to use buffer gases in cold atom systems to serve as the heat baths needed for stabilizing the  broken-symmetry phases
discussed in this work.

\begin{acknowledgments}
\section{Acknowledgments}
We would like to thank Ehud Altman, Vladimir Kalnizky, Gil Refael, and Ari Turner for illuminating discussions and David Cohen and Yan Katz for technical support.
N. L. acknowledges support from the European Research Council (ERC) under the European Union Horizon 2020 Research and Innovation Programme (Grant Agreement No. 639172), and from the Israeli Center of Research Excellence (I-CORE) ``Circle of Light''. M. R. gratefully acknowledges the support of the European Research Council (ERC) under the European Union Horizon 2020 Research and Innovation Programme (Grant Agreement No. 678862) and the Villum Foundation. M. R. and E. B. acknowledge support from CRC 183 of the Deutsche Forschungsgemeinschaft. G. K. G. acknowledges support from Israel Council for Higher Education.
\end{acknowledgments}

\bibliography{Bibliography.bib}

\end{document}


\title{Electronic Floquet Liquid Crystals - Supplemental Material}

\maketitle

\renewcommand\theequation{S\arabic{equation}}
\renewcommand\thefigure{S\arabic{figure}}
\renewcommand\thetable{S\arabic{table}}
\setcounter{equation}{0}
\setcounter{figure}{0}
\setcounter{table}{0}

This supplemental material provides detailed analytical derivations discussed in the paper and explains key aspects of the numerical simulations. Throughout the notes we work in natural units in which $\hb=k_B=1$.

\section{Effective two-band Floquet Hamiltonian}\label{sec:RWAHamiltonian}
Here we derive an effective time-independent Hamiltonian in the rotating wave approximation (RWA), $H_{\rm RWA}(\vk)$ for the periodically driven system described by Eq.~(1) in the main text, without the interaction part. Our goal is to find the effective dispersion and eigenstates near the resonance ring.
The time-dependent Hamiltonian is given by $H(\vk,t)=H_0(\vk)+H_\yd(t)$, where $H_0(\vk)=E_0+\vd(\vk)\cdot \vs$, with
\Eq{
\vd(\vk)=\bR{\lm_0 k_x,\lm_0k_y,\frac{|\vk|^2}{2m_\ast}+\frac{E_\yg}{2}},
\label{eq:FreeHamiltonian}
}
and $H_\yd(t)=\vv\cdot \vs \cos(\W t)$, where $\vv=(0,0,V)$. In the RWA, we shift the lower (upper) bands by $\pm\W/2$ and then neglect rapidly oscillating terms. The operator that shifts the energies reads
\Eq{
\cU_{\yR}(\vk,t)=e^{\frac{i\W t}{2} \nd (\vk)\cdot \vs},
}
where $\nd(\vk)$ is a unit vector in the direction $\vd(\vk)$.
The transformed Hamiltonian, $H_{\rm R}(\vec{k}, t)$, then reads
\Eq{
H_{\rm R}(\vec{k}, t) = \cU_{\rm R}(\vec{k}, t) H(\vec{k,t}) \cU^\dagger_{\rm R}(\vec{k}, t) - i \cU_{\rm R} \frac{\partial \cU^\dagger_{\rm R}}{\partial t}.
}
Note that $\cU_{\yR}$ commutes with $H_0$, but does not commute with the time-derivative and the driving field. Using the explicit form of $H(\vk,t)$, we arrive at
\Eq{
H_{\yR}=E_0+(|\vd|-\W/2)\nd\cdot\vs+\vv'(t)\cdot \vs \cos(\W t),
\label{eq:RWAprime}
}
where $\vv'(t)=\vv \cos(\W t)-\nd\times\vv \sin(\W t)+\nd(\nd\cdot \vv)(1-\cos(\W t))$, following from the Rodrigues' formula of rotation of $\vv$ around $\nd$. The last term in Eq.~\eqref{eq:RWAprime} has constant in time parts and parts oscillating with the frequency $\W$ and $2\W$. Following RWA, we omit the time-oscillating terms, leading to
\Eq{
H_{\rm RWA}(\vk)=E_0+\vd_\yF(\vk)\cdot \vs,
\label{eq:RWAHamiltonian}
}
where $\vd_\yF=(|\vd|-\W/2)\nd+\half[\vv-\nd(\nd\cdot \vv)]$. The spectrum of $H_{\rm RWA}$ then reads, $\ve_{\vk\pm}=E_0\pm \ve_\vk$, where
\Eq{
\ve_\vk=\sqrt{(|\vd|-\W/2)^2+|\vv-\nd(\nd\cdot \vv)|^2/4}.
}
For simplicity of notations, throughout we set $E_0=0$. To leading order in $V/\W$ and $\lm_0^2 m/E_\yg$, the minimum of the upper band is positioned at the resonance ring, given by $|\vd(\vk)|=\W/2$. The radius of the resonance ring is given by $k_\yR=\sqrt{m_\ast\dl E}$ and the quasienergy gap at $|\vk|=k_\yR$ reads $\D_\yF=2V\lm_0k_\yR/\W$. We thus approximate the quasienergy as
\Eq{
\ve_{\vk}\eqa \sqrt{\bR{\frac{k^2}{2m_\ast}-\frac{\dl E}{2}}^2+\bR{\frac{\D_\yF k}{2k_\yR}}^2},
\label{eq:ApproximateEnergy}
}
where $k\eqv|\vk|$.

The eigenstates of $H_{\rm RWA}$ read
\Eq{
\ket{\f_{\vk\pm}'}=\frac{1}{\sqrt{2}}\bR{\sqrt{1\pm\nz\cdot \nd_\yF}\ket{\aup}\mp e^{i\q_\vk}\sqrt{1\mp \nz\cdot \nd_\yF}\ket{\adn}}.
\label{eq:EigenstatesRWA1}
}
where $\q_\vk=\tan\inv(k_y/k_x)$ and $\nd_\yF$ is a unit vector in the direction of $\vd_\yF$, defined in Eq.~\eqref{eq:RWAHamiltonian}.


At the resonance ring, $\nd_\yF$ approximately lies in the $x-y$ plane such that $\nz\cdot \nd_\yF\eqa 0$, therefore
\Eq{
\ket{\f_{\vk{\pm}}'}\at{|\vk|=k_\yR}=\frac{1}{\sqrt{2}}(\ket{\aup}\mp e^{i\q_\vk}\ket{\adn}).
\label{eq:EigenstatesRWA2}
}
To find the time-dependent eigenstates of $H(\vk,t)$ within the RWA in the lab frame, we apply the inverse transformation, $\cU_{\yR}\dg$ on the states, $\ket{\f_{\vk{\pm}}'}$. Approximating, $\cU_{\yR}\dg\eqa e^{-\frac{i\W t}{2}\s^z}$, we arrive at
\Eq{
\ket{\f_{\vk{\pm}}(t)}\at{|\vk|=k_\yR}=\frac{1}{\sqrt{2}}(e^{-i\W t}\ket{\aup}\mp e^{i\q_\vk}\ket{\adn}).
\label{eq:FloquetStatesResonance}
}

\section{Derivation of $U_\yc$}
In this section we solve the mean-field equation [Eq.~(2) in the main text] in the limit $\vh(t)\to 0$.
The solution in this limit provides the minimal interaction strength required for non-zero magnetization, $U_\yc$ [Eq.~(10) in the main text]. In our analysis we assume contact interactions, $\cV_\vk=U/\vp$, and take only the first harmonic of the circularly polarized magnetization field,
\Eq{
\vh(t)=h_1e^{i\W t}(\nx-i\ny)/2 +c.c.,
\label{eq:MFmanetization}
}
as discussed in the main text [see Eq.~(8)], where $h_1$ is real and positive. In the rotating frame of reference such a term reads $\cU_{\yR}(\vh(t)\cdot \vs) \cU_{\yR}\dg=h_1 \s^x$.

We first rewrite Eq.~(2) in the main text in terms of the single particle states and the distribution function, $f_{\vk\n}$,
\Eq{
\vh(t)=-U\sum_{\n=\pm}\int \frac{d^2\vk}{(2\p)^2} \braoket{\f_{\vk\n}(t)}{\vs}{\f_{\vk\n}(t)}f_{\vk\n}.
\label{eq:SelfConst0}
}
We approximate $\ket{\f_{\vk\n}(t)}$ by eigenstates of the rotating-wave approximated mean-field Hamiltonian [Eq.~(3) in the main text] in the lab frame, given by $H_{\rm MF,RWA}(\vk,t)=\cU_\yR\dg H_{\rm MF,RWA}'(\vk) \cU_\yR$, where
\Eq{
H_{\rm MF,RWA}'(\vk)= H_{\rm RWA}(\vk)+h_1\s^x;
\label{eq:RWAMFHamiltonian}
}
$H_{\rm RWA}(\vk)$ is given in Eq.~\eqref{eq:RWAHamiltonian}.

Next, we transform Eq.~\eqref{eq:SelfConst0} given in a vector form, into a scalar equation for a single mode amplitude, $h_1$, defined in Eq.~\eqref{eq:MFmanetization}. To this end, we transform the Floquet states into a rotating frame of reference, leading to
\Eq{
\vh(t)=-U\sum_{\n=\pm}\int \frac{d^2\vk}{(2\p)^2} \braoket{\f_{\vk\n}'}{\cU_\yR\vs\cU_\yR\dg}{\f_{\vk\n}'}f_{\vk\n}.
\label{eq:SelfConst1}
}
Here  $\ket{\f'_{\vk\n}}\eqv\cU_\yR(\vk,t)\ket{\f_{\vk\n}(t)}$ are the eigenstates of $H_{\rm MF,RWA}'(\vk)$, given in Eq.~\eqref{eq:RWAMFHamiltonian}.
Using the Rodrigues' formula, we find,
$\cU_{\rm R}(\vec{k},t)\vs \cU_{\rm R}\dg(\vec{k},t)=\vs \cos(\W t)+\nd\times\vs \sin(\W t)+\nd(\nd\cdot\vs)(1-\cos(\W t))$. We extract only the components proportional to $e^{i\W t}$, and multiply both sides of Eq.~\eqref{eq:SelfConst1} by $(\nx+i\ny)/\sqrt{2}$. Approximating  $\nd\eqa\nz$, we find
\Eq{
\frac{h_1}{\sqrt{2}}=-U\sum_{\n=\pm}\int \frac{d^2\vk}{(2\p)^2} \braoket{\f_{\vk\n}'}{\frac{\s^x+i\s^y}{\sqrt{2}}}{\f_{\vk\n}'}f_{\vk\n},
\label{eq:SelfConst2}
}
where $f_{\vk\n}$ is the particle distribution function. As $H_{\rm MF,RWA}'(\vk)$ is symmetric to reflections of the $y$-axis, the expectation value of $\s^y$ vanishes, leading to
\Eq{
h_1=-U\sum_{\n=\pm}\int \frac{d^2\vk}{(2\p)^2} \braoket{\f_{\vk\n}'}{\s^x}{\f_{\vk\n}'}f_{\vk\n}.
\label{eq:SelfConst2a}
}


To evaluate the integral in Eq.~\eqref{eq:SelfConst2a}, in what follows we assume a low-excitation steady-state corresponding to almost-full lower Floquet band and low density of electron and hole excitations in the bottom of the upper and top of the lower Floquet bands, respectively. We then split the integral into two contributions, $h_1=h_1^{\rm fb}+h^{\rm ex}_1$. Here, $h^{\rm fb}_1$ is the contribution of the full lower Floquet band,
\Eq{h_1^{\rm fb}=-U\int \frac{d^2\vk}{(2\p)^2}\braoket{\f'_{\vk-}}{\s^x}{\f'_{\vk-}},
\label{eq:SelConstFs}
}
and $h^{\rm ex}_1$ is the contribution of the electron and hole excitations,
\Eq{
h_1^{\rm ex}=-U\sum_{\n}\n\int \frac{d^2\vk}{(2\p)^2}\braoket{\f'_{\vk\n}}{\s^x}{\f'_{\vk\n}}\dl f_{\vk\n},
\label{eq:SelConstEx}
}
where $\dl f_{\vk+}\eqv f_{\vk+}$ corresponds to electrons and $\dl f_{\vk-}\eqv 1-f_{\vk-}$ to holes. 

We begin with $h_1^{\rm ex}$. Recall that the Floquet states and quasienergies result from the solution of the mean-field Hamiltonian [Eq.~\eqref{eq:RWAMFHamiltonian}], and therefore depend (implicitly) on $h_1$. As we are interested in the regime near the critical value ($U\sim U_\yc$), where $h_1$ is small, we expand the states $\{\ket{\phi'_{\vec{k}\nu}}\}$ in the integrand in Eq.~\eqref{eq:SelConstEx} in powers of $h_1$ around $h_1=0$. The zeroth-order term in $h_1$ is proportional to $\braoket{\f'_{\vk\n}}{\s^x}{\f'_{\vk\n}}\dl f_{\vk\n}|_{h_1=0}$. This contribution vanishes, as it consists of the momentum integral over an odd-parity function (arising from the symmetry to reflections of the $x$-axis, exhibited by $H_{\rm MF,RWA}'$ at $h_1=0$).
The contribution to linear order in $h_1$ includes two terms: the first one is proportional to $h_1\!\left[\frac{\dpa\braoket{\f'_{\vk\n}}{\s^x}{\f'_{\vk\n}}}{\dpa h_1}\dl f_{\vk\n}\right]_{h_1=0}$ and the second to $h_1\!\left[\frac{\dpa\dl f_{\vk\n}}{\dpa h_1}\braoket{\f'_{\vk\n}}{\s^x}{\f'_{\vk\n}}\right]_{h_1=0}$. In the limit of low filling, the momentum integral over the first term is proportional to the density of particles in the upper band, while the momentum integral over the second term is proportional to the inverse of the density (see below). Therefore, in the low-density limit, considered throughout, we neglect the first term (arising from the dependence of the eigenstates on $h_1$) with respect to the second one (which captures the change of the distribution function due to $h_1$).

In order to compute $\frac{\dpa\dl f_{\vk\n}}{\dpa h_1}|_{h_1=0}$, we approximate the distribution function by the Fermi function
\Eq{
\dl f_{\vk\pm}\eqa[1+e^{(\ve_{\vk}-\D_\yF/2-\m_{\ye/\yh})/T_{\ye/\yh}}]\inv.
\label{eq:FermiFunction}
}
Here we used the particle-hole symmetry of the system, where $\ve_\vk$ is the dispersion relation of the upper Floquet band [cf. Eq.~\eqref{eq:ApproximateEnergy} for $h_1=0$ case]. Note that the effective temperature, $T_{\ye/\yh}$, and the effective chemical potential, $\m_{\ye/\yh}$, must be even functions of $h_1$, as the setting $h_1\to-h_1$ inverts the position of the band minimum in the momentum space, but does not change the overall energetics of the system. 
Therefore, the linear-order dependence of $\dl f_{\vk\n}$ on $h_1$ results predominantly from the dependence of $\ve_\vk$ on $h_1$. We find this dependence using first-order perturbation theory, $\ve_{\vk}=\ve_{0,k}+h_1\braoket{\f'_{0,\vk+}}{\s^x}{\f'_{0,\vk+}} +\cO(h_1^2)$, where $\ket{\f'_{0,\vk\n}}\eqv\ket{\f'_{\vk\n}}|_{h_1=0}$, and $\ve_{0,k}\eqv \ve_\vk|_{h_1=0}$. Finally, using the chain rule, we arrive at $\frac{\dpa\dl f_{\vk\n}}{\dpa h_1}|_{h_1=0}=\frac{\dpa\dl f^0_{k\n}}{\dpa \ve_{0,k}}\braoket{\f'_{0,\vk+}}{\s^x}{\f'_{0,\vk+}}$, where $\dl f^0_{k\n}\eqv\dl f_{\vk\n}|_{h_1=0}$ [see Eq.~\eqref{eq:FermiFunction}]. To make the notations more transparent, throughout we distinguish between rotation-symmetric functions (dependent only on the momentum amplitude) with index $k$, and functions of momentum amplitude and angle with index $\vk$.

We substitute the result of the expansion in small $h_1$ back into Eq.~\eqref{eq:SelConstEx}, yielding
\Eq{
h_1^{\rm ex}=-h_1U\sum_{\n}\int \frac{d^2\vk}{(2\p)^2} \frac{\dpa \dl f^0_{k\n}}{\dpa \ve_{0,k}} |\braoket{\f'_{0,\vk+}}{\s^x}{\f'_{0,\vk+}}|^2 .
\label{eq:SelfConst3}
}
To evaluate the integral in Eq.~\eqref{eq:SelfConst3}, we 
use polar coordinates $\vec{k} = k(\cos\theta, \sin\theta)$, yielding
\Eq{
h_1^{\rm ex}=-h_1U\sum_{\n}\int \frac{k dk}{2\p} \frac{\dpa \dl f^0_{k\n}}{\dpa \ve_{0,k}} \overline{\s^2_k} ,
\label{eq:SelfConst4}
}
where $ \overline{\s^2_k} \eqv \int \frac{d\q}{2\p}|\braoket{\f'_{0,\vk+}}{\s^x}{\f'_{0,\vk+}}|^2$. To further simplify the expression, we approximate $\overline{\s^2_k}$ by its value at the resonance ring, $\overline{\s^2_{k_\yR}}$. Corrections to this approximation lead to higher order terms in the density of excitations, and hence are small in the low-density limit. We  transform the integral over the magnitude of the momentum $k$, in Eq.~(\ref{eq:SelfConst4}) to an integral over energy $\ve$, 
by introducing the density of states $D_\yF(\ve)\eqv \int \frac{d^2\vk}{(2\p)^2} \dl(\ve-\ve_{0,k}+\D_\yF/2)$. These transformations lead to
\Eq{
h_1^{\rm ex}\eqa -h_1U \overline{\s^2_{k_\yR}}\sum_{\n=\pm}\int_{0}^{\infty} d\ve D_\yF(\ve)\dpa_\ve \dl f^0_{\ve\n},
\label{eq:SelfConst5}
}
where $\dl f_{\ve\pm}^0=[1+e^{(\ve-\m_{\ye/\yh})/T_{\ye/\yh}}]\inv$, cf. Eq.~\eqref{eq:FermiFunction}.
We estimate the density of states near the band bottom by
\Eq{
D_\yF(\ve)\eqa D_0\sqrt{\D_\yF/\ve},
\label{eq:DOS}
}
where $D_0$ is a constant depending on the parameters of the Floquet bands. In terms of the bare parameters of the model, we evaluate $D_0= m_\ast/2\p$. We use the eigenstates given in Eq.~\eqref{eq:EigenstatesRWA2} to evaluate, $\overline{\s^2_{k_\yR}}=\half$.

\begin{figure}
  \centering
  \includegraphics[width=8.6cm]{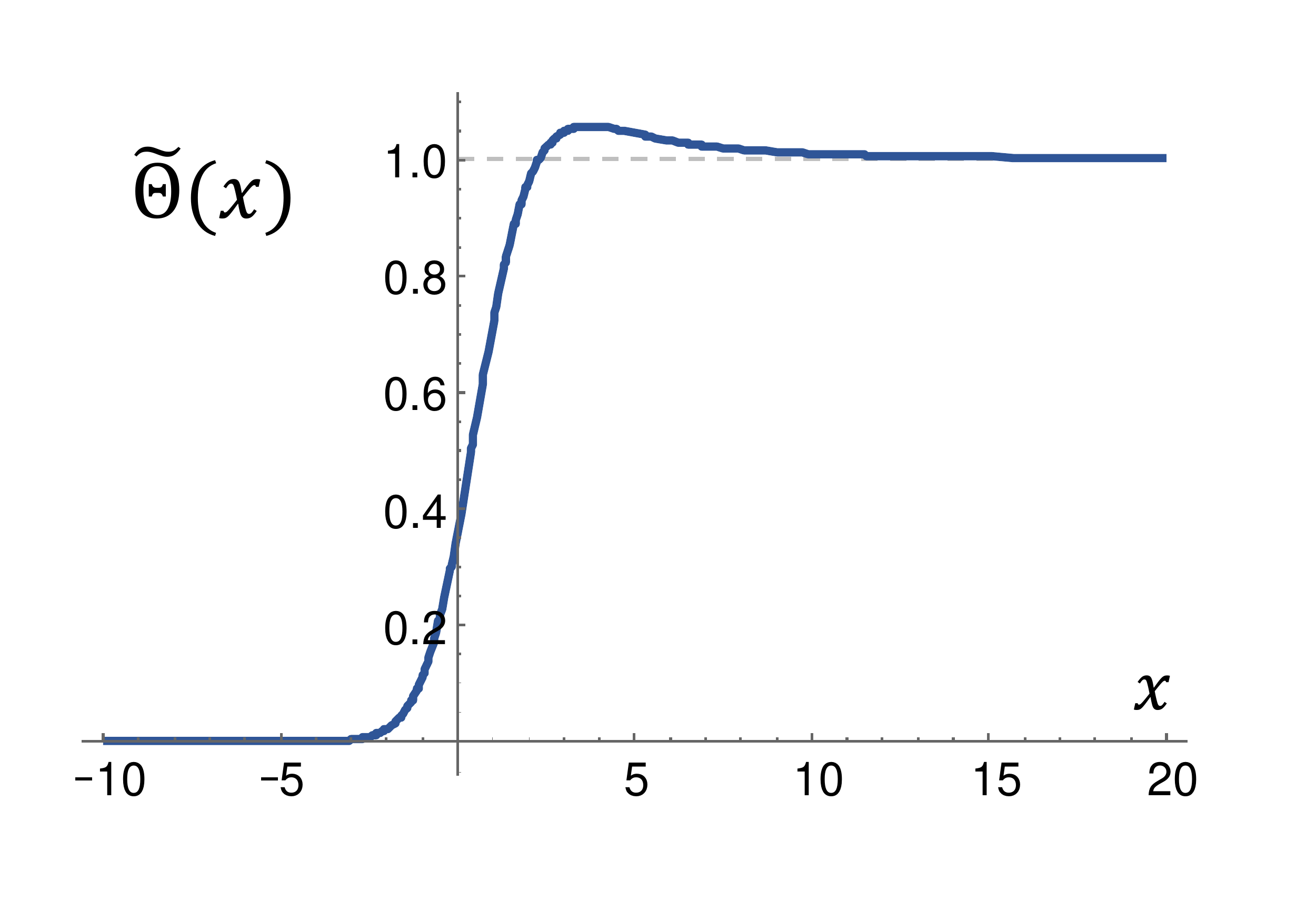}\\
  \caption{A plot of the smeared step function $\tilde \Q(x)\eqv\frac{\p}{2}{\rm Li}_{-1/2}(-e^x){\rm Li}_{1/2}(-e^x)$, appearing in Eq.~(10) in the main text.
 \label{fig:ThetaFPlot}}
\end{figure}

To perform the energy integral in Eq.~\eqref{eq:SelfConst5}, we define a dimensionless integration variable. For $\n=+$ term, we define $x=\ve/T_\ye$. The integral then reads
$\int_0^\infty d\ve D_\yF(\ve)\dpa_\ve\dl f_{\ve+}^0= D_\yF(T_\ye)\int_0^\infty \frac{dx}{\sqrt{x}}\dpa_x \frac{1}{1+e^{x-(\m_\ye/T_\ye)}}=\sqrt{\p}D_\yF(T_\ye){\rm Li}_{-1/2}(-e^{\m_\ye/T_\ye})$, where ${\rm Li}_{s}(z)$ is the polylogarithm function\cite{Lewin1981}. To bring this result into the form of Eq.~(10), we replace the term proportional to the temperature by $\sqrt{T_\ye}=-n_\ye/[\sqrt{\p\D_\yF}D_0{\rm Li}_{1/2}(e^{-\m_\ye/T_\ye})]$. The last relation follows from the definition of the electron density $n_{\ye}=\int d\ve D_\yF(\ve) \dl f^0_{\ve+}=-\sqrt{\p}T_\ye D_\yF(T_\ye){\rm Li}_{1/2}(e^{-\m_\ye/T_\ye})$. We repeat the same calculation for the lower band ($\n=-$), to arrive at
\Eq{
h_1^{\rm ex}= h_1 U U_{\rm ex}\inv \bR{\frac{\tilde \Q(\m_\ye/T_\ye)}{n_\ye/\cA_\yR}+\frac{\tilde \Q(\m_\yh/T_\yh)}{n_\yh/\cA_\yR}}.
\label{eq:SelfConst6}
}
Here $U_{\rm ex}\inv=2 \overline{\s^2_{k_\yR}} \D_\yF D_0^2/\cA_\yR$ and $\tilde \Q(x)\eqv\frac{\p}{2}{\rm Li}_{-1/2}(-e^x){\rm Li}_{1/2}(-e^x)$, see Fig.~\ref{fig:ThetaFPlot}.

Now, we turn to the evaluation of $h_1^{\rm fb}$ given in Eq.~\eqref{eq:SelConstFs}. In contrast to the integral in Eq.~\eqref{eq:SelConstEx}, which is limited to an area in $k$-space near the resonance ring, the integral in Eq.~\eqref{eq:SelConstFs} is defined over the entire Brillouin zone. The dependence on $h_1$ arises from the dependence of Floquet states $\{\ket{\f'_{\vk\n}}\}$ on $h_1$. Using first order perturbation theory, we find $\ket{\f'_{\vk-}}=\ket{\f'_{0,\vk-}}-h_1\frac{\braoket{\f'_{0,\vk+}}{\s^x}{\f'_{0,\vk-}}}{2\ve_{k}}\ket{\f'_{0,\vk+}}$. Employing this expansion of the Floquet states, we express the linear order term in $h_1$ of Eq.~\eqref{eq:SelConstFs} as
\Eq{
h_1^{\rm fb}=h_1 U\int \frac{d^2\vk}{(2\p)^2} \frac{|\braoket{\f'_{0,\vk-}}{\s^x}{\f'_{0,\vk+}}|^2}{\ve_{k}}.
\label{eq:SelConstFs2}
}
This integral is independent of the steady-state distribution [by the definition, see Eq.~\eqref{eq:SelConstFs}]. We denote the value of this integral by $U_{\rm fb}\inv=\int \frac{d^2\vk}{(2\p)^2} |\braoket{\f'_{0,\vk-}}{\s^x}{\f'_{0,\vk+}}|^2/\ve_{k}$.

To evaluate $U_{\rm fb}$, we first note that the mean-field Hamiltonian, $H_{\rm MF}(\vk,t)$ (see Eq.~(3) in the main text) describes the system only near the $\G$-point, while the integral in the definition of $U_{\rm fb}$ is over the entire Brillouin zone. To address this issue, we impose a cutoff in the integral at the momentum $\Lm$, as a stand-in for the Brillouin zone edge. We evaluate the wavefunctions using Eq.~\eqref{eq:EigenstatesRWA1}, where we approximate $\nz\cdot \nd_\yF\eqa (k^2/2m_\ast-\dl E/2)/\ve_{k}$, with $\ve_{k}$ given by Eq.~\eqref{eq:ApproximateEnergy}.
Substituting in Eq.~\eqref{eq:SelConstFs2} we find
\Eq{
U_{\rm fb}\inv=\frac{m_\ast}{2\p}\bS{\log\bR{\frac{4E_{\rm BW}}{\dl E-k_\yR^2/m_\ast}}-1},
}
where we denote $E_{\rm BW}=\ve_k|_{k=\Lm}$.

Summing Eqs.~\eqref{eq:SelfConst6} and \eqref{eq:SelConstFs2} we arrive at
\Eq{
h_1=h_1 U \bS{ U_{\rm ex}\inv \bR{\frac{\tilde\Q(\m_\ye/T_\ye)}{n_\ye/\cA_\yR}+\frac{\tilde\Q(\m_\yh/T_\yh)}{n_\yh/\cA_\yR}}+U_{\rm fb}\inv}.
\label{eq:SelfConst7}
}
Note that $h_1$ appears on the both sides of Eq.~\eqref{eq:SelfConst7}. Dividing by $h_1$ and by $U$, we obtain the expression for $U_\yc$ given in Eq.~(10) in the main text.

The calculation outlined in this section helps to find the critical value of $U$ by expanding the self-consistent equation [Eq.~(2) in the main text] to the linear order in the order parameter. Expanding this equation to higher orders in $h_1$ reveals how the order parameter grows as a function of $U$.

\section{The extended rate model\label{sec:ExtendedRateModel}}
In this section, we give a detailed description of the extended rate model. The results of the model are used to fit the numerical data of $\m_\ye/T_\ye$ in Fig.~3 in the main text and to estimate the heating rates in the discussion section. We partially follow the analysis of Ref.~\onlinecite{Esin2019}. Throughout this section, we consider the ``paramagnetic phase'', with no spontaneous symmetry breaking, corresponding to $U\le U_\yc$. 
In our analysis, we assume that the system reached a steady state with a low-density of electrons ($n_\ye$) in the upper Floquet band (UFB) and holes ($n_\yh$) in the lower Floquet band (LFB), $n_\ye,n_\yh\ll\cA_\yR$. We verified numerically that the distributions of each of the bands can be well approximated by Fermi functions with effective parameters, see Fig.~\ref{fig:FitToFD}.

\begin{figure}
  \centering
  \includegraphics[width=8.6cm]{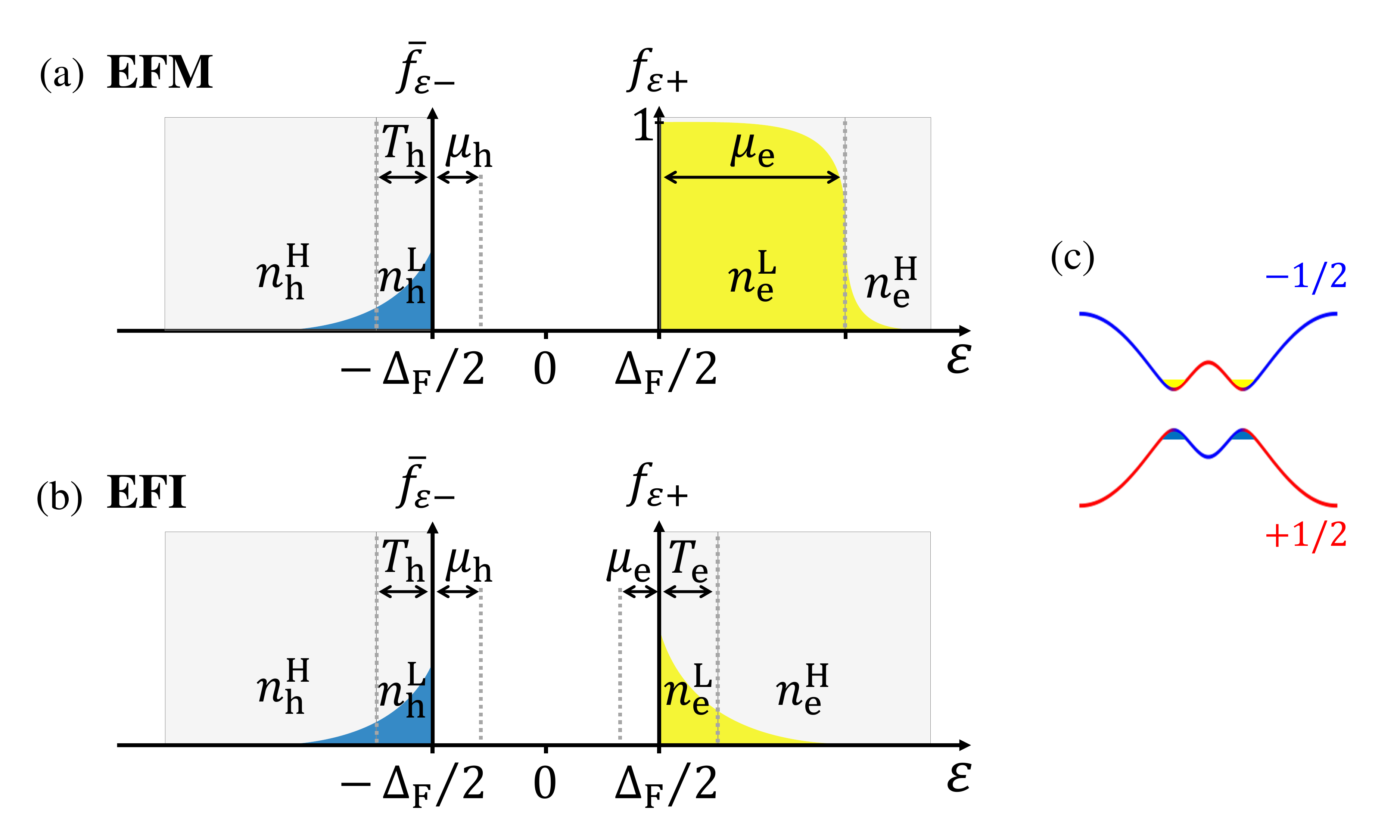}\\
  \caption{Population at the bottom of the UFB and top of the LFB in the EFM and EFI phases. The yellow area indicates population of electrons and the blue area of holes. (a). Distribution in the EFM phase. The electrons exhibit a sharp Fermi-surface at the effective chemical potential $\m_\ye$. We denote densities of subpopulations occupying levels below and above the Fermi surface by $n_\ye^\yL$ and $n_\ye^\yH$, respectively. (b). Distribution of electrons and holes in the EFI phase. The effective chemical potential of the electrons is in the Floquet gap, giving rise to a Maxwell distribution, given by Eq.~\eqref{eq:MBDistributionApprox_elect}. In this case, the low and high-quasienergy populations occupy states below and above the effective temperature. The population of holes in an electron-doped system is always in the non-degenerate phase described by Eq.~\eqref{eq:MBDistributionApprox_hole}. (c). Schematic drawing of the Floquet bands. The red/blue colored sectors indicates the character of the original valence/conduction bands. The color code helps to visually estimate the power $n$ of the factor $(V/\W)^{2n}$ multiplying a scattering rate for the $l$th order Floquet Umklapp process. The power is given by $n=|l+m_i-m_f|$, where $m_i=1/2$ when the state before the scattering is in the red sector and $m_i=-1/2$ when it is in the blue sector, similarly, $m_f$ denotes the sector after the scattering. 
  \label{fig:ExtendedRateModelPop}}
\end{figure}

Here we consider only an electron-doped case ($\D n>0$); the analysis of the hole-doped system is similar due to particle-hole symmetry of the Hamiltonian. We distinguish between two cases: (i) The electrons form a degenerate Fermi gas ($\m_\ye/T_\ye\gg1$) featuring two concentric Fermi surfaces. We refer to this case as the electron Floquet metal (EFM) phase (see Fig.~\ref{fig:ExtendedRateModelPop}). In this phase we evaluate energy integrals using the Sommerfeld expansion \cite{Ashcroft2003}.
(ii) Non-degenerate distribution of electrons ($\m_\ye/T_\ye<0$) referred to as the electron Floquet insulator (EFI) phase. In this phase, we approximate the distribution by the Maxwell's law
\Eq{
f_{\ve+}=z_\ye e^{-\ve/T_\ye},
\label{eq:MBDistributionApprox_elect}
}
where $z_\ye=e^{-|\m_\ye|/T_\ye}$ is the electron fugacity and $\ve$ is accounted from the UFB bottom.
The two phases are separated by a crossover regime, $\m_\ye\eqa T_\ye$, where our analysis does not apply.
Yet, as we show below, an analytic interpolation between the EFM and the EFI phases gives a good agreement with the numerical data (see Fig.~3 in the main text). We also show below that for the system considered in this paper, the holes always form a non-degenerate distribution in the electron-doped system. Therefore we approximate the distribution of holes by $\bar f_{\ve-}\eqv 1-f_{\ve-}$, where
\Eq{
\bar f_{\ve-}=z_\yh e^{\ve/T_\yh}.
\label{eq:MBDistributionApprox_hole}
}
Here $\ve$ is accounted from the LFB top and  $z_h=e^{-|\m_\yh|/T_\yh}$ is the hole fugacity.

In what follows, we seek four equations for four variables, $\m_\ye$, $\m_\yh$, $T_\ye$, and  $T_\yh$.
Two of the equations [Eqs.~\eqref{eq:Doping} and \eqref{eq:ContinuityNe}] determine the total densities of electrons and holes, $n_\ye$ and $n_\yh$ (identical to the derivation of Eq.~(7) in the main text). The other two equations [Eqs.~\eqref{eq:RateEqUFB} and \eqref{eq:RateEqLFB}], are rate equations for the densities of subpopulations (defined below) of the electron and hole distributions. The rate equations include the key processes leading to the steady-state distribution due to electron-phonon, electron-photon and electron-electron interactions, see Fig.~\ref{fig:ExtendedRateModel}. Our goal is to find the electron and hole densities and the densities of their subpopulations as functions of the rates in the steady state, and use them to find the chemical potentials and temperatures of the hole and electron distribution.
The total densities are related to the parameters of the distribution through
\Eq{
n_\ye=\int \frac{d^2\vk}{(2\p)^2} f_{\vk+}\quad , \quad n_\yh=\int \frac{d^2\vk}{(2\p)^2} (1-f_{\vk-}).
\label{eq:TotalDensities}
}
We perform the momentum integral, when the electronic population is deep in the EFM phase, to estimate
\Eq{
n_\ye\eqa 2\m_\ye D_\yF(\m_\ye),
\label{eq:DensityEFM}
}
where $D_\yF$ is given in Eq.~\eqref{eq:DOS}. In the EFI phase, we perform the integral in Eq.~\eqref{eq:TotalDensities} using the Maxwell's distribution [see Eq.~\eqref{eq:MBDistributionApprox_elect}], leading to
\Eq{
n_\ye\eqa z_\ye \sqrt{\p} T_\ye D_\yF(T_\ye).
\label{eq:DensityEFI}
}
Likewise, we estimate the hole density [using Eqs.~\eqref{eq:TotalDensities} and \eqref{eq:MBDistributionApprox_hole}] by
\Eq{
n_\yh\eqa z_\yh \sqrt{\p} T_\yh D_\yF(T_\yh).
\label{eq:DensityHoles}
}
Later in this section, we use the estimates of the densities in terms of the temperature and chemical potential to find the electron and hole temperatures as functions of the parameters of the electronic dispersion in the material, the periodic drive, and the heat-baths. 

\subsection{Equations determining the steady-state densities of the electrons and holes}


Our first equation relates the total densities of electrons and holes with the doping,
\Eq{
n_\ye-n_\yh=\D n.
\label{eq:Doping}
}
While the doping sets the difference between the densities of electrons and holes, it does not resolve the density of each of them. Indeed, even at zero doping, non-zero densities of electrons and holes can be created by the drive due to excitation processes. In fact, the populations of the bands are determined by the balance between the processes transferring electrons from the LFB to the UFB (which we will refer to as heating processes), and processes transferring particles from the UFB to LFB (which we will refer to as cooling processes). The heating processes arise from electron-photon, electron-phonon and electron-electron scattering. We capture the effect of each of these processes by a single parameter $\G_\ell$, $\G_{\ys}$, and $\G_{\rm ee}$, cf. Fig.~\ref{fig:ExtendedRateModel}a-c, which we estimate in Secs.~\ref{sec:PotonHeating}-\ref{sec:AugerHeating}.

The interband cooling processes are dominated by electron-hole recombination assisted by phonons. We expect the recombination rate to be proportional to the densities of electrons, $n_\ye$ and holes, $n_\yh$  (cf. Fig.~\ref{fig:ExtendedRateModel}d). Therefore, we estimate $\dot n_\ye|_{\rm cool}=-\Lm_{\rm inter}n_\ye n_\yh$, see Sec.~\ref{sec:PononCooling} for the evaluation of $\Lm_{\rm inter}$. The interplay of these processes is captured by the rate equation for $n_\ye$,
\Eq{
\dot n_\ye=\G_{\ell}+\G_{\ys}+\G_{\rm ee}-\Lm_{\rm inter}n_\ye n_\yh,
\label{eq:ContinuityNe}
}
satisfying $\dot n_\ye=0$ in the steady state. Note that from the conservation of particles, $\dot n_\yh=-\dot n_\ye$.
In the steady state ($\dot n_\ye=0$), Eq.~\eqref{eq:ContinuityNe} yields
\Eq{
n_\ye n_\yh=\ka,
\label{eq:DefinitionofKa}
}
where we defined,
\Eq{
\ka\eqv \frac{\G_{\ell}+\G_\ys+\G_{\rm ee}}{\Lm_{\rm inter}}.
\label{eq:Balance_parameter}
}
Combining Eqs.~\eqref{eq:Doping} and \eqref{eq:DefinitionofKa}, we arrive at  the total densities of electrons and holes in the UFB and LFB,
\Eq{
n_{\ye/\yh}=\sqrt{(\D n/2)^2+\ka}\pm \D n/2,
\label{eq:ElectronHoleDensities}
}
similar to Eq.~(7) in the main text.

\subsection{Equations determining the effective temperatures of the electron and hole distributions \label{sec:RateEquations}}
Eq.~\eqref{eq:ElectronHoleDensities} yields the total density of electrons and holes, but provides no information on how the particles are distributed within each of the bands. In order to obtain the quasienergy-resolved structure of the distribution in the UFB, we split the electronic population to two subpopulations \cite{Esin2019}. We define a subpopulation occupying low-quasienergy levels close to the bottom of the UFB as
\Eq{
n_\ye^\yL=\int_{\D_\yF/2}^{\D_\yF/2+E_{\ys}} D_\yF(\ve-\D_\yF/2)f_{\ve+}d\ve
\label{eq:DefinitionN_eL}
}
where $D_\yF$ is the density of states, given in Eq.~\eqref{eq:DOS}. We furthermore define a complementary subpopulation corresponding to electrons in quasienergy levels not included in Eq.~\eqref{eq:DefinitionN_eL}. The density of this subpopulation is given by $n_\ye^\yH=n_\ye-n_\ye^\yL$.
In the EFM, we choose $E_{\ys}=\m_\ye$ and in the EFI $E_{\ys}=T_\ye$, see Figs.~\ref{fig:ExtendedRateModelPop}a and b for an illustration.
A different choice of $E_{\ys}$ in the EFI phase, leads to renormalization of the coefficients in the phenomenological model (see below), but does not change the final result. The splitting into subpopulations makes it possible to study intraband processes in terms of rate equations for the densities of the subpopulations.
It is also convenient to define the density of unoccupied states in the low-quasienergy subset of the UFB, denoted by $\bar n_\ye^\yL$. The definition of $\bar n_\ye^\yL$ is similar to Eq.~\eqref{eq:DefinitionN_eL}, where we replace $f_{\ve+}$ by $1- f_{\ve+}$.
When the electrons are deep in the EFM phase, we evaluate their subpopulations [using Eq.~\eqref{eq:DefinitionN_eL}] by
\Eq{
n_\ye^\yH\eqa \bar n_\ye^\yL\eqa \log(2) T_\ye D_\yF(\m_\ye) \,\, {\rm and} \,\, n_\ye^\yL\eqa n_\ye,
\label{eq:EstimateSubpopEFM}
}
where $n_\ye$ is given in Eq.~\eqref{eq:DensityEFM}.
In turn, when the electrons exhibit the EFI phase, we evaluate
\Eq{
n_\ye^\yL \eqa {\rm erf(1)} n_\ye, \,\, n_\ye^\yH\eqa {\rm erfc(1)} n_\ye \,\, {\rm and} \,\, \bar n_\ye^\yL\eqa 2T_\ye D_\yF(T_\ye),
\label{eq:EstimateSubpopEFI}
}
where ${\rm erf(1)}\eqa 0.843$, ${\rm erfc(1)=1-erf(1)}$ and $n_\ye$ is given in Eq.~\eqref{eq:DensityEFI}.

Likewise, we split the distribution of holes in the LFB into two subpopulations. A subpopulation near the top of the LFB within a quasienergy window $\ve\in[-\D_\yF/2-T_\yh,-\D_\yF/2]$. The density of this subpopulation is given by
\Eq{
n_\yh^\yL=\int_{-\D_\yF/2-T_\yh}^{-\D_\yF/2} D_\yF(-\ve-\D_\yF/2)(1-f_{\ve-})d\ve,
\label{eq:DefinitionN_hL}
}
and the complementary subpopulation, whose density is $n_\yh^\yH=n_\yh-n_\yh^\yL$.
We also define the density of populated states near the top of the LFB, $\bar n_\yh^\yL$, defined by Eq.~\eqref{eq:DefinitionN_hL} with $1-f_{\ve-}$ replaced by $f_{\ve-}$.
Explicitly performing the integral in Eq.~\eqref{eq:DefinitionN_hL} on the distribution of holes [given in Eq.~\eqref{eq:MBDistributionApprox_hole}], we evaluate
\Eq{
n_\yh^\yL \eqa {\rm erf(1)} n_\yh, \,\, n_\yh^\yH\eqa {\rm erfc(1)} n_\yh \,\, {\rm and} \,\, \bar n_\yh^\yL\eqa 2T_\yh D_\yF(T_\yh),
\label{eq:EstimateSubpopHoles}
}
where $n_\yh$ is given in Eq.~\eqref{eq:DensityHoles}.

In what follows, we express $\dot n^\yL_\ye$ and $\dot n^\yL_\yh$ in terms of the incoming and outgoing rates, and find a balance between them in the steady state by requiring $\dot n_\ye^\yL=0$ and $\dot n_\yh^\yL=0$.
We begin with the equation for $\dot n_\ye^\yL$.
The subpopulations of the UFB are subjected to intraband relaxation processes transferring particles between high- and low-quasienergy sectors within the band. Essentially, these are phonon-assisted electron-hole pair annihilation processes with the rate estimated by $\dot n_\ye^\yL|_{\rm relax}=\Lm_{\rm intra}(T_\ye)n_\ye^\yH\bar n_\ye^\yL$ (see Fig.~\ref{fig:ExtendedRateModel}e). As we show in Sec.~\ref{sec:PononThermalization}, the relaxation rate in the UFB depends on the temperature of the electrons.
The incoming rate of electrons into the low-quasienergy sector is balanced by the phonon-assisted interband recombination rate, serving as a sink of electrons from this sector. The interband recombination rate is proportional to the density of electrons in the low-quasienergy sector in the UFB and density of holes in the LFB, with approximately the same coefficient $\Lm_{\rm inter}$, as the total interband relaxation appearing in Eq.~\eqref{eq:ContinuityNe}. We thus estimate the interband recombination rate by $\dot n_\ye^\yL|_{\rm inter}=-\Lm_{\rm inter}n_\ye^\yL n_\yh$.



Our treatment also includes equilibration between electron and hole distributions by  electron-electron scattering. Electron-electron scattering tends to equalize the temperatures of the populations in the UFB and LFB, without changing their total densities (see Fig.~\ref{fig:ExtendedRateModel}f). As the population of each of the bands is assumed to be in a local equilibrium, which is well-described by the Fermi-function, scattering of two electrons in the same band can not affect it. To account for the interband equilibration, it is convenient to define the average temperature of the UFB and LFB populations  $T=(T_\ye+T_\yh)/2$ and their temperature difference, $\D T=T_\ye-T_\yh$.

First, assume that the electrons and holes have the same temperature ($\D T=0$). In this case, the thermalization rate is zero, as the populations are already in equilibrium. Next, consider a small temperature difference of the populations, $\D T$. If $\D T>0$, the electron-electron scattering induces electron-hole pair annihilation in the UFB and electron-hole pair creation in the LFB, effectively cooling down the UFB and heating up the LFB. In contrast, when $\D T<0$, electron-hole pairs are created in the UFB and annihilated in the LFB. When the thermalization is significant, we expect the  temperatures of the two bands to be almost the same, i.e., $|\D T|\ll T$.
Such pair creation and annihilation in the two bands occurs in a quasienergy window of width $\sim T$.
Notice that in the numerical simulation (Figs.~2 and 3 in the main text), the equilibration between electron and hole distributions due to electron-electron scattering was not taken in account. We later justify this (see Sec.~\ref{sec:SolutionExtendedRate2}) by showing that this effect does not significantly affect the majority population.

Following from the above arguments, we expect the electron-electron scattering rate to be linear in $\D T$ to the leading order. We estimate the equilibration rate by $\dot n_\ye^\yL|_{\rm ee}=\g_{\rm ee}\D T n_\ye^\yH\bar n_\ye^\yL n_\yh^\yH \bar n_\yh^\yL$. This rate serves as a source of particles for the low-quasienergy sector in the UFB when $\D T>0$, and a sink when $\D T<0$. It is proportional to the densities in the UFB of occupied states (above $E_\ys$) and unoccupied states (below $E_\ys$) available for annihilation, and densities in the LFB of occupied states (above $-\D_\yF/2-T_\yh$) and unoccupied states (below $-\D_\yF/2-T_\yh$) available for a pair creation. A further justification for this form will be discussed in Sec.~\ref{sec:AugerThermalization}, where we estimate the value of $\g_{\rm ee}$ in terms of the system's parameters. 
Combining all the terms above, we arrive at the full rate equation for $n_\ye^\yL$, which reads
\Eq{
\dot n_\ye^\yL=\Lm_{\rm intra}(T_\ye) n_\ye^\yH \bar n_\ye^\yL-\Lm_{\rm inter} n_\ye^\yL n_\yh+\g_{\rm ee}\D T n_\ye^\yH \bar n_\ye^\yL  n_\yh^\yH \bar n_\yh^\yL.
\label{eq:RateEqUFB}
}

Our fourth equation describes thermalization and relaxation in the LFB, expressed through an equation for $\dot n_\yh^\yL$.
The intraband and interband rates are estimated by $\dot n_\yh^\yL|_{\rm intra}=\Lm_{\rm intra}(T_\yh)n_\yh^\yH \bar n_\yh^\yL$ and  $\dot n_\yh^\yL|_{\rm inter}=-\Lm_{\rm inter} n_\yh^\yL n_\ye$. Here we approximate $\Lm_{\rm inter}$ and $\Lm_{\rm intra}$ by the same coefficients as in Eq.~\eqref{eq:RateEqUFB} evaluated at the hole temperature $T_\yh$, as follows from the particle-hole symmetry of the Hamiltonian.
In analogy to the discussion above Eq.~\eqref{eq:RateEqUFB}, we estimate the electron-electron thermalization rate by $\dot n_\yh^\yL|_{\rm ee}=-\g_{\rm ee}\D T n_\yh^\yH\bar n_\yh^\yL n_\ye^\yH \bar n_\ye^\yL$. 
Combining all the incoming and outgoing rates we arrive at the full rate equation for $n_\yh^\yL$, which reads
\Eq{
\dot n_\yh^\yL=\Lm_{\rm intra}(T_\yh)n_\yh^\yH \bar n_\yh^\yL-\Lm_{\rm inter} n_\yh^\yL n_\ye-\g_{\rm ee}\D T n_\yh^\yH\bar n_\yh^\yL n_\ye^\yH \bar n_\ye^\yL.
\label{eq:RateEqLFB}
}

In order to solve Eqs.~\eqref{eq:RateEqUFB} and \eqref{eq:RateEqLFB} for the chemical potentials and effective temperatures in the steady state [for total densities $n_\ye$ and $n_\yh$ fixed by  Eq.~\eqref{eq:ElectronHoleDensities}], we first need to estimate how the densities and the coefficients appearing in these equations (such as $\Lm_{\rm inter}$, $\Lm_{\rm intra}$ and $\g_{\rm ee}$) depend on the parameters of the steady state.
In the sections below, we perform these estimations. In Sec.~\ref{sec:SolutionExtendedRate12}, we extract the expressions for the chemical potentials and temperatures as a function of $\ka$, $\D n$ and the parameters of the system.

\subsection{Evaluation of the rates and Floquet-Fermi's golden rule}
\begin{figure}
  \centering
  \includegraphics[width=8.6cm]{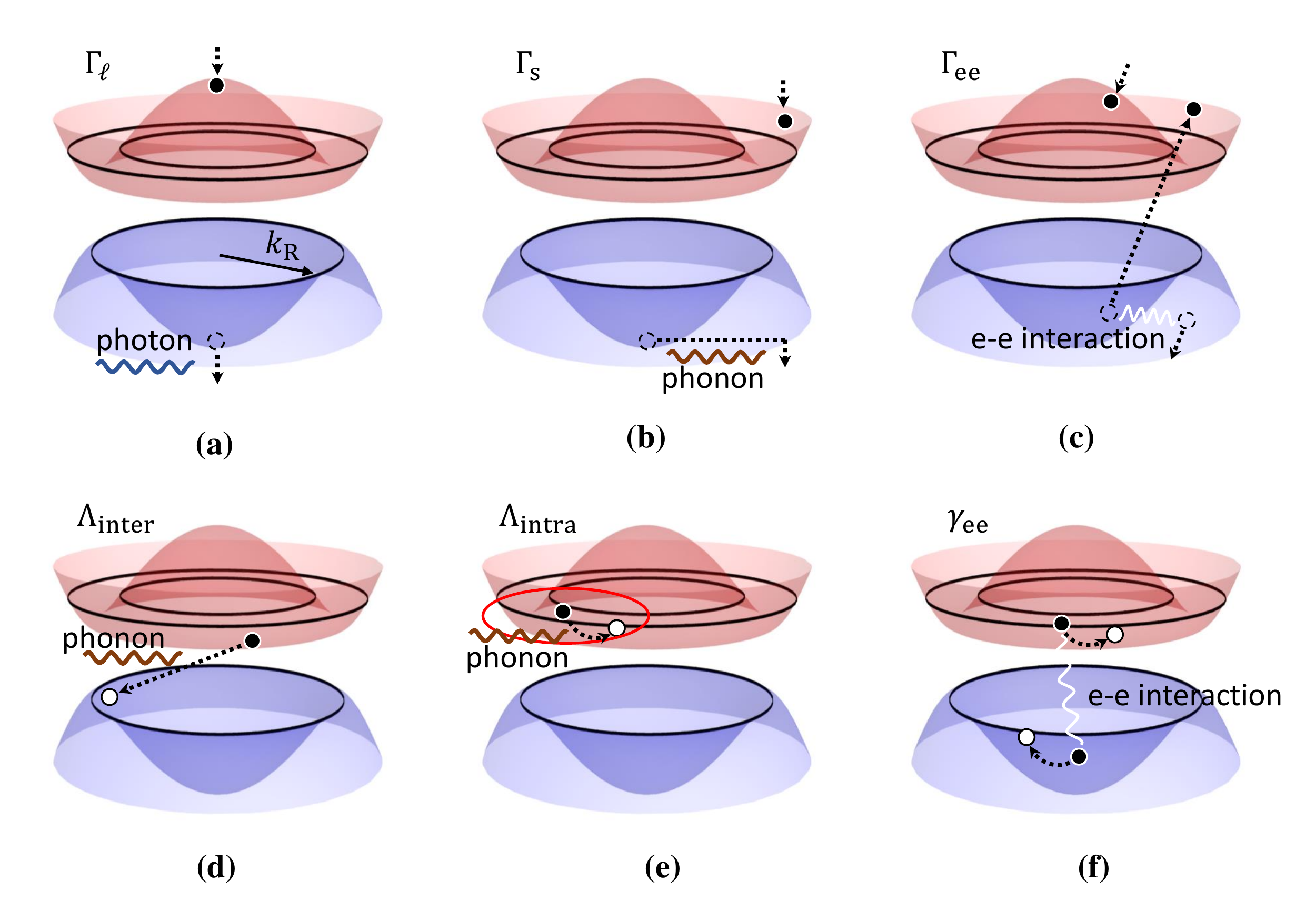}\\
  \caption{ Key processes contributing to the steady-state distribution in the nearly symmetric case, $U\lesssim U_\yc$. We consider here an EFM phase for the electrons in the UFB, associated with two concentric Fermi surfaces (represented by a double black ring), and a non-degenerate distribution of holes in the LFB, with a maximum of the distribution shown by a single black ring. The heating processes are summarized in panels (a),(b), and (c); Panels (d),(e), and (f) demonstrate cooling and thermalization processes. (a). Photon-mediated excitation processes, dominant inside the resonance ring, with approximately zero-momentum transfer. (b). Phonon-mediated excitations, predominantly transferring electrons between states inside and outside the resonance ring. (c). Floquet-Auger heating, corresponding to a transition of two electrons occupying the interior and exterior of the resonance ring, to states outside and inside the resonance ring, respectively (see Fig.~\ref{fig:InteractionProcesses}). (d). Phonon-assisted interband recombination of electrons occupying the Fermi sea in the UFB and holes in the LFB. Assuming a large Floquet gap ($\D_\yF\gg 2k_\yR v_\ys$) the phonon density of states supports scattering between states in the UFB and LFB of any two points close to the resonance ring. (e). Phonon-assisted intraband relaxation. In the low-temperature limit, $T<2k_\yR v_\ys$, the phonon density of states only allows for small-momentum transitions with a momentum transfer below $\sim T/v_\ys$. Therefore, the allowed states to scatter into are within a circle of radius $\sim T/v_\ys$ centered around the initial electron position (indicated by the red circle)
  (f). Intraband thermalization by electron-electron interactions. The main role of these processes is equilibration of temperatures of the electron and hole distributions. The total density of particles in each of the bands after the collision is preserved.
 \label{fig:ExtendedRateModel}}
\end{figure}

Before we evaluate the scattering rates for the processes presented in Fig.~\ref{fig:ExtendedRateModel}, we first review the Fermi's golden rule for transitions between Floquet states \cite{Seetharam2015, Rudner2020}. For brevity, we use a notation $\vk\n$ to indicate a state with momentum $\vk$ and Floquet band $\n$.

We first discuss the rate for scattering of an electron between states $\vk\n$ and $\vk'{\n'}$ due to a collision with a phonon ($p=\ys$) or photon ($p=\ell$) [resulting from the coupling Hamiltonian in Eq.~(5) in the main text]. We assume the phonon bath is at zero temperature. The scattering rate is given by
\EqS{
&(\dot f_{\vk\n})_{p,{\vk'{\n'}}}=2\p \sum_l \cP_p^{(l)}(\vk\n,\vk'{\n'})\times\\
&[\ro_p(\vq,\w_l)F(\vk'{\n'},\vk\n)-\ro_p(\vq,-\w_l)F(\vk{\n},\vk'{\n'})],
\label{eq:FermiGoldenRule1}
}
where
$\cP^{(l)}_{p}(\vk\n,\vk'{\n'})=\abs{\sum_m\braoket{\f_{\vk'\n'}^{m+l}}{\cM_p(\vq,\w_l)}{\f_{\vk\n}^{m}}}^2$ and
\Eq{
F(\vk\n,\vk'{\n'})=f_{\vk\n}\bar f_{\vk'\n'},
\label{DefinitionOfF}
}
where $\bar f\eqv 1-f$.
The in-plane momentum and energy transfers are given by $\vq=\vk'-\vk$ and $\w_l=\ve_{\vk'\n'}-\ve_{\vk\n}+l\W$. We consider the following electron-phonon coupling
\Eq{
\cM_\ys(\vq,\w)=g_\ys |\vq|/\sqrt{\w}\I,
\label{eq:ES_Coupling}
}
where $\I$ is the identity matrix, and consider electron-photon coupling of two polarizations,
\Eq{
\cM_\ell^{(1)}=g_\ell\s^{x} \quad, \quad \cM_\ell^{(2)}=g_\ell\s^{y}.
\label{eq:EL_Coupling}
}
We take the density of states of three-dimensional phonons as a function of the in-plane momentum $\vq$ (the in-plane momentum of the phonon is the momentum transfer of the electron), $\ro_\ys(\w,\vq)=\ro_\ys^0\w/\sqrt{\w^2-v_\ys^2|\vq|^2}$ when $\w>v_\ys|\vq|$ and zero otherwise. For the heat bath of photons, we consider $\ro_\ell(\w,\vq)=\dl^{(2)} (\vq) \ro^0_\ell$.

The scattering rate due to electron-electron interaction depends on occupations of four electronic states. Consider an event in which two electrons occupying states $\vk\n$ and $\bp\m$, scatter into $\vk'{\n'}$ and  $\bp'{\m'}$. The net rate of such an event and its reversed process reads
\EqS{
&(\dot f_{\vk\n})_{\rm ee}=2\p \sum_l \cP^{(l)}_{\rm ee}(\vk\n,\bp\m;\vk'{\n'},\bp'{\m'})\times\\
& F_{\rm ee}(\vk\n,\bp\m;\vk'{\n'},\bp'{\m'})\dl(\D\ve_1+\D\ve_2+l\W),
\label{eq:FermiGoldenRule2}
}
where $\cP_{\rm ee}^{(l)}=|\sum_{jmn}\braoket{\f^{l-j+m}_{\vk\n}\f^{j+n}_{\bp\m}}{\hat\cH_{\rm int}}{\f^{m}_{\vk'\n'}\f^n_{\bp'\m'}}|^2$, \Eq{
F_{\rm ee}=f_{\vk'\n'}\bar f_{\vk\n}f_{\bp'\m'}\bar f_{\bp\m} -f_{\vk\n}\bar f_{\vk'\n'}f_{\bp\m}\bar f_{\bp'\m'},
\label{eq:DefinitionOfFee}
}
 and
$\D\ve_1=\ve_{\vk'\n'}-\ve_{\vk\n}$, $\D\ve_2=\ve_{\bp'\m'}-\ve_{\bp\m}$. Throughout, we consider contact interactions, $\hat\cH_{\rm int}=\sum_{\vk_1\vk_2\vk_3}(U/a^2) \hat c\dg_{\vk_1+\vk_3\aup }\hat c_{\vk_1\aup}\hat c\dg_{ \vk_2-\vk_3\adn}\hat c_{ \vk_2\adn}$, where $\hat c\dg_{\vk \aup(\adn)}$ creates an electron of momentum $\vk$ and pseudospin $\aup(\adn)$ [see discussion below Eq.~(1) in the main text]. Using the above form of  $\hat\cH_{\rm int}$, the expression for the squared matrix element can be written as
\EqS{
\cP_{\rm ee}^{(l)}=\frac{U^2}{4a^4}|\sum_{jmn}\braket{\f^{l-j+m}_{\vk\n}}{\f^{m}_{\vk'\n'}} \braket{\f^{j+n}_{\bp\m}}{\f^n_{\bp'\m'}}&-\\
-\sum_{\a=x,y,z}\braoket{\f^{l-j+m}_{\vk\n}}{\s^\a}{\f^{m}_{\vk'\n'}} \braoket{\f^{j+n}_{\bp\m}}{\s^\a}{\f^n_{\bp'\m'}}&|^2,
\label{eq:EEMatrixElement}
}
when $\vk+\bp=\vk'+\bp'$ and $\cP_{\rm ee}^{(l)}=0$ otherwise. Here $\s^\a$ is a Pauli matrix in the pseudospin basis. Clearly, the RHS of  Eq.~\eqref{eq:EEMatrixElement} vanishes when the two colliding electrons have parallel pseudospins, and is maximal when the pseudospins are antiparallel. This property directly follows from the contact interactions between the electrons and the Pauli principle.
Once we established the Floquet-Fermi's golden rule formalism, we are in the position to evaluate the heating, cooling and thermalization rates.


\subsection{Heating rates \label{sec:HeatingRates}}

We begin our discussion with heating processes, transferring electrons from the LFB to UFB. For zero-temperature heat baths, energy required for the excitations is provided by the Floquet-Umklapp mechanism, dominated by $l=1$ terms in Eqs.~\eqref{eq:FermiGoldenRule1} and \eqref{eq:FermiGoldenRule2}.
In the estimation of heating rates we will  approximate $f_{\vk-}\eqa 1$ and $f_{\vk+}\eqa 0$, which is exact in the limit $\ka\to 0$. As the majority of excitation processes occur away from the resonance ring, we simplify the analysis by taking the limit $V/\W\to0$. In this limit, the Floquet wavefunctions are not analytic at $|\vk|=k_\yR$, yet obtain a simple structure away from the resonance ring. The zeroth order in $(V/\W)$ reads
\Eq{
\ket{\f_{\vk\pm}(t)}=e^{\frac{i\W t}{2}}\condf{\ket{\Y_{\vk\mp}}e^{\pm\frac{ i\W t}{2}}&,|\vk|<k_\yR\\\ket{\Y_{\vk\pm}}e^{\mp\frac{ i\W t}{2}}&,|\vk|>k_\yR},
\label{eq:FloquetStatesAprox1}
}
where $\ket{\Y_{\vk\pm}}$ are the eigenstates of $H_0(\vk)$ [see Eq.~\eqref{eq:FreeHamiltonian}], given by
\Eq{
\ket{\Y_{\vk\pm}}=\frac{1}{\sqrt{2}}\bR{\sqrt{1\pm\nz\cdot \nd(\vk)}\ket{\aup}\mp e^{i\q_\vk}\sqrt{1\mp \nz\cdot \nd(\vk)}\ket{\adn}}.
}
To determine the power of $V/\W$ multiplying the scattering rates for the higher order terms, we use a method described in Fig.~\ref{fig:ExtendedRateModelPop}c.
Note, that these approximate expressions of the wavefunctions can not be used to analyze the distributions near the Floquet gap. Therefore, we use a different approximation when we discuss relaxation and thermalization processes in Secs.~\ref{sec:PononCooling}-\ref{sec:AugerThermalization}.

\subsubsection{Photon-assisted excitation rate \label{sec:PotonHeating}}

We begin with the photon-assisted interband excitations. These are vertical Floquet-Umklapp processes [the initial and final momentum of the photons are identical, see the definition of the photon density of states below Eq.~\eqref{eq:EL_Coupling}], predominantly transferring electrons inside the resonance ring. Processes outside the ring can be neglected in the limit $V/\W\to 0$, as they are suppressed by a factor of $\sim (V/\W)^4$ (see Fig.~\ref{fig:ExtendedRateModelPop}c for an explanation).
In what follows, we estimate the dominant contribution, illustrated in Fig.~\ref{fig:ExtendedRateModel}a.

We define the rate of change (ROC) of the density in the UFB as $\dot {n}_\ye|_{\rm photon}=\G_\ell$, where
\Eq{
\G_\ell=\int \frac{a^2d^2\vk d^2\vk'}{(2\p)^4}(\dot f_{\vk+})_{\ell,\vk'-}.
\label{eq:ROCphotonAssisted}
}
We approximate $(\dot f_{\vk+})_{\ell,\vk'-}$ [given in Eq.~\eqref{eq:FermiGoldenRule1}] by $(\dot f_{\vk+})_{\ell,\vk'-}\eqa 2\p \cP^{(1)}_{\ell}(\vk+,\vk-)\ro_\ell^0\dl(\vk-\vk')$, when $|\vk|<k_\yR$ and zero otherwise. The sum of squared matrix elements for two photon polarizations [Eq.~\eqref{eq:EL_Coupling}] evaluated in the states $\ket{\Y_{\vk\n}}$ [Eq.~\eqref{eq:FloquetStatesAprox1}] inside the resonance ring ($|\vk|<k_\yR$) is given by $\cP^{(1)}_{\ell}(\vk+,\vk-)=\sum_{i=1,2} |\braoket{\Y_{\vk-}}{\cM_\ell^{(i)}}{\Y_{\vk+}}|^2=g_\ell^2(1+\nz\cdot\nd(\vk))$. We further approximate $\nz\cdot \nd\eqa 1$ to leading order in $\lm_0k_\yR/E_\yg$, yielding $\cP^{(1)}_{\ell}\eqa 2g_\ell^2$ and perform the trivial momentum integral in Eq.~\eqref{eq:ROCphotonAssisted} to arrive at
\Eq{
\G_\ell=g_\ell^2\ro^0_\ell\cA_\yR a^2/(4\p^3).
}
Recall that $\cA_\yR=\p k_\yR^2$ is the area in the reciprocal space enclosed by the resonance ring.


\subsubsection{Phonon-assisted excitation rate\label{sec:PononHeating}}
Next, we estimate the phonon-assisted excitation rate (for an illustration see Fig.~\ref{fig:ExtendedRateModel}b). The ROC of the density in the UFB due to phonon-assisted excitation, $\dot n_\ye|_{\rm phonon}=\G_\ys$, reads
\Eq{
\G_\ys=\int \frac{a^2 d^2\vk d^2\vk'}{(2\p)^4}(\dot f_{\vk+})_{\ys,\vk'-},
\label{eq:ROCphononAssisted}
}
where we use the approximate expression
\Eq{
(\dot f_{\vk+})_{\ys,\vk'-} \eqa 2\p \cP_{\ys}^{(1)}(\vk+,\vk'-)
\ro_\ys(\vk-\vk',\w),
\label{eq:Fdot_phonons}
}
and $\w=\ve_{\vk'-}-\ve_{\vk+}+\W$, see Eq.~\eqref{eq:FermiGoldenRule1}.
We evaluate the expectation value in $\cP_\ys^{(1)}$ [defined below Eq.~\eqref{eq:FermiGoldenRule1}] using the electron-phonon coupling [Eq.~\eqref{eq:ES_Coupling}] and
the Floquet states given in Eq.~\eqref{eq:FloquetStatesAprox1}.
The result splits to four different rates, depending whether the momenta $\vk$ and $\vk'$ are inside or outside the resonance ring.

First, consider transitions where both momenta $\vk$ and $\vk'$ are inside the resonance ring. The rate of such transitions is suppressed by $\sim (\lm_0 k_\yR/E_\yg)^2$ as the electron-phonon coupling [which we assume has a diagonal form in the pseudospin basis, see Eq.~\eqref{eq:ES_Coupling}] connects almost orthogonal pseudospin states. Furthermore, transitions inside the resonance ring are unfavorable as the electron-phonon coupling favours short-wavelength phonons, while the wavelength corresponding to the momentum transfer inside the resonance ring is at least $\sim (2k_\yR)\inv$. Therefore, in what follows we neglect  transitions where both momenta are inside the resonance ring.

Next, we consider transitions for which $\vk'$ is inside and $\vk$ is outside the resonance ring (or otherwise, $\vk'$ outside and $\vk$ inside the resonance ring). These processes require an absorption of a drive-photon leading to suppression of their rate by a factor of $(V/\W)^2$ [see Fig.~\ref{fig:ExtendedRateModelPop}c for explanation]. Assuming a narrow-band semiconductor (analogous to the assumption in our numerical analysis, see Tab.~\ref{tab:Parameters}) we approximate the energy transfer, appearing in the definition of $\cP^{(1)}_\ys$ and $\ro_\ys$ in Eq.~\eqref{eq:Fdot_phonons}, by $\w\sim \W$, leading to $\cP_{\ys}^{(1)}(\vk+,\vk'-) \eqa \frac{g_\ys^2|\vk-\vk'|^2}{\W}\bR{\frac{V}{\W}}^2$ [cf. Eq.~\eqref{eq:ES_Coupling}] and  $\ro_\ys(\vq,\W)\eqa\ro_\ys^0$.
Explicitly integrating over $(\dot f_{\vk+})_{\ys,\vk'-}$ in Eq.~\eqref{eq:ROCphononAssisted}, we arrive at
\Eq{
\G_{\ys}\eqa\frac{ g_\ys^2\ro_\ys^0 a^2\cA_\yR \Lm^4}{8\p^2\W}\bR{\frac{V}{\W}}^2,
\label{eq:Gamma_s}
}
where $\Lm\sim \p/a$ is the large momentum cutoff for phonon transitions. Note that the processes considered here, involve collisions with large-energy phonons (of energy $\sim \W$). Such phonons might not be supported in materials in which the Debye frequency is smaller than $\W$.

An alternative scenario, which describes many periodically driven semiconductor systems, is the case of a large electronic bandwidth compared to the drive frequency. In this case, the bandstructure supports transitions transferring electrons between momentum states inside and outside the resonance ring (similar to the scenario discussed above), assisted by low-energy phonons. In particular, the allowed energy of the phonon is within the range $\w\in[\w_0,\w_\yD]$, where $\w_\yD$ is the Debye energy of the semiconductor and $\w_0$ is the minimal energy allowed by the kinematic constraints. For simplicity, we assume the parabolic dispersion near the $\G$-point (with an effective mass $m_*$) extends to energies $\gtrsim \W$.
We also assume the limit $k_\yR/\sqrt{2m_*\W}\to 0$. In this limit, the momentum transfer $|\vq|=|\vk-\vk'|$, approximately equals the momentum of the electron after the transition, $|\vk|$. It follows from the above, the minimal allowed phonon energy $\w_0$ is given by $\w_0=v_\ys |\vq_0|$, where $|\vq_0|=\sqrt{2m_*(\W-\w_0)}$ is the maximal allowed momentum transfer in the process.

To evaluate $\cP_{\ys}^{(1)}(\vk+,\vk'-)$ and the density of states, $\ro_\ys$ [appearing in Eq.~\eqref{eq:Fdot_phonons}], we estimate the energy transfer by $\w\eqa \W-\ve_{\vk+}$. Recall that $\vk$ denotes momentum far from the resonance ring, where the dispersion is approximately parabolic, $\ve_{\vk+}\eqa |\vk|^2/(2m_*)$. Therefore, we approximate
$\cP_{\ys}^{(1)}(\vk+,\vk'-) \eqa \frac{g_\ys^2|\vk|^2}{\W-\ve_{\vk+}}\bR{\frac{V}{\W}}^2$ and $\ro_\ys\eqa \frac{\ro_\ys^0(\W-\ve_{\vk+})}{\sqrt{(\W-\ve_{\vk+})^2-v_\ys^2 |\vk|^2}}$.

We are now ready to evaluate the integral in Eq.~\eqref{eq:ROCphononAssisted}. The $\vk'$-integral is in the domain $|\vk'|<k_\yR$. The integral over $\vk$ is performed in the range $|\vk|\in [|\vq_D|,|\vq_0|]$, where $\vq_D$ is found from $\w_D=\W-\ve_{\vq_{D}+}$ (leading to $|\vq_D|=\sqrt{2m_*(\W-\w_\yD)}$.)
 Transforming to polar coordinates where $k=|\vk|$ and $k'=|\vk'|$, we obtain
$\G_\ys=\frac{g_\ys^2\ro^0_\ys a^2}{\p}\bR{\frac{V}{\W}}^2\int_0^{k_\yR} dk'k'\int _{|\vq_D|}^{|\vq_0|} dk k^3/\sqrt{(\W-\ve_{\vk+})^2-v_\ys^2 k^2}$.
Explicitly integrating over $k$ and $k'$ in the limit $\W\gg\w_\yD\gg m_*v_\ys^2$, we arrive at
\Eq{
\G_\ys\eqa\frac{2c_\yD g_\ys^2\ro^0_\ys a^2\cA_\yR m_*^2 \W}{\p^2}\bR{\frac{V}{\W}}^2,
}
where $c_\yD=\log\bR{\frac{\sqrt{2m_* v_\ys^2 \W}}{\w_\yD-\sqrt{\w_\yD^2-2m_*v_\ys^2\W}}}$.

Finally, consider processes where both $\vk$ and $\vk'$ are outside the resonance ring. These processes are suppressed by $(V/\W)^4$ as they require absorption of two virtual photons [see Fig.~\ref{fig:ExtendedRateModelPop}c]. Furthermore, they will be suppressed by $\sim(2m_* \lm_0/\Lm)^2$ as such transitions connect almost orthogonal pseudospin states. We thus neglect the contribution of these processes to the phonon-assisted excitation rate.

\subsubsection{Floquet-Auger excitation rate \label{sec:AugerHeating}}
Here, we estimate the rate of transfer of electrons from the LFB to the UFB due to electron-electron scattering. The main processes contributing to such transfer are electron-pair excitations from the LFB to UFB \cite{Seetharam2019}. For the estimation of this rate, we assume a state with a full LFB and empty UFB (we therefore neglect processes which result in net transfer of electrons from the UFB to the LFB). 
The leading processes resulting in a transfer of two electrons from the LFB to the UFB involve a change of the total quasienergy by $\W$ via the Floquet-Umklapp mechanism (illustrated in Fig.~\ref{fig:ExtendedRateModel}c).
 The ROC of the density due to these processes is given by $\dot n_\ye|_{\rm ee}=\G_{\rm ee}$, where
\Eq{
\G_{\rm ee}=2\int \frac{a^4 d^2\vk d^2\bp d^2 \vq}{(2\p)^6}(\dot f_{\vk+})_{\rm ee}.
\label{eq:ROCElectronElectron}
}
Here $\vk$ and $\bp$ are the momenta of the final states in the UFB and  $\vk'=\vk+\vq$, $\bp'=\bp-\vq$ denote the initial momenta in the LFB; the factor $2$ accounts for a pair of electrons excited in each collision.
We estimate $(\dot f_{\vk+})_{\rm ee}$, given in Eq.~\eqref{eq:FermiGoldenRule2}, by
$(\dot f_{\vk+})_{\rm ee}\eqa2\p \cP_{\rm ee}^{(1)}(\vk+,\bp+,\vk'-,\bp'-)\dl (\D\ve_1+\D\ve_2+\W)$ [see definitions of $\D \ve_{1,2}$ below Eq.~\eqref{eq:FermiGoldenRule2}], and evaluate $\cP_{\rm ee}^{(1)}$ using Eq.~\eqref{eq:EEMatrixElement} and the approximate Floquet wavefunctions given in Eq.~\eqref{eq:FloquetStatesAprox1}.
As the approximate wavefunctions are non-analytic at $|\vk|=k_\yR$, we divide the analysis of the scattering rate to 9 distinct cases. In each case, the momenta $\vk$, $\vk'$, $\bp$ and $\bp'$ are either inside or outside the resonance ring.

\begin{figure}
  \centering
  \includegraphics[width=8.6cm]{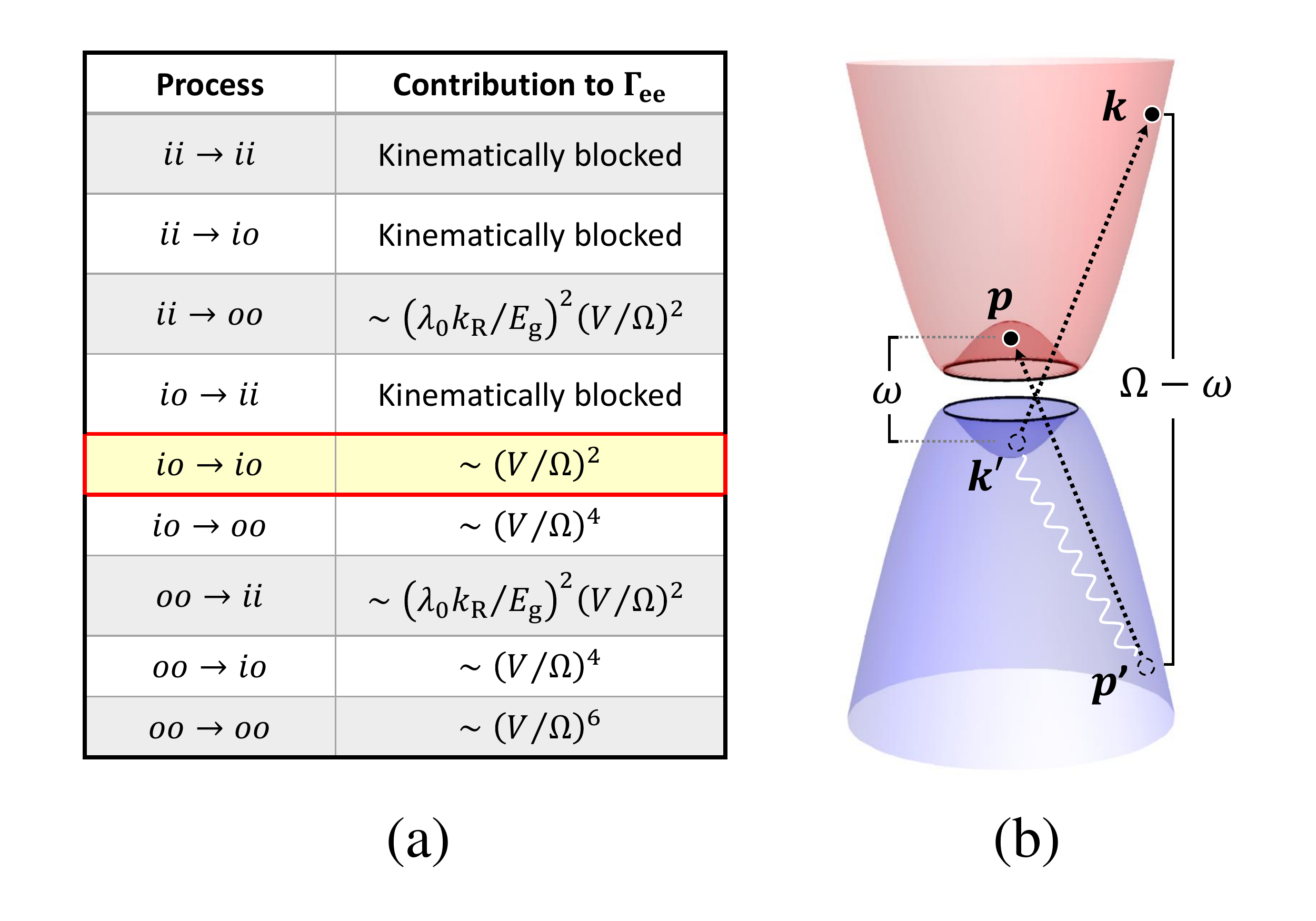}\\
  \caption{ Floquet-Auger excitation rate. (a). A summary of possible scattering processes and their estimated contribution to $\G_{\rm ee}$, divided to cases. The first column shows the possible types of processes; $i$ and $o$ indicate momenta inside and outside the resonance ring. The first two letters correspond to initial momenta of the two electrons (in the LFB) and the last two letters correspond to momenta after the collision (in the UFB.) The second column shows the estimated weight of each process in terms of powers of $(V/\W)$. We denote by ``kinematically-blocked'', processes that can not satisfy the energy and momentum conservation conditions due to the bandstructure constraints. (b). Schematic illustration of the dominant collision, $io\to io$, highlighted in panel (a). The momenta $\vk'$ and $\bp'$ denote the states of two electrons in the LFB before the collision, and the momenta $\vk$ and $\bp$ denotes the states in the UFB after the collision. The energy differences, $\w$ and $\W-\w$ are defined as $\w=\ve_{\bp+}-\ve_{\vk'-}$ and $\W-\w=\ve_{\vk+}-\ve_{\bp'-}$.
 \label{fig:InteractionProcesses}}
\end{figure}

Fig.~\ref{fig:InteractionProcesses}a shows a table of all possible cases and their estimated contribution to $\G_{\rm ee}$. We consider the limit $\dl E\ll \W$ which kinematically constraints some of the processes. For example, processes where the two initial and the two final momenta are inside the resonance ring (corresponding to $ii\to ii$ in Fig.~\ref{fig:InteractionProcesses}a) are kinematically blocked, as there is no such process for which the total energy change (which is limited by $2\dl E$ by the Floquet bandstructure) can be equal $\W$. Processes corresponding to $ii\to oo$ and $oo\to ii$ involve scattering of two electrons starting and ending in the same band in terms of the undriven band structure, cf. Eq.~\eqref{eq:FloquetStatesAprox1}. These processes are Pauli-suppressed by a factor of $\sim (\lm_0k_\yR/E_\yg)^2$ as the pseudospins of the electrons involved in the collision are almost parallel, see discussion following Eq.~\eqref{eq:EEMatrixElement}.

The most significant contribution to the rate [proportional to $(V/\W)^2$] corresponds to transitions of two electrons initially occupying one state inside the resonance ring and one state outside of it, scattered into two states in the UFB where one is inside and one is outside the resonance ring (denoted by $io\to io$ in Fig.~\ref{fig:InteractionProcesses}a). An illustration of such a process is shown in Fig.~\ref{fig:InteractionProcesses}b. We choose the momenta $\vk'$ and $\bp$ inside the resonance ring and $\bp'$ and $\vk$ outside the resonance ring. To account for processes where ($\vk'$, $\bp$) and ($\bp'$, $\vk$) are exchanged, we multiply the rate by $2$. We consider the process for which $\D\ve_1=\ve_{\vk'-}-\ve_{\vk+}<0$, and $\D\ve_2=\ve_{\bp'-}-\ve_{\bp+}>0$. The amplitude of such a process is suppressed by a factor of $(V/\W)^2$, see Fig.~\ref{fig:ExtendedRateModelPop}c.  We also assume that the amplitude of the process where $\D\ve_1>0$, and $\D\ve_2<0$ constructively interfere with the previous case such that the matrix element for both processes is twice the matrix element of each one of them.
Following from the above, we approximate $\cP^{(1)}_{\rm ee}\eqa 2(2U V/a^2\W)^2$. 
Notice that there is an additional scenario included in $io\to io$ processes, corresponding to $|\vk|,|\vk'|<k_\yR$ and $|\bp|,|\bp'|>k_\yR$. This scenario gives much smaller contribution to the rate since it includes transitions between  almost orthogonal pseudospin states. We thus neglect it in our analysis.

To evaluate the integral in Eq.~\eqref{eq:ROCElectronElectron} we introduce a new variable, $\w$, which splits the energy-conservation $\dl$-function as follows, $\dl(\D \ve_1+\D\ve_2+\W)=\int d\w \dl (\D\ve_3+\w)\dl (\D\ve_4+\W-\w)$. Here, we defined $\D\ve_3\eqv\ve_{\vk'-}-\ve_{\bp+}$ and $\D\ve_4\eqv\ve_{\bp'-}-\ve_{\vk+}$, such that $\D \ve_1+\D \ve_2=\D\ve_3+\D\ve_4$. Therefore, $\w$ denotes the energy transfer inside the resonance ring and $\W-\w$ the energy transfer outside the resonance ring (see Fig.~\ref{fig:InteractionProcesses}b). It is also useful to define the momentum transfer inside the resonance ring, $\vq'=\bp-\vk'$. By conservation of the total momentum, it equals the momentum transfer outside the resonance ring, $\vq'=\bp'-\vk$. The bandstructure near the $\G$-point restricts the values of $\w$ and $|\vq'|$ to $\w<\dl E$ and $|\vq'|<2k_\yR$. Therefore, given the constraint $-\D \ve_4=\W-\w$, we can approximate $\D \ve_4\eqa 2\ve_{\bp'-}\eqa-2\ve_{\vk+}$, since $\dl E\ll\W$.

We use the last approximation to replace $\D \ve_4$ by $-2\ve_{\vk+}$ in the energy-conservation $\dl$-function in Eq.~\eqref{eq:ROCElectronElectron}, and integrate over $\vk$, yielding $\int d^2\bk\dl(\D \ve_4+\W-\w)\eqa \p m_\ast$. Remarkably, due to the nearly parabolic dispersion outside the resonance ring with a density of states which is constant as a function of energy, the result of the $\vk$-integral is independent of $\W-\w$. Next, we perform the $\w$-integral over $\dl(\D \ve_3+\w)$, yielding $1$. Finally, we perform the trivial $\bp$ and $\vk'$-integrals over a constant function remained from the previous integrals, yielding a factor of $\cA_\yR$ for each of the variables. Collecting all the factors, we arrive at
\EqS{
\G_{\rm ee}\eqa\frac{ \cA_\yR^2 U^2m_\ast}{2\p^4}\bR{\frac{V}{\W}}^2.
\label{eq:AugerRate}
}

\subsection{Phonon-assisted interband recombination rate \label{sec:PononCooling}}

Here, we discuss phonon-assisted interband relaxation processes.
Interband relaxation serves as a sink mechanism for the excitations created by the heating processes (described in Sec.~\ref{sec:HeatingRates}). Electrons transferred to the UFB by the heating processes first undergo fast intraband relaxation followed by multiple-phonon emission, transferring electrons to the bottom of the UFB. Then, they recombine with the holes in the LFB, which were, in turn, transferred to the top of the UFB by similar processes. Here we focus on such interband electron-hole recombination processes predominantly mediated by phonons (see Fig.~\ref{fig:ExtendedRateModel}d). We approximate the states involved in the scattering by the RWA states at the resonance ring, see Eq.~\eqref{eq:FloquetStatesResonance}.

We expect the ROC of the density due to interband-recombination to be proportional to the densities of the electrons and holes in the UFB and LFB respectively. We thus define $\dot n_\ye|_{\rm inter}=-\Lm_{\rm inter}n_\ye n_\yh$, where
\Eq{
\Lm_{\rm inter}=-\frac{1}{n_\ye n_\yh}\int \frac{a^2d^2\vk d^2\vk'}{(2\p)^4}(\dot f_{\vk+})_{\ys,\vk'-}.
\label{eq:Lm_inter_int}
}
Here $(\dot f_{\vk+})_{\ys,\vk'-}$ is given by Eq.~\eqref{eq:FermiGoldenRule1} where we take only the $l=0$ term and estimate the energy transfer by $\sim \D_\yF$, leading to
\Eq{
(\dot f_{\vk,+})_{\ys,\vk'-}\eqa -2\p \cP_{\ys}^{(0)}(\vk+,\vk'-)\ro_\ys(\vq,\D_\yF)F(\vk+,\vk'-).
\label{eq:approx_f}
}
The function $F(\vk+,\vk'-)$ is non zero only when $|\vk|,|\vk'|\eqa k_\yR$. Therefore, we estimate $|\vq|=|\vk'-\vk|\eqa 2 k_\yR|\sin\bR{\D \q/2}|$, where $\D\q=\q-\q'$, and $\q,\q'$ are the angles of the vectors $\vk$ and $\vk'$ with respect to the $\nx$ axis. We evaluate $\cP_{\ys}^{(0)}$ by the matrix elements of the electron-phonon coupling [Eq.~\eqref{eq:ES_Coupling}] estimated at $\w=\D_\yF$ with Floquet states given in Eq.~\eqref{eq:FloquetStatesResonance}, leading to $\cP_{\ys}^{(0)}\eqa (4g^2_\ys k^2_\yR/\D_\yF)\sin^4\bR{\D\q/2}$. In the analysis we consider the limit, $\D_\yF\gg 2 v_\ys k_\yR$, which allows us to treat the density of states of the phonons as constant, $\ro_\ys(\vq,\D_\yF)\eqa \ro_\ys^0$.



Now, we turn to the evaluation of the integral in Eq.~\eqref{eq:Lm_inter_int}. The approximate form of $(\dot f_{\vk+})_{\ys,\vk'-}$, discussed in Eq.~\eqref{eq:approx_f} and further approximations below it, lead to a separation of the integrand into a product of radial and angular terms of $\vk$ and $\vk'$. We thus transform $d^2\vk\to d\q kdk$ and $d^2\vk'\to d\q'k'dk'$. We first integrate over the the radial components, which only appear in $F(\vk+,\vk'-)$, leading to $\frac{1}{(2\p)^2}\int kdk k'dk' F(\vk+,\vk'-)= n_\ye n_\yh$.
Then we integrate over the rest of the components, which have only angular dependencies, proportional to $\sin^4(\D\q/2)$. We thus arrive at the following expression,
\Eq{
\Lm_{\rm inter}\eqa\frac{3\p g_\ys^2\ro_\ys^0a^2 k_\yR^2}{\D_\yF}.
\label{eq:Lambda_inter}
}

\subsection{Phonon-assisted intraband relaxation rate \label{sec:PononThermalization}}
Next, we estimate the phonon-assisted intraband relaxation rate. Phonon-assisted intraband relaxation is the dominant mechanism for reducing the effective temperature of the excitations created by heating processes (described in  Sec.~\ref{sec:HeatingRates}) within each of the bands. Namely, it transfers electrons in the UFB to the bottom of the band and holes in the LFB to the top of the band. To account for the relaxation rate, we divide the population of electrons into two subsets occupying high and low-quasienergy states of electron densities $n_\ye^\yH$ and $n_\ye^\yL$, respectively [cf. Eqs.~\eqref{eq:DefinitionN_eL} and  \eqref{eq:DefinitionN_hL}]. For concreteness, in what follows we consider an EFM phase, associated with a sharp Fermi surface for the distribution of the electrons (the analysis in the EFI phase is similar). In this case, the separation between high- and low-quasienergy subsets for the distribution of the electrons is at the effective chemical potential measured relative to the UFB bottom, $\m_\ye$ [see an extended discussion in Sec.~\ref{sec:RateEquations} and Fig.~\ref{fig:ExtendedRateModelPop}a for an illustration].


For the EFM phase, intraband relaxation in the UFB is essentially electron-hole pair annihilation across the effective chemical potential with the rate $\dot n_\ye^\yL|_{\rm intra}=\Lm_{\rm intra}n_\ye^\yH\bar n_\ye^\yL$, where $\bar n_\ye^\yL$ is the density of unoccupied states in the low-quasienergy subset. The rate $\Lm_{\rm intra}$ is given by
\Eq{
\Lm_{\rm intra}=\frac{1}{ n_\ye^\yH \bar n_\ye^\yL}\int \frac{a^2d^2\vk d^2\vk'}{(2\p)^4}(\dot f_{\vk+})_{\ys,\vk'+},
\label{eq:Lm_intra_int}
}
where the range of $\vk'$-integral is all the states in the high-energy sector, corresponding to $\ve_{\vk'+}>\D_\yF/2+\m_\ye$. Subsequently, the range of $\vk$-integral is all the states in the low-quasienergy sector, $\D_\yF/2<\ve_{\vk+}<\D_\yF/2+\m_\ye$. We estimate $(\dot f_{\vk+})_{\ys,\vk'+}$ [given in Eq.~\eqref{eq:FermiGoldenRule1}] by the $l=0$ term,  thus this is not a Floquet-Umklapp process. The typical energy transfer in these processes equals the effective temperature of the electrons $\sim T_\ye$. 
Therefore, we approximate
\Eq{(\dot f_{\vk+})\eqa 2\p \cP_{\ys}^{(0)}(\vk+,\vk'+)\ro_\ys(\vq,T_\ye)F(\vk+,\vk'+).
\label{eq:approx_f2}
}
Since both of the momenta involved are close to the band minimum at $|\vk|,|\vk'|=k_\yR$, the momentum transfer is given by $|\vq|\eqa 2k_\yR|\sin\bR{\D\q/2}|$, where $\D \q$ is the angle between the vectors $\vk$ and $\vk'$. For low temperatures $T_\ye\ll v_\ys k_\yR$, the density of states $\ro_\ys(\vq,T_\ye)$, restricts the momentum transfers within a small circle in the momentum space,  $|\vq|<T_\ye/v_\ys$, illustrated by a red circle in Fig.~\ref{fig:ExtendedRateModel}e. 
To further simplify the expression in Eq.~\eqref{eq:approx_f2}, we use the paraxial approximation of small angles, $|\vq|\eqa k_\yR\D\q$. We estimate $\cP_{\ys}^{(0)}$ by the matrix elements of the electron-phonon coupling [Eq.~\eqref{eq:ES_Coupling}] estimated at $\w=T_\ye$, using the Floquet states given in Eq.~\eqref{eq:FloquetStatesResonance}. The above approximations result in  $\cP_{\ys}^{(0)}\eqa g_\ys^2k_\yR^2\D\q^2/T_\ye$, and $\ro_{\ys}(\vq,T_\ye)\eqa \ro^0_\ys T_\ye/\sqrt{T_\ye^2-(v_\ys k_\yR\D\q)^2}$, where $|\D\q|<\q_\ys$; $\q_\ys=T_\ye/(k_\yR v_\ys)$.

Now, we are ready to evaluate the integral in Eq.~\eqref{eq:Lm_intra_int}; we will use polar coordinates. The integrand splits to a product of terms dependent on the radial and angular components of the momentum. The integral over the radial part yields $\frac{1}{(2\p)^2}\int kdkk'dk' F(\vk+,\vk'+)=n_\ye^\yH \bar n_\ye^\yL$. In turn, the angular part yields $\int_{-\q_\ys}^{\q_\ys} d\D\q \D\q^2/\sqrt{1-(\D\q/\q_\ys)^2}=\p\q_\ys^3/2$. Combining the above results of the angular and radial integrals, we arrive at
\Eq{
\Lm_{\rm intra}\eqa\frac{\p g_\ys^2 \ro_\ys^0 a^2 T_\ye^2}{2k_\yR v_\ys^3}.
\label{eq:Lambda_intra}
}
A similar analysis  for  the EFI phase yields the same estimate [Eq.~\eqref{eq:Lambda_intra}]. Subsequently, an analysis for holes gives the same result upon a replacement $T_\ye\to T_\yh$ (due to the particle-hole symmetry of the Floquet bands).

\subsection{Equilibration of the electron and hole distributions via electron-electron scattering \label{sec:AugerThermalization}}
Here we estimate the rate of intraband thermalization processes induced by electron-electron scattering (depicted in Fig.~\ref{fig:ExtendedRateModel}f).
We consider scattering processes in which one of the incoming electrons is in the UFB, and one in the LFB, and likewise for the outgoing electrons. Thus, no interband particle transfer is involved. These processes tend to equalize the temperatures of the electron and hole populations as a result of the collisions between them.
We do not consider intraband scattering, where both electrons are in the UFB or LFB, as such processes do not contribute if the distribution within the band can be described by a Fermi function (i.e., in a phase-space-local equilibrium).
For simplicity, we consider a situation where the electron and hole populations in the two bands have close temperatures, such that the difference  between the temperatures, $\D T=T_\ye-T_\yh$, is much smaller than the mean temperature, $T=(T_\ye+T_\yh)/2$, $|\D T|\ll T$.

We define the thermalization rate by $\dot n_\ye^\yL|_{\rm ee}=\g_{\rm ee}\D T n_\ye^\yH\bar n_\ye^\yL n_\yh^\yH\bar n_\yh^\yL$, see the discussion above Eq.~\eqref{eq:RateEqUFB}, where
\Eq{
\g_{\rm ee}=\frac{\int a^4 d^2\vk d^2 \vk' d^2\bp d^2\bp'(\dot f_{\vk+})_{\rm ee}\dl^{(2)}(\vk+\bp-\vk'-\bp')}{(2\p)^6\D T n_\ye^\yH\bar n_\ye^\yL n_\yh^\yH\bar n_\yh^\yL}.
\label{eq:Thermalization1}
}
Here $(\dot f_{\vk+})_{\rm ee}$ is given by Eq.~\eqref{eq:FermiGoldenRule2}, where we take only $l=0$. We assume that $\vk$ and $\vk'$ describe states in the UFB and $\bp$ and $\bp'$ in the LFB, i.e., $(\n,\n',\m,\m')=(+,+,-,-)$, according to the notations in Eq.~\eqref{eq:FermiGoldenRule2}. For concreteness, we consider an EFM phase for electrons in the UFB (the analysis for the EFI phase is similar). The holes in the LFB form a non-degenerate distribution.
The range of $\vk'$-integral is all the states in the high-energy sector of the UFB, corresponding to $\ve_{\vk'+}>\D_\yF/2+\m_\ye$. Subsequently, the range of $\vk$-integral is all the states in the low-quasienergy sector of the UFB, $\D_\yF/2<\ve_{\vk+}<\D_\yF/2+\m_\ye$. The $\bp$ and $\bp'$-integrals are in the range $-\D_\yF/2-T<\ve_{\bp-}<-\D_\yF/2$ and $\ve_{\bp'-}<-\D_\yF/2-T$, according to the definition of the low and high-quasienergy sectors in the LFB, see the text surrounding Eq.~\eqref{eq:DefinitionN_hL} and Fig.~\ref{fig:ExtendedRateModelPop}.

In the limit $\D T/T\to 0$, we approximate  $(\dot f_{\vk+})_{\rm ee}$ by the leading (linear) order in $\D T$ such that $\g_{\rm ee}$ is a constant in $\D T$ by the definition [Eq.~\eqref{eq:Thermalization1}]. To extract the first-order term in $\D T$ from $(\dot f_{\vk+})_{\rm ee}$, we describe the electron and hole distributions as Fermi functions and use this description in the expression for  $F_{\rm ee}(\vk+,\bp-;\vk'+\bp'-)$ [given in Eq.~\eqref{eq:DefinitionOfFee}]. In particular, we use the property $1-\frac{1}{e^x+1}=e^x\bR{\frac{1}{e^x+1}}$ to switch between $f$ and $\bar f$-terms in Eq.~\eqref{eq:DefinitionOfFee}, leading to
\Eq{
F_{\rm ee}=(1-e^{\D\ve_1/T_\ye}e^{\D\ve_2/T_\yh})F(\vk+,\vk'+)F(\bp-,\bp'-),
\label{eq:ApproximateFee1}
}
where $F(\vk\n,\vk'\n')$ is defined in Eq.~\eqref{DefinitionOfF}.
The term in the brackets in the RHS of Eq.~\eqref{eq:ApproximateFee1} is linear to the leading order in $\D T$, i.e., $1-e^{\D\ve_1/T_\ye}e^{\D\ve_2/T_\yh}=\D \ve_1 \D T/T^2+\cO(\D T^2)$ (since by energy conservation $\D\ve_1+\D\ve_2=0$). We thus take only the zeroth order in $\D T$ of the rest of the terms in the RHS, to arrive at
\Eq{
F_{\rm ee}(\vk+,\bp-;\vk'+\bp'-)\eqa\frac{\D\ve_1\D T}{T^2} F(\vk+,\vk'+)F(\bp-,\bp'-),
\label{eq:ApproximateFee2}
}
where all the Fermi functions in $F$ are evaluated at $\D T=0$. 

Next, we evaluate $\cP^{(0)}_{\rm ee}$, employing  Eq.~\eqref{eq:EEMatrixElement} and Floquet states given in Eq.~\eqref{eq:FloquetStatesResonance}, which leads to
\EqS{
&\cP^{(0)}_{\rm ee}(\vk+,\bp-;\vk'+,\bp'-)\eqa \\
&\eqa(U/a^2)^2 \cos^2[(\q-\f)/2]\cos^2[(\q'-\f')/2].
\label{eq:PeeEstimated}
}
Here, we used polar coordinates, $\vk=(k\sin(\q),k\cos(\q))$, $\vk'=(k'\sin(\q'),k'\cos(\q'))$, $\bp=(p\sin(\f),p\cos(\f))$, and $\bp'=(p'\sin(\f'),p'\cos(\f'))$.

Once we estimated all the ingredients appearing in the definition of $(\dot f_{\vk+})_{\rm ee}$ [Eq.~\eqref{eq:FermiGoldenRule2}], we are in the position to evaluate the integral in Eq.~\eqref{eq:Thermalization1}.
To this end, we first replace each of the two-dimensional momenta integrals by the energy and angle integrals using polar representation. For example, $\int \frac{d^2\vk}{(2\p)^2}=\int d\ve_k D_\yF(\ve_k)\int\frac{d\q}{2\p}$, where $D_\yF(\ve_k)$ is the density of states [given in Eq.~\eqref{eq:DOS}].

\begin{figure}
  \centering
  \includegraphics[width=8.6cm]{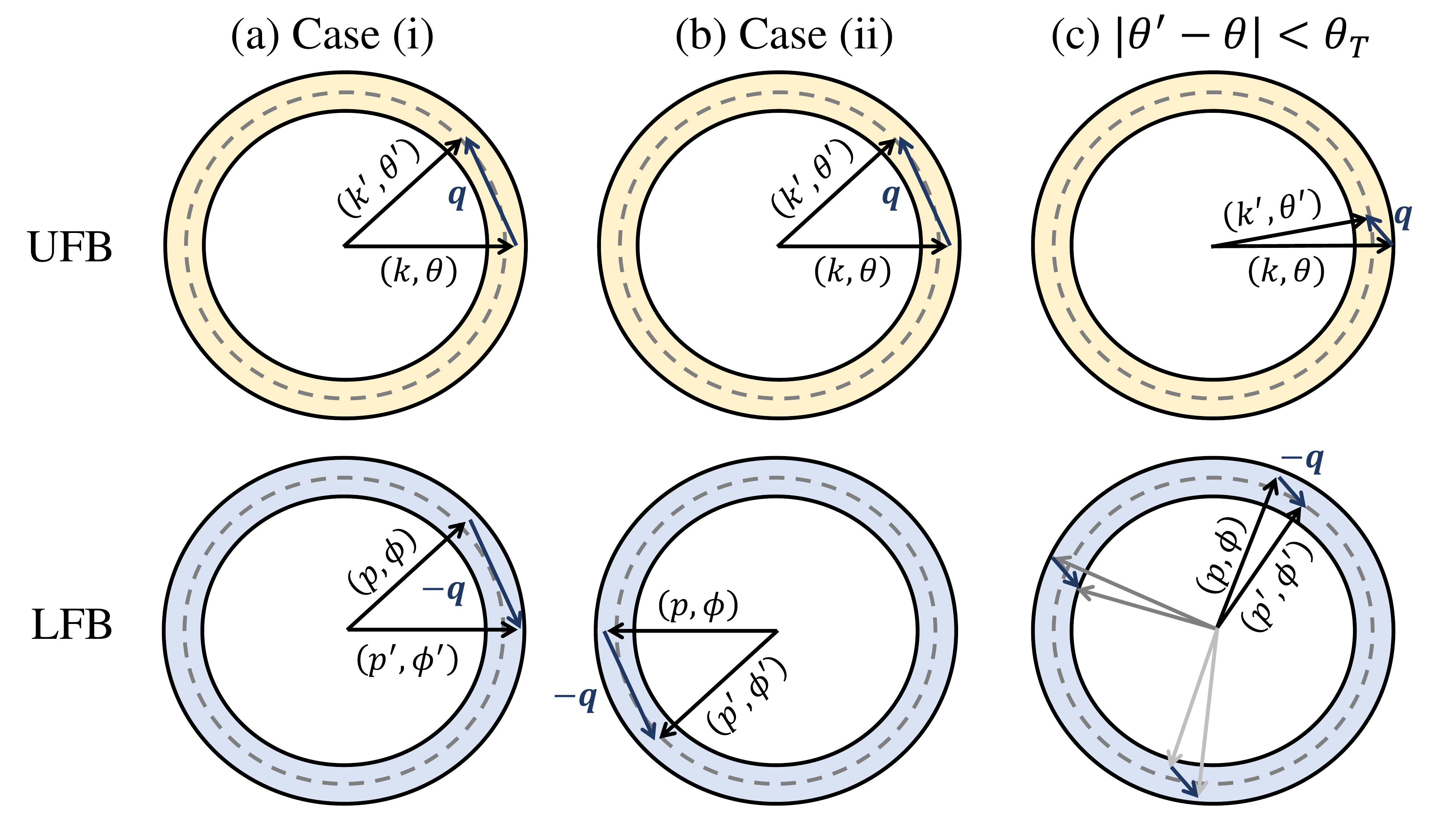}\\
  \caption{ Electron-electron scattering processes leading to equilibration of the electron and hole distributions. Yellow (top) and blue (bottom) rings indicate the area of states in the reciprocal space available for scattering near the resonance ring (gray dashed line), the width of the ring is approximately $\sim k_\yR \q_T$. The black arrows in the upper circle represent the momentum of the electron in the UFB before and after the scattering event, with a momentum transfer $\vq$ indicated by blue arrow. Similarly, the black arrows in the figure at the bottom correspond to momentum of an electron in the LFB before and after the transition, with a momentum transfer $-\vq$, due to total momentum conservation. (a). Possible scattering event for the case (i), for which $\f\eqa \q'$ and $\f'\eqa \q$. (b). Possible scattering event for the case (ii), for which $\f\eqa \p+\q$ and $\f'\eqa\p+ \q'$. This case is Pauli suppressed as it scatters between almost orthogonal pseudospin states [cf. Eq.~\eqref{eq:PeeEstimated}]. (c). Possible scattering events when $|\q'-\q|<\q_T$. In this case, there are many possible configurations of the angle $\f'$, all satisfying momentum conservation. \label{fig:AugerThermalization}}
\end{figure}

We start with the angular part of the integral. We fix two of the angles, $\q$ and $\q'$, and find the other two, $\f$ and $\f'$ by momentum conservation. If the scattering is exactly at the resonance ring, i.e., $k=k'=p=p'=k_\yR$, conservation of the total momentum  implies either (i) $\f=\q'$ and $\f'= \q$, or (ii) $\f=\p+\q$ and $\f'= \p+\q'$, see Fig.~\ref{fig:AugerThermalization}a and b. In practice, the momentum amplitude fluctuates around $k_\yR$, due to the thermal and doping-induced width of the distributions, leading to fluctuations of the angles, $\f=\f_0+\dl\f$ and $\f'=\f'_0+\dl\f'$, where $\f_0$ and $\f'_0$ are the solutions for $\f$ and $\f'$ in terms of $\q$ and $\q'$ which we found in the cases (i) and (ii) above. To find how  $\dl \f$ and $\dl\f'$ depend on the fluctuations in the amplitudes (defined as $\dl k=k-k_\yR$ and similarly for other momenta), we expand the conservation of momentum in the $\nx$ and $\ny$ directions to the first order in the fluctuations. The linearized equations for the case (i) read
\begin{subequations}
\begin{eqnarray}
&&\dl \f k_\yR \sin(\q')=\dl k-\dl p'+(\dl p-\dl k') \cos(\q')\\
&&\dl \f k_\yR\cos(\q')=k_\yR\dl\f'+(\dl k'-\dl p)\sin(\q'),
\end{eqnarray}
\label{eq:ThermalizationMomentumConservation}%
\end{subequations}
where we set $\q=0$, as the system is rotational symmetric.
The momentum-conserving $\dl$-function for the case (i) then reads
\Eq{
\dl_1^{(2)}(\vk+\bp-\vk'-\bp')=\frac{\dl(\f-\q'-\dl \f_1)\dl(\f'-\q-\dl \f'_1)}{k_\yR^2 |\sin(\q')|},
\label{eq:MometumConservation}
}
where $\dl \f_1$ and $\dl \f'_1$ obtained from the solution to Eq.~\eqref{eq:ThermalizationMomentumConservation} for $\dl \f$ and $\dl \f'$ as functions of $\dl k,\dl k',\dl p,\dl p'$ and $\q'$.
Next, we consider the case (ii). Here, the momentum conserving $\dl$-functions reads
\EqS{
&\dl_2^{(2)}(\vk+\bp-\vk'-\bp')=\\
&=\frac{\dl(\f-\p-\q-\dl \f_2)\dl(\f'-\p-\q'-\dl \f'_2)}{k_\yR^2 |\sin(\q')|}.
\label{eq:MometumConservation2}
}
In the case (ii), $\dl \f_2$ and $\dl \f'_2$ are obtained from equations similar to  Eq.~\eqref{eq:ThermalizationMomentumConservation} with an exchange $\dl p \leftrightarrow \dl p'$, $\dl \f \leftrightarrow \dl \f'$.
The full momentum conserving $\dl$-function is a sum of Eqs.~\eqref{eq:MometumConservation} and \eqref{eq:MometumConservation2}.

The cases (i) and (ii) above, apply when $|\vk'-\vk|\gg \dl k,\dl k',\dl p,\dl p'$. We introduce a cutoff parameter, $\q_T$ such that for $|\q'|>\q_T$, the momentum conservation is ensured by the sum of Eqs.~\eqref{eq:MometumConservation} and \eqref{eq:MometumConservation2}.
(Recall that $|\vk'-\vk|\eqa k_\yR|\q'|$ for $\q=0$ and small $\q'$.) This cutoff depends on the average values of $\dl k,\dl k',\dl p$ and $\dl p'$, therefore it is in general a function of $T$ and $\m_\ye$. However as we shall show later, the result only weakly (logarithmically) depends on the cutoff, therefore its $T$ and $\m_\ye$ dependence is not significant. For $|\q'|<\q_T$, the momentum can be conserved for many values of $\f'$, see Fig.~\ref{fig:AugerThermalization}c. In this case, we fix $\f'$ and $\q=0$ and find the corresponding angles $\f$ and $\q'$ from the momentum conservation. Defining $\f=\f'+\dl \f$, where $\dl \f$ is small (of the order of $\q'$), we arrive at the equations for $\dl \f$ and $\q'$,
\begin{subequations}
\begin{eqnarray}
&&\dl \f k_\yR \sin(\f')=\dl k-\dl k'+(\dl p-\dl p') \cos(\f')\\
&&\dl \f k_\yR\cos(\f')=k_\yR\q'+(\dl p'-\dl p)\sin(\f').
\end{eqnarray}
\label{eq:ThermalizationMomentumConservation2}%
\end{subequations}
The momentum conserving $\dl$-function then reads
\Eq{
\dl_3^{(2)}(\vk+\bp-\vk'-\bp')
=\frac{\dl(\f-\f'-\dl \f_3)\dl(\q'- \q'_3)}{k_\yR^2 |\sin(\f')|}.
\label{eq:MometumConservation3}
}
where $\dl \f_3$ and $\q'_3$ are the solutions to Eqs.~\eqref{eq:ThermalizationMomentumConservation2}. Eq.~\eqref{eq:MometumConservation3} is valid above approximately same cutoff $|\f'|>\q_T$.




We are now ready to perform the angular integral in Eq.~\eqref{eq:Thermalization1}. The only angle-dependent parts in this integral are the momentum-conserving $\dl$-functions [Eqs.~\eqref{eq:MometumConservation}, \eqref{eq:MometumConservation2} and \eqref{eq:MometumConservation3}] and $\cP_{\rm ee}^{(0)}$ given in Eq.~\eqref{eq:PeeEstimated}. We define an angular integral over each of the $\dl$-functions, $I_{\rm ang}=I^{(1)}_{\rm ang}+I^{(2)}_{\rm ang}+I^{(3)}_{\rm ang}$, where $I^{(1,2)}_{\rm ang}=\int_{|\q'|>\q_T}\frac{d\q d\q'd\f d\f'}{(2\p)^4}\dl_{1,2}^{(2)}\cP^{(0)}_{\rm ee}$, and $I^{(3)}_{\rm ang}=
\int_{|\q'|<\q_T,|\f'|>\q_T}\frac{d\q d\q'd\f d\f'}{(2\p)^4}\dl_3^{(2)}\cP^{(0)}_{\rm ee}$.
The integral over $\q$ can be trivially performed in each of $I^{(i)}_{\rm ang}$, yielding $2\p$. Next, we perform the integrals over the $\dl$-functions, leading to
\Eq{
I^{(1)}_{\rm ang}=\frac{U^2}{a^4}\int\frac{d\q'}{(2\p)^3}\frac{\cos^2(\frac{\q'+\dl\f_1}{2})\cos^2(\frac{\q'-\dl\f'_1}{2})}{k_\yR^2|\sin(\q')|}
\label{eq:Iang1}
}
\Eq{
I^{(2)}_{\rm ang}=\frac{U^2}{a^4}\int\frac{d\q'}{(2\p)^3}\frac{\sin^2(\frac{\dl\f_2}{2})\sin^2(\frac{\dl\f'_2}{2})}{k_\yR^2|\sin(\q')|}
\label{eq:Iang2}
}
\Eq{
I^{(3)}_{\rm ang}=\frac{U^2}{a^4}\int\frac{d\f'}{(2\p)^3}\frac{\cos^2(\frac{\f'+\dl\f_3}{2})\cos^2(\frac{\q'_3-\f'}{2})}{k_\yR^2|\sin(\f')|}
\label{eq:Iang3}
}
The $\q'$-integral in Eqs.~\eqref{eq:Iang1} and \eqref{eq:Iang2} and $\f'$-integral in Eq.~\eqref{eq:Iang3} are in the range $[\q_T,2\p-\q_T]$.

The last integral [Eq.~\eqref{eq:Iang3}] is nonzero when $|\q'_3|<\q_T$ [recall that $\q'_3$ is obtained from Eqs.~\eqref{eq:ThermalizationMomentumConservation2}]. We choose $\q_T$ to be large enough for this condition to be satisfied for any allowed value of $\dl k,\dl k',\dl p,\dl p',\dl \f$ and $\f'$.
Choosing too large value of $\q_T$ will lead to an underestimate for the total rate, as some of the values of $\q'$ will not be included in the calculation.

Turning back to Eqs.~\eqref{eq:Iang1}-\eqref{eq:Iang3}, we neglect small angles $\dl \f_1$ and $\dl \f'_1$ with respect to $\q'$ in Eq.~\eqref{eq:Iang1}, and $\dl \f_3$ and $\q_3'$ with respect to $\f'$ in Eq.~\eqref{eq:Iang3}. We also neglect $I_{\rm ang}^{(2)}$ as its value is proportional to the squares of small angles $\propto\dl \f_2^2\dl \f_2'^2$.
Therefore, we arrive at
$I_{\rm ang}\eqa\frac{2 U^2}{(2\p)^3 a^4k_\yR^2}\int d\q'\frac{\cos^4(\q'/2)}{|\sin(\q')|}$, where $\q'\in[\q_\yT,2\p-\q_\yT]$. This results, in the limit of small $\q_\yT$, in $I_{\rm ang}\eqa U^2\frac{2\log(2/\q_\yT)-1}{(2\p)^3 a^4 k_\yR^2}$.

We note that in the calculation of the angular integral we omitted parameter regimes in which the angular integral can not be separated from the amplitude integral, by introducing the cutoff $\q_T$. As the angular phase space corresponding to this regime is small, its omission only weakly affects the final result.

Now, we turn to the integration over the energy-dependent terms arising from $F_{\rm ee}$ [Eq.~\eqref{eq:ApproximateFee2}] and $\dl(\D\ve_1+\D\ve_2)$. Recall, that we consider an EFM phase for the electrons in the UFB, and describe the distribution of holes by Eq.~\eqref{eq:MBDistributionApprox_hole}.
We define $I_{\rm en}$ as the energy integral over the energy-dependent terms, given by
\EqS{
I_{\rm en}=&\int_{0}^{\m_\ye} d\ve_k \int_{\m_\ye}^{\infty} d\ve_{k'} \int_{-T}^{0} d\ve_p\int_{-\infty}^{-T} d\ve_{p'}\\
&D_\yF(\ve_k)D_\yF(\ve_{k'})D_\yF(-\ve_p)D_\yF(-\ve_{p'})\frac{\D T (\ve_{k'}-\ve_k)}{T^2}\\
&\dl(\ve_{k'}-\ve_k+\ve_{p'}-\ve_p)f_{\ve_{k'}+}\bar f_{\ve_{k}+}f_{\ve_{p'}-}\bar f_{\ve_{p}-},
}
where $\ve_k$ and $\ve_{k'}$ are accounted from the bottom of the UFB and $\ve_p$ and $\ve_{p'}$ are accounted from the top of the LFB (and are negative).
Assuming the electrons in the UFB are deep in the EFM regime, $\m_\ye/T\gg1$, we approximate $D_\yF(\ve_k)$ and $D_{\yF}(\ve_{k'})$ by $D_\yF(\m_\ye)$ and take the lower limit of the $\ve_k$-integral to $-\infty$. We set $\bar f_{\ve_p-}\eqa z_\yh e^{\ve_p/T}$ and $f_{\ve_{p'}-}\eqa 1$, assuming $z_h\ll1$, while $\bar f_{\ve_k+}=\frac{1}{e^{-(\ve_k-\m_\ye)/T}+1}$ and $f_{\ve_{k'}+}=\frac{1}{e^{(\ve_{k'}-\m_\ye)/T}+1}$. To evaluate the integral, we shift $\ve_k$ and $\ve_{k'}$ by $\m_\ye$ and rescale all the energies by $T$, $\ve_k=\m_\ye-T x$, $\ve_{k'}=\m_\ye+T y$, $\ve_{p}=-T z$, and $\ve_{p'}=-T w$, leading to
\EqS{
I_{\rm en}=C_1 C_2\frac{\D T}{T^2} n_\ye^\yH \bar n_\ye^{\yL}  n_\yh^\yH \bar n_\yh^\yL,
}
where
\EqS{
C_1=& \int_{0}^{\infty} dx \int_{0}^{\infty} dy \int_0^{1} dz\int_{1}^{\infty} dw
\frac{y+x}{\sqrt{zw}}\times\\
\times&\dl(y+x+z-w)\frac{1}{e^x+1}\frac{1}{e^y+1}e^{-z}\eqa 0.94,
}
and $C_2=[2\sqrt{\p}{\rm erfc(1)}\log^2(2)]\inv \eqa 3.73$. In this calculation we used explicit expressions of the densities of electron and hole subpopulations given in Eqs.~\eqref{eq:EstimateSubpopEFM} and \eqref{eq:EstimateSubpopHoles}.

Combining the angular and energy integrals, we arrive at $\g_{\rm ee}=\frac{(2\p)^3 a^4 I_{\rm ang} I_{\rm en}}{\D T  n_\ye^\yH \bar n_\ye^{\yL}  n_\yh^\yH \bar n_\yh^\yL}$ leading to
\Eq{
\g_{\rm ee}\eqa \frac{C_\g U^2}{k_\yR^2 T^2},
\label{eq:gamma_therm}
}
where $C_\g=[2\log(2/\q_\yT)-1]C_1C_2$.
The same calculation for the electrons in the EFI phase and for the holes, leads to a similar expression for $\g_{\rm ee}$, up to a numerical factor $\cO(1)$.

\subsection{Solution of the extended rate equations \label{sec:SolutionExtendedRate12}}
Now, we are ready to solve Eqs.~\eqref{eq:RateEqUFB} and \eqref{eq:RateEqLFB}. Our goal is to find the dependence of the effective temperatures and chemical potentials of the electrons and holes on the speed of sound, $v_\ys$, the doping, $\D n$ and the ``balance parameter'', $\ka$ [defined in Eq.~\eqref{eq:Balance_parameter}].

\subsubsection{Electron and hole temperatures for $\g_{\rm ee}=0$ \label{sec:SolutionExtendedRate}}

First, we assume $\g_{\rm ee}=0$, which corresponds to the model used for the full numerical simulation of the kinetic equation. Later, we show that if we include a finite $\g_{\rm ee}$ in the model, the distribution of the electrons, whose density is significantly larger then the density of the holes, will be almost unaffected. As a result, we expect the predicted phase boundary in Fig.~3 in the main text, to remain almost unchanged in the presence of electron-electron scattering processes which equilibrate the electron and hole populations. For $\g_{\rm ee}=0$, each temperature, $T_{\ye}$ or $T_{\yh}$, is obtained from its respective equation, Eq.~\eqref{eq:RateEqUFB} or Eq.~\eqref{eq:RateEqLFB}.

In what follows, we obtain $T_\ye$ from the solution to Eq.~\eqref{eq:RateEqUFB}, which for $\g_{\rm ee}=0$ reads
\Eq{
\Lm_{\rm intra} (T_\ye)n_\ye^\yH \bar n_\ye^\yL=\Lm_{\rm inter}n_\ye^\yL n_\yh.
\label{eq:ExRateEqsSol1}
}
First, we assume an EHM phase.
Our strategy in the analysis, is to replace the densities of subpopulations of electrons by their expressions, given in Eq.~\eqref{eq:EstimateSubpopEFM}. Then  instead of $\m_\ye$, we substitute its expression in terms of $n_\ye$
\Eq{
\m_\ye\eqa n_\ye^2/(4D_0^2\D_\yF),
\label{eq:MeAsFunctionOfNe}
}
obtained by inverting Eq.~\eqref{eq:DensityEFM},  where  $D_0=m_*/2\p$ [see definition below Eq.~\eqref{eq:DOS}]. We  substitute $n_\yh=\ka/n_\ye$ [cf. Eq.~\eqref{eq:DefinitionofKa}] and explicit expressions of $\Lm_{\rm inter}$ and $\Lm_{\rm intra}$ [Eq.~\eqref{eq:Lambda_inter} and \eqref{eq:Lambda_intra}] in Eq.~\eqref{eq:ExRateEqsSol1} to extract,
\Eq{
T_{\ye}=\bR{\frac{6 k_\yR^3 v_\ys^3 \ka n_\ye^2}{4 \log^2(2) \D_\yF^3 D_0^4}}^{1/4}.
\label{eq:EffectiveTempEFM}
}
Note that in Eq.~\eqref{eq:EffectiveTempEFM}, $T_\ye$ is expressed as a function of $n_\ye$. Recall, however, that $n_\ye$ is an explicit function of $\D n$ and $\ka$, as follows from Eq.~\eqref{eq:ElectronHoleDensities}.
Therefore, Eq.~\eqref{eq:EffectiveTempEFM} expresses $T_\ye$ as a function of $\kappa$ and $\D n$. The power $1/4$ in Eq.~\eqref{eq:EffectiveTempEFM} arises from the LHS term in  Eq.~\eqref{eq:ExRateEqsSol1}. This term is proportional to $T_{\ye}^4$, as follows from the definitions of $\Lm_{\rm intra}$ [see Eq.~\eqref{eq:Lambda_intra}],  $n_\ye^\yH$, and $\bar n_\ye^\yL$ [see Eq.~\eqref{eq:EstimateSubpopEFM}].
Finally, we divide the expression for $\m_\ye$ given in Eq.~\eqref{eq:MeAsFunctionOfNe}  by $T_{\ye}$ given in Eq.~\eqref{eq:EffectiveTempEFM} to arrive at
\Eq{
\frac{\m_\ye}{T_{\ye}}=\bR{\frac{C_{\rm EFM}\, n_\ye^6}{\D_\yF D_0^4 k_\yR^3 v_\ys^3 \ka}}^{1/4},
\label{eq:MTScaling1}
}
where $C_{\rm EFM}=\log^2(2)/384\eqa 0.00125$. In Eq.~\eqref{eq:MTScaling1}, $n_\ye$ can be expressed as a function of $\D n$ and $\ka$, using Eq.~\eqref{eq:ElectronHoleDensities}.

Next, we solve Eq.~\eqref{eq:ExRateEqsSol1} in the EFI phase. Here, we replace the electron subpopulation densities by expressions given in Eq.~\eqref{eq:EstimateSubpopEFI} and substitute $n_\yh=\ka/n_\ye$ [cf.  Eq.~\eqref{eq:DefinitionofKa}].
We then extract an expression for $T_{\ye}$ from the resulting equation, leading to
\Eq{
T_{\ye}=\bR{\frac{3 {\rm erf(1)}}{\rm erfc(1)}\frac{ k_\yR^3 v_\ys^3 \ka}{\D_\yF^{3/2}D_0 n_\ye}}^{2/5}.
\label{eq:EffectiveTempEFI}
}
Next, we find $z_\ye$, using
\Eq{
z_\ye=n_\ye/(\sqrt{\p}T_{\ye} D_\yF(T_{\ye}))
\label{eq:ExpressionForZe}
}
obtained from Eq.~\eqref{eq:DensityEFI}.
We substitute Eq.~\eqref{eq:EffectiveTempEFI} in Eq.~\eqref{eq:ExpressionForZe} to obtain
\Eq{
z_\ye\eqa\bR{\frac{C_{\rm EFI} \, n_\ye^6}{ \D_\yF D_0^4 k_\yR^3 v_\ys^3 \ka}}^{1/5},
\label{eq:MTScalingEFI}
}
where $C_{\rm EFI}={\rm erfc(1)}/(3\p^{5/2}{\rm erf(1)})\eqa 0.0036$. Recall that $\m_\ye/T_{\ye}$ can be extracted from $z_\ye$ using the relation $\m_\ye/T_{\ye}=\log(z_\ye)$.

We note that the phenomenological analysis outlined in this chapter is aimed to only capture the right scaling of $T_\ye$ and $\m_\ye$ as functions of the parameters of the system (i.e., $\ka$, $\D n$, $v_\ys$, etc.), and not the values of the numerical coefficients $C_{\rm EFM}$ and $C_{\rm EFI}$. We use these coefficients as fitting parameters in Fig.~3 in the main text. The best fit is obtained when $C_{\rm EFM}=C_{\rm EFI}\eqv  C$.

Next, we consider the distribution of holes in the LFB.
The analysis of Eq.~\eqref{eq:RateEqLFB} in the regime $\g_{\rm ee}=0$ is similar to the analysis of Eq.~\eqref{eq:ExRateEqsSol1} in the EFI phase, above. The density of holes and their subpopulations are given by Eqs.~\eqref{eq:DensityHoles} and \eqref{eq:EstimateSubpopHoles}. In this case, $T_\yh$ and $z_\yh$ are given by
\Eq{
T_{\yh}=\bR{\frac{3 {\rm erf(1)}}{\rm erfc(1)}\frac{ k_\yR^3 v_\ys^3 \ka}{\D_\yF^{3/2}D_0 n_\yh}}^{2/5},
\,
z_\yh\eqa\bR{\frac{C_{\rm EFI}\, n_\yh^6}{\D_\yF D_0^4 k_\yR^3 v_\ys^3 \ka}}^{1/5}.
\label{eq:MTScalingHoles}
}

Finally, we verify that Eq.~\eqref{eq:MTScalingHoles} is consistent with an assumption of the non-degenerate hole distribution, when the system is doped with electrons. To this end, we show that  $z_\yh\to 0$ when we take $\ka\to 0$ for fixed $\D n>0$. In this limit, $n_\yh\eqa \ka/\D n$, see Eq.~\eqref{eq:ElectronHoleDensities}. Substituting the latter relation in  Eq.~\eqref{eq:MTScalingHoles}, we find $z_\yh\sim \ka$. Therefore, the nondegenerate-hole-gas assumption is consistent; indeed $z_\yh\to 0$ as $\ka\to 0$. Interestingly, note that this assumption is not consistent with Eq.~\eqref{eq:MTScalingHoles} when the system is substantially doped with holes, i.e., when $\D n\ll-\ka$. In this case, $n_\yh\eqa |\D n|$, leading to $z_\yh\sim \ka^{-1/5}$, i.e.,  $z_\yh\to \infty$ as $\ka\to 0$. In fact, when the system is substantially hole doped, the holes form a degenerate distribution analogous to the EFM phase for electrons in an electron-doped system.

\subsubsection{Electron and hole temperatures in the presence of electron-electron scattering  \label{sec:SolutionExtendedRate2}}

\begin{figure}
  \centering
  \includegraphics[width=8.6cm]{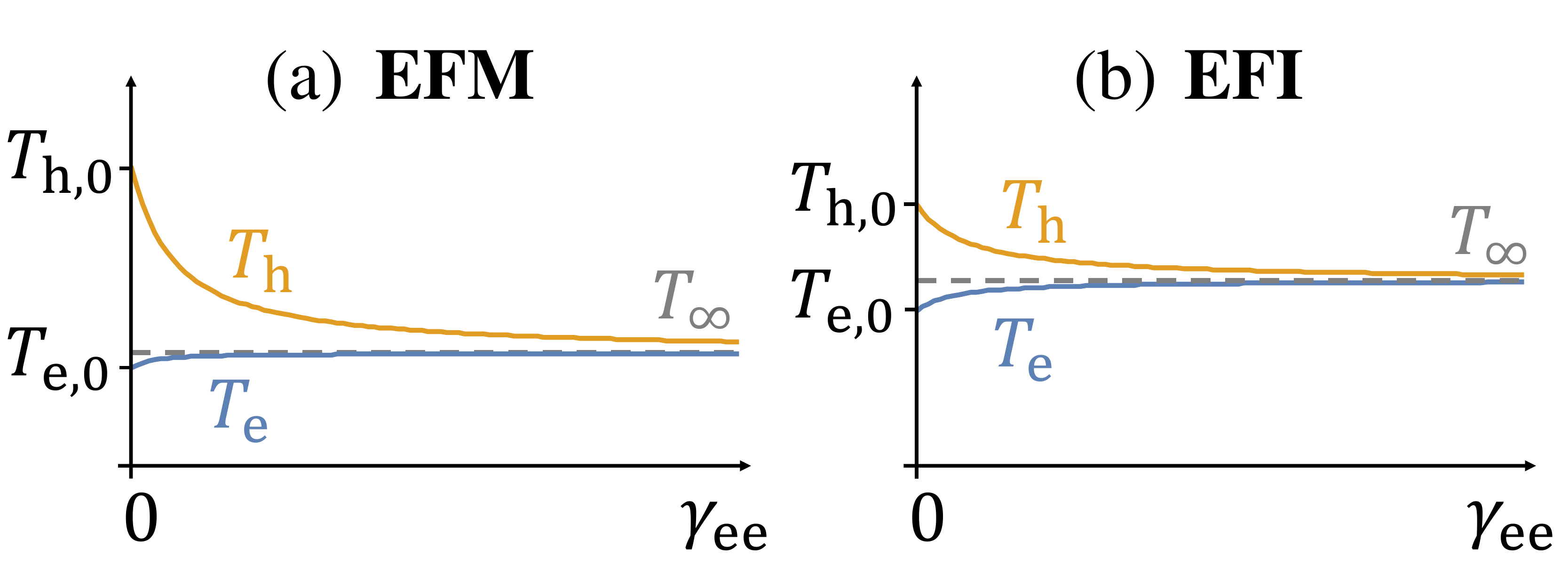}\\
  \caption{ Equilibration of the electron and hole distribution due to electron-electron scattering. The temperatures of electrons in the UFB ($T_\ye$) and holes in the LFB ($T_\yh$) are obtained from the solution to Eqs.~\eqref{eq:RateEqUFB} and \eqref{eq:RateEqLFB}, as a function of $\g_{\rm ee}$ for typical system's parameters in (a) the EFM phase and (b) EFI phase. The temperatures $T_{\ye,0}$ and $T_{\yh,0}$ obtained at $\g_{\rm ee}=0$ are given in Eqs.~\eqref{eq:EffectiveTempEFM}, \eqref{eq:EffectiveTempEFI} and \eqref{eq:MTScalingHoles}. The temperature for $\g_{\rm ee}\to \infty$, $T_\infty$, [given in Eqs.~\eqref{eq:TinfEFM} and \eqref{eq:TinfEFI}] is indicated by the dashed line.   \label{fig:Thermalization}}
\end{figure}

In subsection~\ref{sec:SolutionExtendedRate} we considered the case $\gamma_{\rm ee}=0$, for which electron-electron interactions do not equilibrate the electron and hole distributions. Now, we consider the regime where the electron-electron scattering is significant, i.e., when $\g_{\rm ee}$ is finite. In this regime, both Eqs.~\eqref{eq:RateEqUFB} and \eqref{eq:RateEqLFB} depend on $T_\ye$ and $T_\yh$ and need to be solved jointly. Fig.~\ref{fig:Thermalization} shows $T_\ye$ and $T_\yh$ obtained from a numerical solution of Eqs.~\eqref{eq:RateEqUFB} and \eqref{eq:RateEqLFB} in the EFM and EFI regimes. Here, we treat $\g_{\rm ee}$  as a constant varied from $\g_{\rm ee}=0$ to large values of $\g_{\rm ee}$, where interband thermalization is significant.
When $\g_{\rm ee}=0$, the electron and hole temperatures are found in Sec.~\ref{sec:SolutionExtendedRate}. Here, we denote by $T_{\ye,0}$ the temperature of the distribution of the electrons given in Eq.~\eqref{eq:EffectiveTempEFM} if the electrons exhibit an EFM phase and Eq.~\eqref{eq:EffectiveTempEFI} if they exhibit an EFI phase. Likewise, we denote by $T_{\yh,0}$ the temperature of the distribution of holes for $\g_{\rm ee}=0$, given in Eq.~\eqref{eq:MTScalingHoles}.

As $\g_{\rm ee}$ is increased, the difference between the temperatures of the two distribution is reduced until, in the limit $\g_{\rm ee}\to \infty$, they reach a common temperature, $T_\infty$. Predictably, the minority population is more affected by this equilibration, while the temperature of the majority population is only slightly changed. For large values of $\gamma_{\rm ee}$, as $T_\yh$ tends towards $T_\infty$, the difference between $T_\yh(\gamma_{\rm ee})$ and $T_{\yh,0}$ is proportional to $\sim 1/\g_{\rm ee}$. In what follows, we solve analytically Eqs.~\eqref{eq:RateEqUFB} and \eqref{eq:RateEqLFB} in the limit $\g_{\rm ee}\to \infty$, to obtain an analytical expression for $T_{\rm \infty}$ in the EFM and EFI regimes. In this limit,
the electron and hole temperatures are almost the same, $|\D T|\ll T$ [see discussion above Eq.~\eqref{eq:RateEqUFB}]. Assuming this is the case, we can take only zeroth order in $\D T$ in all the terms in Eqs.~\eqref{eq:RateEqUFB} and \eqref{eq:RateEqLFB}, except of the terms proportional to $\g_{\rm ee}$. Note that the term proportional to $\g_{\rm ee}$ is the same in both of these equations (up to a sign), thus cancels out upon addition of the two equations. The sum of Eqs.~\eqref{eq:RateEqUFB} and \eqref{eq:RateEqLFB} then reads,
\Eq{
\Lm_{\rm intra}(T_{\infty})(n_\ye^\yH \bar n_\ye^\yL+n_\yh^\yH \bar n_\yh^\yL)=\Lm_{\rm inter}(n_\ye^\yL n_\yh+ n_\yh^\yL n_\ye).
\label{eq:ExtendedRateSum}
}
Importantly, all the terms in Eq.~\eqref{eq:ExtendedRateSum} depend only on the mean temperature, $T_{\infty}$, defined as $T_\infty=\lim_{\g_{\rm ee}\to \infty} T$, where in this limit, $\D T=0$.
In what follows, we use Eq.~\eqref{eq:ExtendedRateSum} to determine $T_{\infty}$. Similarly, $\D T$ can be found from the difference between Eqs.~\eqref{eq:RateEqUFB} and \eqref{eq:RateEqLFB}.

First, consider an EFM regime for the electrons. To simplify Eq.~\eqref{eq:ExtendedRateSum}, we  neglect the term proportional to $n_\yh^\yH \bar n_\yh^\yL$ relative to $n_\ye^\yH \bar n_\ye^\yL$ in the limit $\ka\to0$. We also approximate $n_\ye^\yL\eqa n_\ye$ and use $n_\yh^\yL={\rm erf(1)}n_\yh$ [cf. Eqs.~\eqref{eq:EstimateSubpopEFM} and \eqref{eq:EstimateSubpopHoles}].
As a result, Eq.~\eqref{eq:ExtendedRateSum} simplifies to
\Eq{
\Lm_{\rm intra}(T_{\infty})n_\ye^\yH \bar n_\ye^\yL=C'\Lm_{\rm inter}n_\ye n_\yh,
\label{eq:ExtendedRateSumApprox}
}
where $C'=1+{\rm erf(1)}\eqa 1.84$. Note that Eq.~\eqref{eq:ExtendedRateSumApprox} is reminiscent of  Eq.~\eqref{eq:ExRateEqsSol1} up to a constant, $C'$. Subsequently, the mean temperature obtained from the solution to Eq.~\eqref{eq:ExtendedRateSumApprox} is related to the solution to Eq.~\eqref{eq:ExRateEqsSol1} by
\Eq{
T_{\infty}=(C')^{1/4}T_{\ye,0},
\label{eq:TinfEFM}
}
where $T_{\ye,0}$ is given in Eq.~\eqref{eq:EffectiveTempEFM}.

Now, we turn to the analysis of the EFI regime. This regime always arises at low-electron doping, which also includes half-filling (where $n_\ye=n_\yh$). Therefore in this case, we can not neglect the hole density with respect to the electron one, in Eq.~\eqref{eq:ExtendedRateSum}, as we did in the analysis of the EFM regime, above.
To extract $T_{\infty}$ from Eq.~\eqref{eq:ExtendedRateSum}, we replace the densities of subpopulations by their expressions given in Eqs~\eqref{eq:EstimateSubpopEFI} and \eqref{eq:EstimateSubpopHoles} in terms of $n_\ye$, $n_\yh$ and $T_{\infty}$ [substituting $T_\infty$ instead of $T_\ye$ and $T_\yh$]. An extraction of $T_{\infty}$ from the resulting equation yields
\Eq{
T_{\infty}=\bR{\frac{2n_\ye}{n_\ye+n_\yh}}^{5/2}T_{\ye,0},
\label{eq:TinfEFI}
}
where $T_{\ye,0}$ is given in Eq.~\eqref{eq:EffectiveTempEFI}.
At half-filling $T_{\infty}=T_{\ye,0}$, and away from the half-filling, as $n_\yh/n_\ye\to 0$, $T_{\infty}\to 2^{2/5} T_{\ye,0}$.

To conclude, in this section we showed that for an electron doped system, and in the presence of strong electron-electron scattering,
the temperature of the distribution of the electrons in the UFB is close to the temperature obtained by setting $\gamma_{\rm ee}=0$, given by Eqs.~\eqref{eq:EffectiveTempEFM} and \eqref{eq:EffectiveTempEFI}. In contrast, the temperature of the distribution of the holes may substantially change due to electron-electron scattering relative to temperature obtained for $\gamma_{\rm ee}=0$.
However the properties of the distribution of holes in the LFB only weakly effect the critical interaction strength, and therefore the phase diagram of the system [cf. Fig.~2a in the main text] (which is mainly set by the properties of the distribution of the electrons in the UFB). Therefore, we expect the phase diagram obtained for $\gamma_{\rm ee}=0$ to give a good description of the phase diagram of the system in the presence of a nonzero value of $\gamma_{\rm ee}$.

\subsection{Interpolation between the EFM and the EFI regimes}
In Sec.~\ref{sec:SolutionExtendedRate}, we obtained expressions for $\m_\ye/T_\ye$ in the EFM and EFI regimes for $\g_{\rm ee}=0$, see Eqs.~\eqref{eq:MTScaling1} and \eqref{eq:MTScalingEFI}. Here we discuss an interpolation between the two equations which we used to fit the numerical data in Fig.~3 in the main text. We assume $C_{\rm EFM}=C_{\rm EFI}\eqv C$, which yields the best fit for the data in Fig.~3, and define
\Eq{
x_\ye\eqv\z n_\ye^6/(v_\ys^3\ka),
\label{eq:DefinitionXe}
}
where $\z=C /(\D_\yF D_0^4 k_\yR^3)$. Using Eq.~\eqref{eq:DefinitionXe}, we rewrite Eqs.~\eqref{eq:MTScaling1} and \eqref{eq:MTScalingEFI} as $\m_\ye/T_\ye\eqa x_\ye^{1/4}$ in the EFM regime and $e^{\m_\ye/T_\ye}\eqa x_\ye^{1/5}$ in the EFI regime. The transition between the two regimes occurs in the crossover area, corresponding to $\m_\ye/T_\ye= \cO(1)$.



Now we introduce an analytic function which interpolates between the values of $\m_\ye/T_\ye$ in the two regimes. This function needs to interpolate between power-law (in the EFM regime) and exponential (in the EFI regime) functions. Therefore, we expect it to be of the form of the complete Fermi-Dirac integral function,
\Eq{
\cF_j(\lm \m_\ye/T_\ye)=x_\ye^\h.
\label{eq:Interpolation}
}
Here $\cF_j(x)=\frac{1}{\G(j+1)}\int_0^{\infty} \frac{t^j }{e^{t-x}+1}dt$,
is the complete Fermi-Dirac integral and $\G(j+1)=j\G(j)$; $\G(1)=1$ is the Euler gamma function. The function in the LHS of Eq.~\eqref{eq:Interpolation} has the following asymptotic values\cite{Wood1992},
\EqL{
&&\lim_{\m_\ye/T_\ye\to-\infty}\cF_j(\lm \m_\ye/T_\ye)=e^{\lm \m_\ye/T_\ye}\\
&&\lim_{\m_\ye/T_\ye\to\infty}\cF_j(\lm \m_\ye/T_\ye)=\frac{(\lm\m_\ye/T_\ye)^{j+1}}{\G(j+2)}.
}
To obtain the right constants in the two asymptotic limits to match the dependencies in the EFM and EFI regimes, we require
$\h/\lm=1/5$, $\h/(j+1)=1/4$, and $\lm^{j+1}=\G(j+2)$. These conditions have a unique solution $j\eqa -0.304$, $\lm\eqa 0.871$ and $\h\eqa 0.174$.

In order to use the interpolation in Fig.~3c in the main text, we need to obtain an expression for $\m_\ye/T_\ye$ as a function of $\D n$. To this end, we invert Eq.~\eqref{eq:Interpolation}, and employ the definition of $x_\ye$ [Eq.~\eqref{eq:DefinitionXe}] to obtain
\Eq{
\m_\ye/T_\ye=(1/\lm)\cF_j\inv\bC{[\z n_\ye^6/(v_\ys^3\ka)]^\h},
\label{eq:FitMT}
}
where $\cF_j\inv\bS{\cF_j(x)}=x$.
Finally, we replace $n_\ye$ by a function of $\D n$ and $\ka$, given in Eq.~\eqref{eq:ElectronHoleDensities}.
We fit the data in Fig.~3c by Eq.~\eqref{eq:FitMT} with a single fitting parameter $\z$. The same fitting parameter is used for all the curves shown in the Fig.~3.
In order to fit the data in Figs.~3b and 3d, we used the equation for $\tilde U_\yc$ [Eq.~(10) in the main text], where we substituted Eq.~\eqref{eq:FitMT} for $\m_\ye/T_\ye$ appearing inside the $\tilde \Q$-function. Eq.~(10) depends also on $\m_\yh/T_\yh$, which from the analysis similar to the one leading to Eq.~\eqref{eq:FitMT}, yields
\Eq{
\m_\yh/T_\yh=(1/\lm)\cF_j\inv\bC{[\z n_\yh^6/(v_\ys^3\ka)]^\h}.
\label{eq:FitMTholes}
}
The fit in Figs. 3b and 3d is performed with the same value of $\z$ as in Fig.~3c. We used $\tilde U_{\rm ex}$ and $\tilde U_{\rm fs}$ as additional fitting parameters.

\subsection{Evaluation of the optimal doping, $\D n_*$}
Finally, we evaluate the optimal doping $\D n_*$, at which $\tilde U_\yc$ exhibits a deep [cf. Figs.~3b and d in the main text]. The optimal doping is defined by
\Eq{
\frac{\dpa \tilde U_\yc}{\dpa \D n}\at{\D n=\D n_*}=0,
\label{eq:OptimalDopingDefinition}
}
where $\tilde U_\yc$ is given in Eq.~(10) in the main text, and we estimate $\m_\ye/T_\ye$ and $\m_\yh/T_\yh$ by Eqs.~\eqref{eq:FitMT} and \eqref{eq:FitMTholes} and $n_\ye$ and $n_\yh$ by Eq.~\eqref{eq:ElectronHoleDensities}. For simplicity, we neglect the term proportional to $\tilde U_{\rm fs}$ as it is constant in $\D n$, and neglect the term proportional to the density of holes, as its contribution is  negligible in the electron-doped system. The approximate expression for $\tilde U_\yc$ then reads
\Eq{
\tilde U_\yc\eqa \tilde n_\ye/\tilde \Q(\m_\ye/T_\ye).
\label{eq:UcApprox}
}
As $\ka$ is a weak function of $\D n$ near the phase boundary (see Fig.~\ref{fig:KappaVsKappa0}), we neglect its derivative with respect to $\D n$. Therefore, instead of finding the minimum with respect to $\D n$, we can find the minimum with respect to $n_\ye$ [which according to Eq.~\eqref{eq:ElectronHoleDensities} is only a function of $\ka$ and $\D n$] for fixed $\ka$. Replacing the $\D n$-derivative in Eq.~\eqref{eq:OptimalDopingDefinition} by $n_\ye$-derivative and using the approximate form of $\tilde U_\yc$ given in Eq.~\eqref{eq:UcApprox}, we arrive at
\Eq{
\frac{\tilde \Q'(\m_\ye/T_\ye)}{\tilde \Q(\m_\ye/T_\ye)}=\frac{\lm}{6\h}\frac{\cF_j'(\lm \m_\ye/T_\ye)}{\cF_j(\lm \m_\ye/T_\ye)},
\label{eq:OptimalDopingMT}
}
where $\tilde \Q'(x)=\frac{d}{dx}\tilde\Q(x)$ and $\cF'_j(x)=\frac{d}{dx}\cF_j(x)$. Numerical solution of Eq.~\eqref{eq:OptimalDopingMT} leads to $\m_\ye/T_\ye\eqa 1.13$ which, using Eq.~\eqref{eq:Interpolation}, corresponds to $x_\ye= c_*$, where $c_*\eqa 2.03$.
Finally, using Eq.~\eqref{eq:DefinitionXe}, we find the density at the optimal doping,
\Eq{
n_{\ye,*}=(c_* v_\ys^3\ka/\z)^{1/6}.
}
The optimal doping as a function of $n_{\ye,*}$ is given by $\D n_*=n_{\ye,*}-\ka/n_{\ye,*}$ [as follows from Eq.~\eqref{eq:ElectronHoleDensities}]. In the limit $\ka\to 0$, this simplifies to $\D n_*\eqa n_{\ye,*}$.

\section{Phase transition in the presence of two time-reversal partners}
In the main text and throughout this supplement, we have discussed a model describing half of the degrees of freedom of a time-reversal symmetric semiconductor [see Eq.~(2) in the main text]. In this section, we generalize this model to a model for a material which is time-reversal symmetric (absent the drive). The model includes two copies considered in Eq.~(2), which are related by time reversal symmetry. This model is described by the Hamiltonian
\Eq{
\hat\cH^{\rm TR}(t)=\sum_{\vk}\hat\vc\dg_{\vk} H^{\rm TR}_0(\vk,t)\hat\vc_{\vk}+
\hat \cH^{\rm TR}_{\rm intra}+
\hat \cH^{\rm TR}_{\rm inter},
\label{eq:TRHamiltonian}
}
where  $\hat \vc_\vk\dg=(\hat c_{\vk\aup1}\dg, \hat c_{\vk\adn1}\dg,\hat c_{\vk\aup2}\dg, \hat c_{\vk\adn2}\dg)$, is a four-dimensional spinor; $\s=\bC{\aup,\adn}$ denotes the pseudospin degree of freedom and $\ta=\bC{1,2}$ are two components related by time reversal. The non-interacting part of the Hamiltonian reads
\Eq{
H_0^{\rm TR}(\vk,t)=\mat{H_0(\vk)+H_\yd(t)&0\\0&H\as_0(-\vk)+H_\yd(t)},
}
where $H_0(\vk)=E_0+(|\vk|^2/2m_*+E_\yg/2)+\lm_0\vk\cdot \vs$ and $H_\yd(t)=V\cos(\W t)\vs^z$, in accordance with the definitions in Eq.~(1) in the main text. We consider contact interaction of electrons of the same time-reversal components,
\Eq{
\hat \cH^{\rm TR}_{\rm intra}=\int d^2\vr \sum_{\ta=1,2}U \hat n_{\aup\ta}(\vr)\hat n_{\adn\ta}(\vr) ,
\label{eq:Hintra}
}
and contact interaction between the components
\Eq{
\hat \cH^{\rm TR}_{\rm inter}=\int d^2\vr U_{12} \hat n_1(\vr) \hat n_2(\vr),
\label{eq:Hinter}
}
where $\hat n_{\s\ta}(\vr)=\int \frac{d^2\vq}{(2\p)^2} e^{i\vq\cdot\vr}\hat c_{\vk+\vq\s\ta}\dg \hat c_{\vk\s\ta}$ and
$\hat n_\ta(\vr)=\hat n_{\aup \ta}(\vr)+\hat n_{\adn \ta}(\vr)-n_0/2$; $n_0$ is the density of electrons at half-filling.

We will study the system described by Eq.~\eqref{eq:TRHamiltonian} using the mean-field approximation. In particular, we consider two independent magnetizations and densities for each of the time-reversal partners ($\a=1,2$),
\begin{subequations}
\begin{eqnarray}
&&\vh_{\a}(t)=-\frac{U}{\vp}\sum_{\vk}\av{\vc_{\vk}\dg(\vs\otimes \ta_{\a})\vc_\vk}\\
&&n_{\a}=\frac{1}{\vp}\int d^2\vr \av{\hat n_{\a}(\vr)},
\end{eqnarray}
\label{eq:OrderParameter}%
\end{subequations}
where $\vp$ is the volume of the system and $\ta_{1(2)}=\mat{1(0)&0\\0&0(1)}$. Such a choice of an order parameter gives rise to the following mean-field Hamiltonian, $\hat \cH^{\rm TR}_{\rm MF}(t)=\sum_{\vk}\hat\vc_{\vk}\dg[ H_{0}^{\rm TR}(\vk,t)+ H_{\rm mag}^{\rm TR}(t)]\hat \vc_{\vk}+\vp U_{\rm 12}n_1 n_2$, where
\Eq{
H_{\rm mag}^{\rm TR}(t)=\mat{\vh_1(t)\cdot \vs&0\\0&\vh_2(t)\cdot \vs}.
}

For simplicity, we will treat the problem as that of an equilibrium system in the rotating frame using
the rotating wave approximation (RWA) (see Sec.~\ref{sec:RWAHamiltonian}) and minimize quasienergy with the mean-field band structure. This situation corresponds to $\ka=0$, i.e., a situation when the  Floquet-Umklapp terms are neglected. Our goal is to find $\vh_1(t)$, $\vh_2(t)$, $n_1$ and $n_2$ such that the ground state of $H_{\rm MF}^{\rm TR}$ will minimize the full interacting Hamiltonian, given in Eq.~\eqref{eq:TRHamiltonian}. 
We 
work in the limit of small doping, $\D n\eqv n_1+n_2\ll\cA_\yR$.





Our choice of the order parameter [Eq.~\eqref{eq:OrderParameter}] leads to a mean-field Hamiltonian in which the two components related by time reversal are decoupled, for a state with fixed $n_1$ and $n_2$. Therefore, each of these components can be analyzed independently. We consider circularly polarized magnetizations for both of them, $\vh_\ta(t)=h_\ta e^{i\W t}(\nx-i\ny)/2+c.c.$ (see discussion following Eq.~(8) in the main text). The phase of $h_\ta$ spontaneously breaks the rotational symmetry individually exhibited by each partner. In what follows we will present the analysis of the first component ($\ta=1$); the analysis of its partner ($\ta=2$) is identical. Our analysis partially follows the analysis presented in Ref.~\onlinecite{Berg2012}. We choose real and positive $h_1$ leading to the magnetization  $h_1\hat \vx$ in the RWA, see  Eq.~\eqref{eq:RWAMFHamiltonian}.
The single-particle bandstructure in the mean-field and RWA is given by  $\ve_{\vk,\pm}=\pm \ve_{\vk}$ [see  Eq.~\eqref{eq:ApproximateEnergy}], where
\Eq{
\ve_{\vk}=\sqrt{\bR{\frac{k^2}{2m_\ast}-\frac{\dl E}{2}}^2+\bR{\frac{\D_\yF k_x}{2k_R}+h_1}^2+\bR{\frac{\D_\yF k_y}{2k_R}}^2}.
}

\begin{figure}
  \centering
  \includegraphics[width=8.6cm]{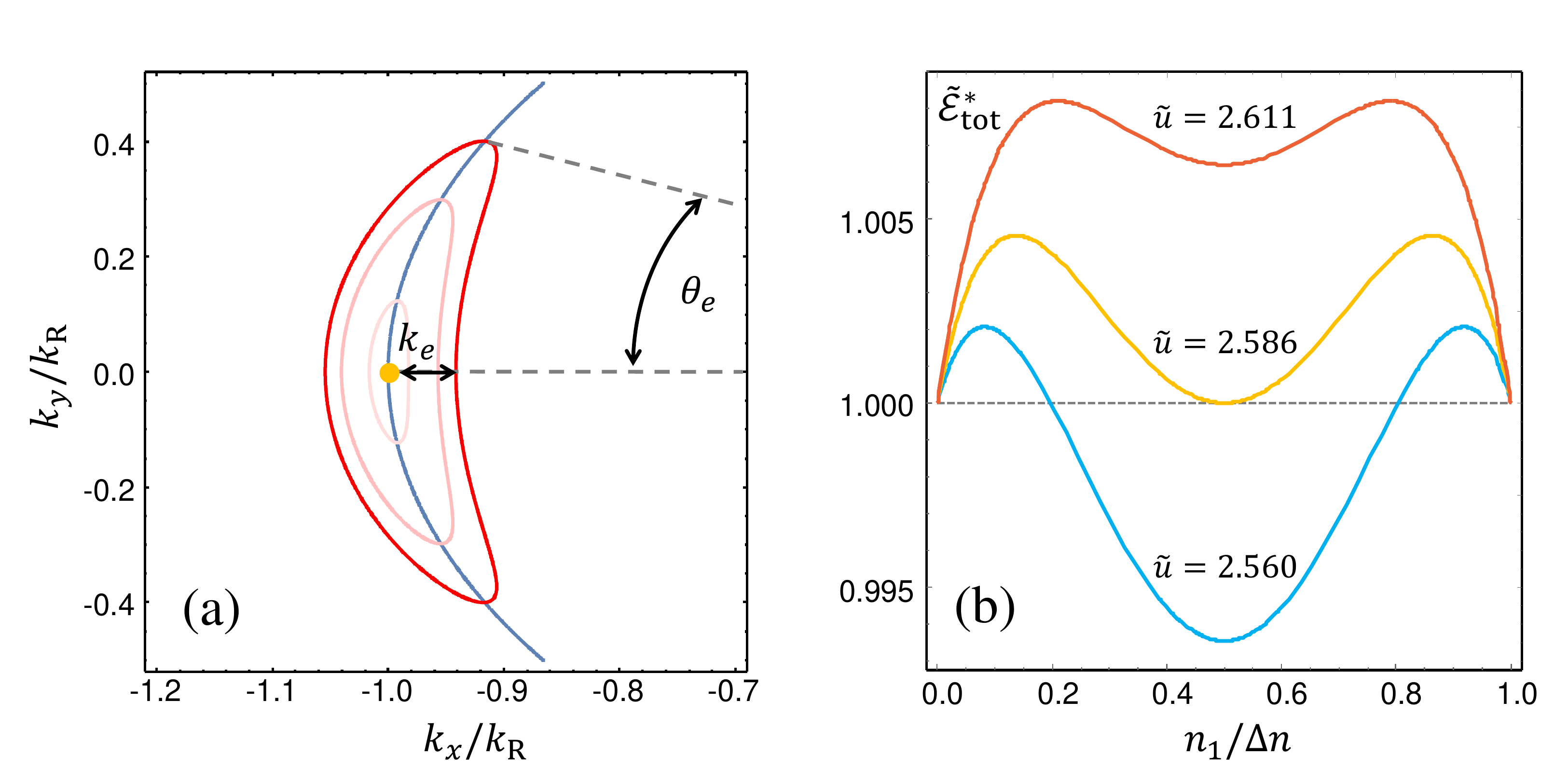}\\
  \caption{(a). Fermi surface of one of the time-reversal partners in the symmetry broken phase, for three densities of electrons. The orange dot indicates the band-minimum point, $\vk_0$ around which we expand the quasienergy. The opening angle ($\q_e$) and the width of the oval-shaped Fermi surface ($k_e$) are indicated on the figure.
  (b). The normalized optimal quasienergy, $\tilde \cE\as_{\rm tot}$, given in Eq.~\eqref{eq:NormalizedOptimalE}, as a function of the normalized density $n_1/\D n$, for three values of the normalized inter-component interaction $\tilde u$. For $\tilde u<4-\sqrt{2}\eqa 2.586$ the total energy has a single global minimum point at $n_1=\D n/2$, corresponding to equal density of particles in the two time-reversal partners. For $\tilde u>4-\sqrt{2}$, the total energy has two global minimum points at $n_1=0$ and $n_1=\D n$. In this case, one of the partners is fully occupied and the other is empty due to strong inter-partner repulsion. The two cases are separated by a critical point at $\tilde u=4-\sqrt{2}$, where the three possibilities coexist.
 \label{fig:TRComp}}
\end{figure}

The Fermi surface, corresponding to a fixed density $n_1$, has a curved elliptic shape centered around the band-minimum point, $\vk_0\eqv -k_\yR \hat {\vx}$, see Fig.~\ref{fig:TRComp}a.
We expand $\ve_{\vk}$ around $\vk_0$ in polar coordinates, $k_x=k \cos(\q)$, $k_y=k\sin(\q)$, for small $\q$ and $\dl k=k-k_R$.
For simplicity, we assume $\D_\yF\ll\dl E$ and $h_1\ll k_R^2/2m_\ast$ leading, up to the second order in $\dl k$ and $\q$, to
\Eq{
\ve_{\vk}\eqa \D_\yF/2-h_1+\frac{\dl E}{\D_\yF}\frac{\dl k^2}{m_\ast}+\frac{h_1}{2}\q^2.
\label{eq:ExpandedEnergy}
}

We define the kinetic energy density by $\cE_{\rm kin}=\int \frac{d^2 \vk}{(2p)^2} (\ve_\vk-h_1\av{\s^x}_\vk)$, where $\av{\s^x}_{\vk}=(\D_\yF k_x/2k_\yR+h_1)/\ve_{\vk}$. In the definition of $\cE_{\rm kin}$ we subtracted the term proportional to $h_1$ to avoid double counting of the magnetization energy, as it will be accounted in the interaction energy density $\cE_{\rm int}$, see below.
The expansion of $\av{\s^x}_{\vk}$ up to the second order in $\dl k$ and $\q$ reads
\EqS{
&\av{\s^x}_{\vk}\eqa -1+\frac{\D_\yF\dl k}{2k_\yR(\D_\yF/2-h_1)}+\\
&+\frac{\dl E \dl k^2}{m_* \D_\yF(\D_\yF/2-h_1)} +\bR{\half+\frac{h_1}{\D_\yF/2-h_1}}\q^2.
\label{eq:ExpandedSigmax}
}
The integral of $\cE_{{\rm kin}}$ is over a curved elliptical area in $k$-space enclosed by the Fermi surface, characterized by the radial length $k_e$ and the angular aperture $\q_e$, see Fig.~\ref{fig:TRComp}a. We parametrize this area by  $\dl k=k_e x \cos(y)$ and $\q=\q_e x \sin(y)$, where $x\in [0,1]$ and $y\in[0,2\p]$. By geometrical constraints, the values of $k_e$ and $\q_e$ are fixed by the density and the magnetization magnitude as follows,
\EqL{
&&k_e\q_e=\frac{4\p n_1}{k_\yR}\\
&&\frac{k_e^2}{\q_e^2}=\frac{\D_\yF}{\dl E}\frac{m_\ast h_1}{2}.
}
Using the new variables, the kinetic energy density reads
$\cE_{\rm kin}=\frac{k_e k_\yR \q_e}{(2\p)^2}\int_{0}^{1}x dx\int_{0}^{2\p}dy (\ve_{\vk}-h_1\av{\s^x}_\vk)$. Integration over $x$ and $y$ [the dependence of $\ve_\vk$ and $\av{\s^x}_\vk$ on $x$ and $y$ is given by Eqs.~\eqref{eq:ExpandedEnergy} and \eqref{eq:ExpandedSigmax}] yields
\Eq{
\cE_{\rm kin}(n_1,h_1)\eqa \frac{\D_\yF n_1}{2}+\sqrt{\frac{h_1}{2\D_\yF}}\frac{\p n_1^2}{m_\ast}+\cO(n_1^2h_1^{3/2}).
\label{eq:EkinPartners}
}

Next, we evaluate the interaction energy density due to interaction of electron-electron interaction given in Eq.~\eqref{eq:Hintra},
$\cE_{\rm int}=\frac{U}{\vp^2}\sum_{\vq\vk\vk'}\av{\hat c\dg_{\vk+\vq\aup}\hat c_{\vk\aup}\hat c\dg_{\vk'-\vq\adn}\hat c_{\vk'\adn}}$. Using the Wick's theorem the interaction energy density simplifies to $\cE_{\rm int}=\frac{U}{4}(n_1^2-\av{\s^x}^2-\av{\s^y}^2-\av{\s^z}^2)$,
where $\av{\s^\a}\eqv\int\frac{d^2\vk}{(2\p)^2} \av{\s^\a}_{\vk}$. We further neglect $\av{\s^y}^2$ and $\av{\s^z}^2$ as they depend on higher powers of $h_1$ and $n_1$. Substituting Eq.~\eqref{eq:ExpandedSigmax} in the definition for $\av{\s^x}$, we find $\av{\s^x}\eqa -n_1+\frac{\p n_1^2}{m_*\sqrt{2h_1\D_\yF }}+\cO(n_1^2h_1^{1/2})$. Therefore, the interaction energy density reads
\Eq{
\cE_{\rm int}(n_1,h_1)\eqa \frac{U\p n_1^3}{m_\ast\sqrt{8h_1\D_\yF}}+\cO(n_1^{4}h_1^{-1}).
\label{eq:EintPartners}
}

The kinetic and interaction energy densities of the time-reversal partner obtain expressions similar to Eqs.~\eqref{eq:EkinPartners} and \eqref{eq:EintPartners} with $n_2$ and $h_2$ replacing $n_1$ and $h_1$.
We define the full energy density of each of the components ($\ta=1,2$) as the sum of the kinetic and the interaction energy densities, $\cE_{\rm full}(n_\ta,h_\ta)=\cE_{\rm kin}(n_\ta,h_\ta)+\cE_{\rm int}(n_\ta,h_\ta)$. Finally, the total energy density includes the full energy densities of both components and the interaction between them [Eq.~\eqref{eq:TRHamiltonian}]. The total energy density of the system then reads
\EqS{
&\cE_{\rm tot}(n_1,n_2,h_1,h_2)=\\
&=\cE_{\rm full}(n_1,h_1)+\cE_{\rm full}(n_2,h_2)+U_{12}\,n_1 n_2.
\label{eq:EtotPartners}
}

Next, we minimize $\cE_{\rm tot}$ with respect to $h_1$, $h_2$ and $n_1$ (given $n_2=\D n-n_1$). Due to the separable form of $\cE_{\rm tot}$, its minimum with respect to $h_1$ and $h_2$ for fixed $n_1$ and $n_2$, coincides with the minimum of its respective terms.
The minimum of $\cE_{\rm full}(n_1,h_1)$ with respect to $h_1$ yields $h_1^{\rm opt}=\half U n_1$. Similarly, the minimum of $\cE_{\rm full}(n_2,h_2)$ with respect to $h_2$ yields $h_2^{\rm opt}=\half U n_2$. Note that $\cE_{\rm tot}$ denotes the total energy density relative to the full lower band. In fact, the energy of the full lower band also depends on $h_1$ and $h_2$. However, this dependence is weak, hence we neglected the contribution of the full lower band in our calculation of $h_1^{\rm opt}$ and $h_2^{\rm opt}$.

Finally, we define the optimal energy density with kinetic energy measured relative to the band bottom as follows, $\cE_{\rm tot}\as(n_1)\eqv\cE_{\rm tot}(n_1,\D n-n_1,h_1^{\rm opt},h_2^{\rm opt})$. The minimum of $\cE_{\rm tot}\as$ with respect to $n_1$ yields the filling of each of the components, $n_1^{\rm opt}$ and $n_2^{\rm opt}=\D n-n_1^{\rm opt}$ in the variational ground state of the system in the rotating frame. Substituting explicit expressions of the energy densities [Eqs.~\eqref{eq:EkinPartners} and \eqref{eq:EintPartners}] and optimal magnetizations ($h_1^{\rm opt}$ and $h_2^{\rm opt}$) into Eq.~\eqref{eq:EtotPartners}, we find
$\cE_{\rm tot}\as(n_1)=\frac{\p(\D n)^2}{m_\ast}\sqrt{\frac{U
\D n}{\D_\yF}}\tilde\cE_{\rm tot}\as(n_1/\D n)+\D n\D_\yF/2$, where
\Eq{
\tilde \cE_{\rm tot}\as(\tilde n)=
\tilde n^{5/2}+(1-\tilde n)^{5/2}+\tilde u\, \tilde n (1-\tilde n),
\label{eq:NormalizedOptimalE}
}
and
$\tilde u\eqv \frac{m_\ast U_{12}}{\p}\sqrt{\frac{\D_\yF}{U\D n}}$.

Fig.~\ref{fig:TRComp}b shows $\cE_{\rm tot}\as$ as a function of the normalized density $n_1/\D n$ for three values of $\tilde u$.
For $\tilde u<4-\sqrt{2}$, the quasienergy has a global minimum at $n_1=\D n/2$, corresponding to an equal share of particles between the two time-reversal components, see Fig.~\ref{fig:TRComp}b. For $\tilde u>4-\sqrt{2}$, the quasienergy density has two global minima at $n_1=0$ and $n_1=\D n$, corresponding to the situation where one of the components is undoped due to the strong inter-partner repulsion.

In conclusion, we showed that a model which includes the time reversal partner of the Hamiltonian discussed in the main text [see Eq~(1)], may still exhibit a symmetry breaking phase. Within our mean-field analysis above, we found that when the repulsion between the time-reversal partners is not too strong, the symmetry breaking in both time-reversal partners minimizes the energy in the rotating frame. If the inter-partner repulsion is stronger than a critical value, only one of the partners becomes doped and develops a symmetry breaking term.
Interestingly, within the mean-field approximation the energy is independent of the relative orientation of the magnetizations of the two time-reversal partners. To determine the optimal relative orientation, one needs to go beyond the mean-field theory we considered. We believe this is an interesting problem for future research. In addition, the analysis presented in this section is limited to $\ka=0$ case. It would be interesting to extend this analysis to finite effective temperature steady states.

\begin{table}
\centering
\begin{tabular}{| c| c || c | c |}
   \hline   \hline
 $A/E_\yg$     & 1/60  & $V/E_\yg$                      & $\sqrt{0.025}$ \\
 $B/E_\yg$     & 1/30  & $\dl E/E_\yg$                  & 1/200        \\
 $A'/E_\yg$    & 1/240 & $v_\ell/(\D_\yF/ k_\yR)$    & 1.6          \\
 $B'/E_\yg$    & 1/120 &                                &              \\
   \hline   \hline
\end{tabular}
\caption{Parameters used in the numerical simulations. \label{tab:Parameters}}
\end{table}

\begin{figure}
  \centering
  \includegraphics[width=8.6cm]{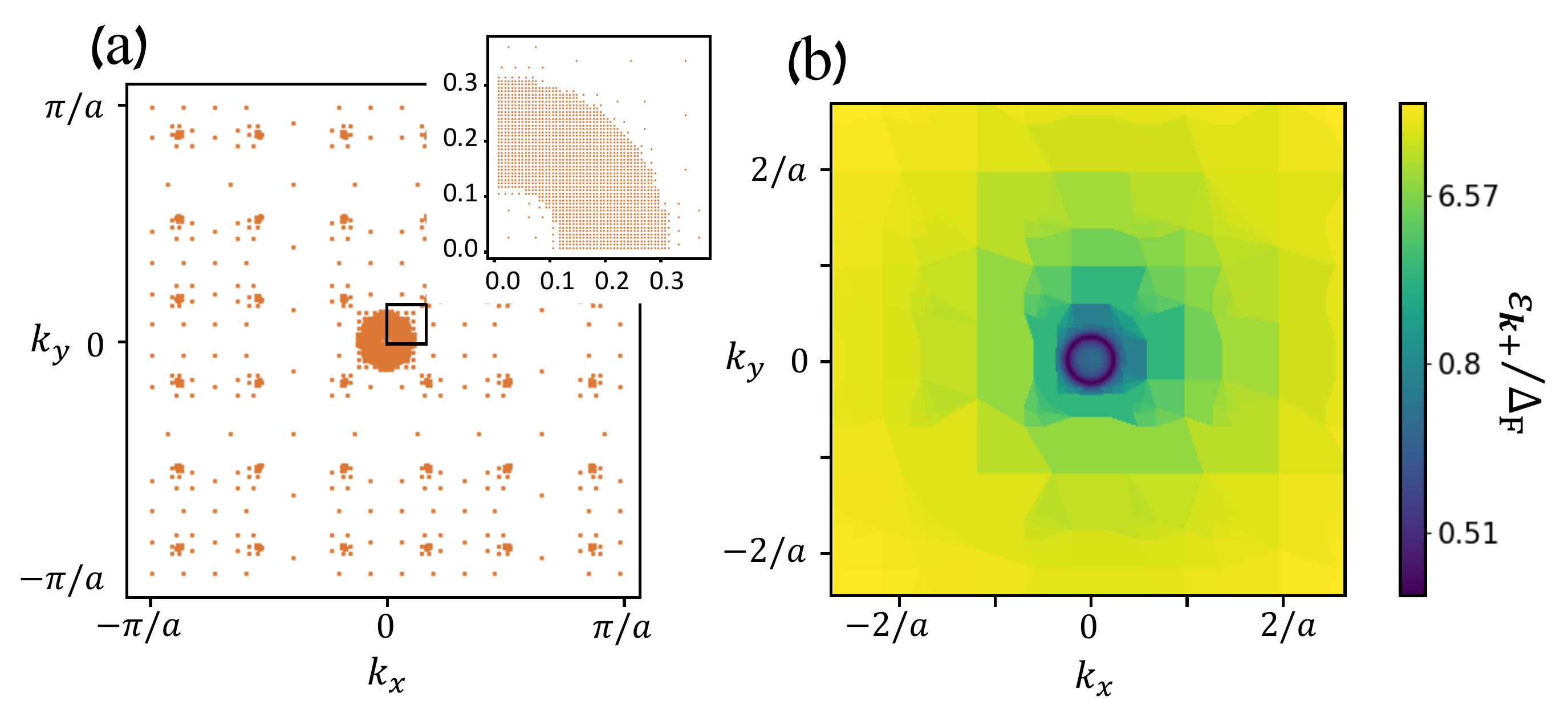}\\
  \caption{(a). The nonuniform momentum grid which we used in the simulation. The highest density of momentum points resides around the resonance curve, $|\vk|=k_R$. The total number of points is $8008$. The inset shows a zoom in on the area with high density of points. (b). The energy of the upper band on the logarithmic scale computed on the nonuniform grid in panel a.
 \label{fig:Numerics}}
\end{figure}

\section{Details of the numerical simulation\label{sec:NumericalDetails}}

In this section we discuss the details of the numerical analysis. The parameters used in the simulation are summarized in Table~\ref{tab:Parameters}. To see a pronounced suppression in $U_\yc$ (see Fig.~2a) we worked at low densities of excitations, requiring a high-resolution grid of $k$-points. To reduce the amount of computational power, we created a nonuniform grid of momentum points with an enhanced density of points around the resonance ring, and total $8008$ $k$-points. Fig.~\ref{fig:Numerics} demonstrates the grid that we used and the Floquet quasienergy levels of the upper Floquet band near the resonance ring. Each point in panel (a) of this figure indicates a center of a grid cell, of area depending on the density of the grid in the vicinity of this cell.
To ensure that the areas of all the cells amount to the total area of the Brillouin zone [$(2\p/a)^2$], we built the nonuniform grid as follows. First, we   partitioned the Brillouin zone to $2^{10}\times 2^{10}$ cells. Then, we recursively combined clusters of $(2\times 2)^n$ cells to form super cells, where the value of $n<10$ controls the density of cells in the vicinity of the cluster. The area of such a cluster equals $(2\p/(2^{10-n}a))^2$.


The $k$-grid is used to define the populations and the rates in the kinetic equation [Eq.~(4) in the main text].
We calculated the rates $I^\ys_{\vk\n}$, $I^\ell_{\vk\n}$,  appearing in this equation, using the Fermi's golden rule,
\Eq{
I^p_{\vk\n}=\sum_{\vk'\n'}(\dot f_{\vk\n})_{p,\vk'\n'},
\label{eq:RateNumerics}
}
where $(\dot f_{\vk\n})_{p,\vk'\n'}$ is given by Eq.~\eqref{eq:FermiGoldenRule1}.
In a realistic realization of the driven system in a solid state setup, we expect the electron-phonon excitation rates to be suppressed by $(V/\W)^2$, as explained in Sec.~\ref{sec:PononHeating}. However, in our simulation, we took a large value of $V/\W$ (see Tab.~\ref{tab:Parameters}) which facilitates large density of states near the Floquet gap. To capture the suppression of the electron-phonon excitations, we multiplied the squared matrix element $\cP_\ys^{(l)}$ appearing in the definition of $(\dot f_{\vk\n})_{\ys,\vk'\n'}$ [see Eq.~\eqref{eq:FermiGoldenRule1}], by a factor $\upsilon_V^{2l}$, where $\upsilon_V=10^{-2}$ represents realistic values of $V/\W$.


\begin{figure}
  \centering
  \includegraphics[width=8.6cm]{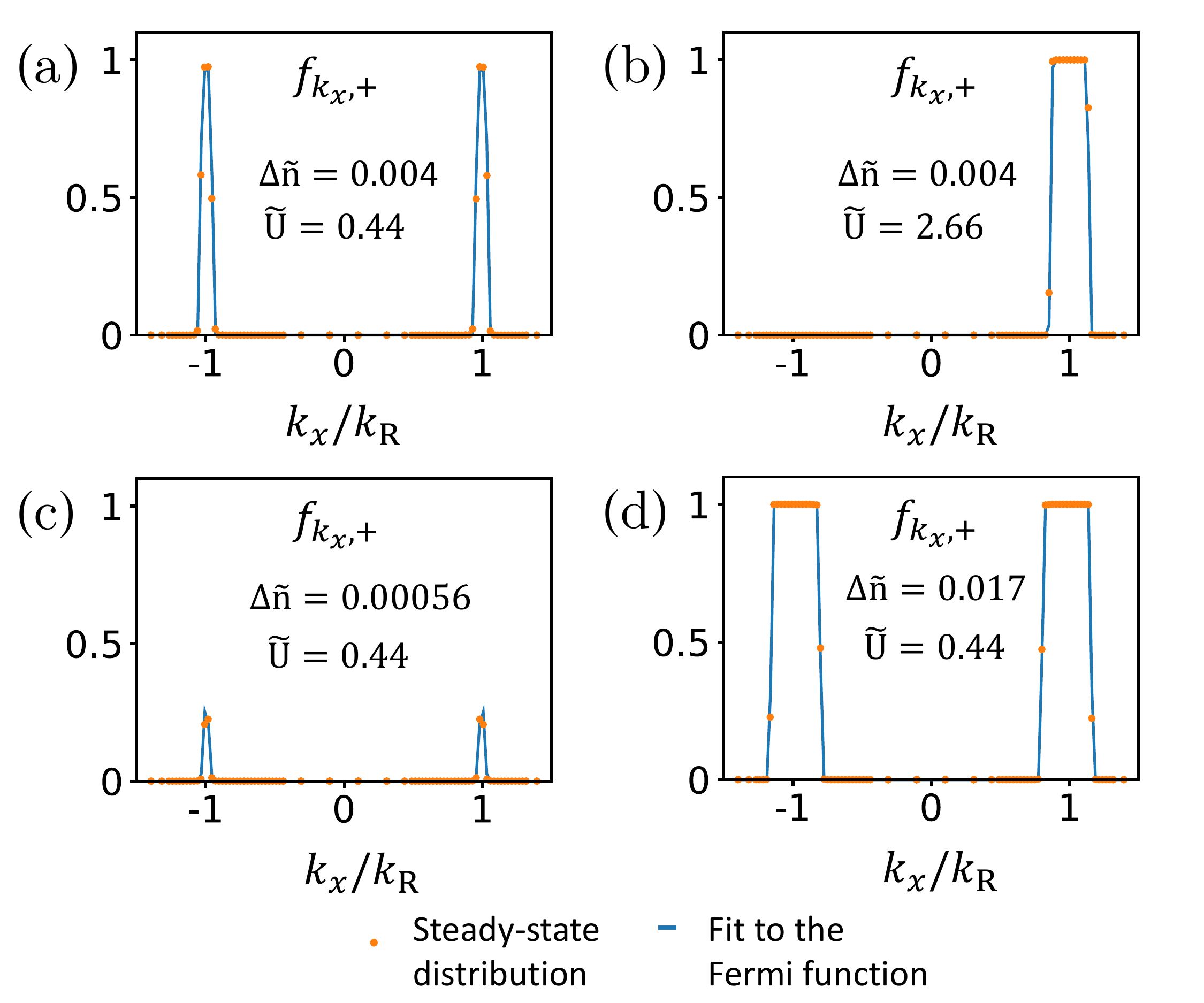}\\
  \caption{Comparison between the steady state distribution of electrons in the upper Floquet band extracted from the numerical simulation, and the fit to the Fermi function, along the cut $k_y=0$. The values of $\D \tilde n$ and $\tilde U$ are indicated on the panels. (a,b). Steady-state distributions at the points indicated by the green and red squares in Fig. 2a in the main text. (c). Low-doping non-degenerate distribution of electrons in the ``paramagnetic'' phase, corresponding to the EFI regime. (d). Strongly doped  ``paramagnetic'' phase, describing a degenerate Fermi distribution in the EFM regime.
 \label{fig:FitToFD}}
\end{figure}

To avoid using four-point terms for the electron-electron scattering collision integral in the numerical simulation of the kinetic equation [Eq.~\eqref{eq:FermiGoldenRule2}], we used an approximate expression for the collision integral of the form $I^{\rm ee}_{\vk\n}=r_\yA U^2I^\ell_{\vk\n}$. Such a collision integral mimics a source term due to Electron-electron interaction-activated excitation rate, described in Sec.~\ref{sec:AugerHeating}. The parameter $r_\yA$ sets the relative strength of this rate. Note that our numerical simulation, with the above simplification for the collision integral, does not take in account equilibration of the electron and hole
distributions via electron-electron scattering. We argue in Sec.~\ref{sec:SolutionExtendedRate2} that including this process will not qualitatively change the results of the numerical simulation.


In the first iteration of the algorithm we set the initial value of the magnetization field to have a form of the in-plane rotating field $\vh^{(0)}(t)=h^{(0)}_1(\nx-i\ny)e^{i\W t}+c.c.$, with $|h_1^{(0)}|/\D_\yF\sim 5\times 10^{-7}$.
We then solved the kinetic equation [given in Eq.~(4) in the main text] in the steady-state ($\dot f_{\vk\n}=0$), using the Newton-Raphson method and obtained the distribution $f_{\vk\n}$ after the first iteration. We used $f_{\vk\n}$ to calculate the magnetization vector $\vh^{(1)}(t)$ for the next iteration of the algorithm, using Eq.~(2) in the main text. As the momentum integral appearing in this equation is very sensitive to finite size effects, we evaluated the magnetization using an adaptive integration method. To this end, we used a fit of the steady-state distribution by two Fermi functions for the electron and hole excitations. Fig.~\ref{fig:FitToFD} shows the comparison of the fits to the numerically obtained distributions at several points in the phase diagram in Fig.~2a in the main text. We then numerically integrate Eq.~(2) over the fitting functions, to obtain the magnetization, $\vh^{(1)}(t)$. Finally, we substituted $\vh^{(1)}(t)$ in the mean-field Hamiltonian [Eq.~(3)] and used it for the next iteration of the algorithm. The iteration loop terminates when the change in amplitude of the left-handed circular component of the magnetization reaches a threshold value. We verified that other components have also converged.

\begin{figure}
  \centering
  \includegraphics[width=8.6cm]{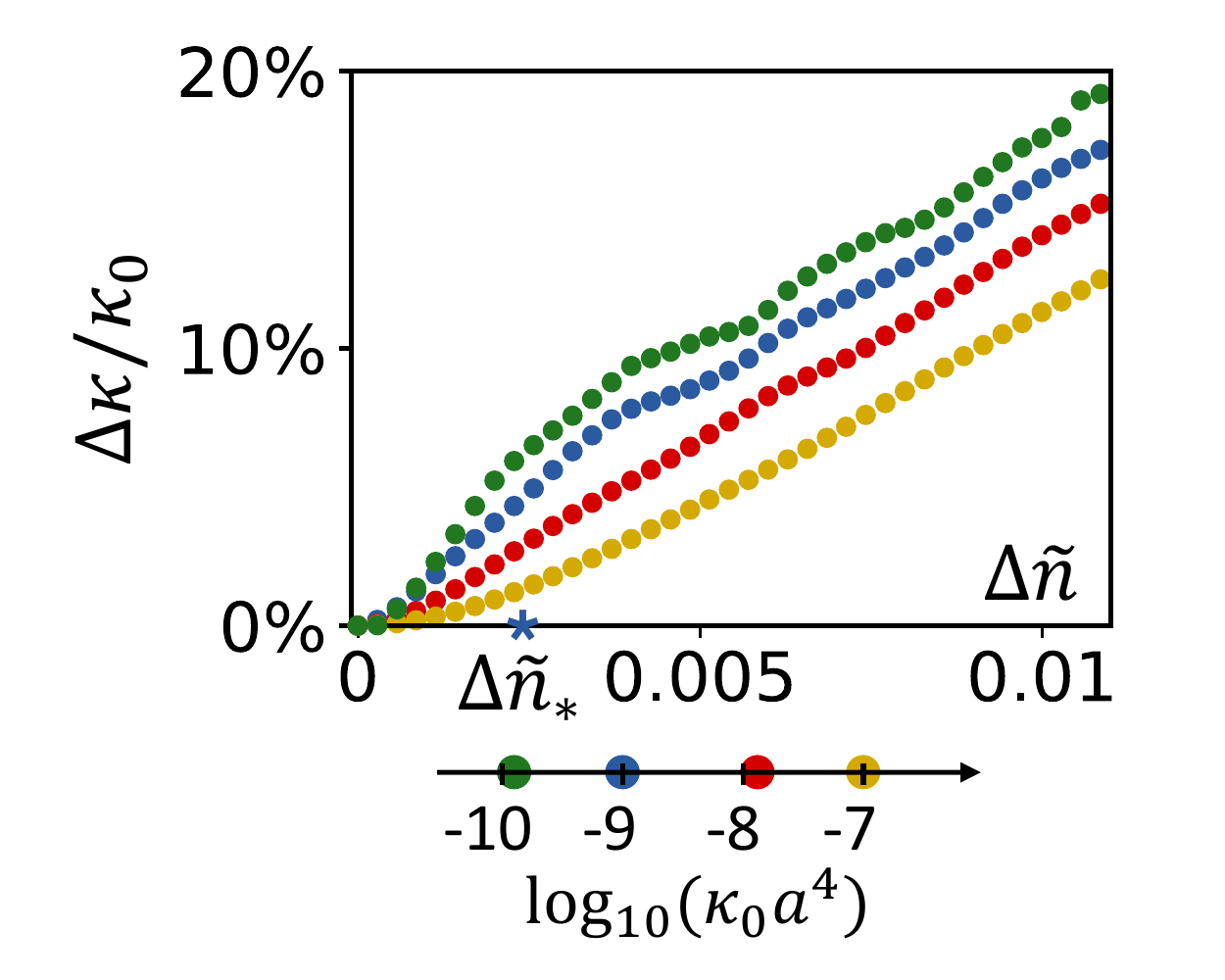}\\
  \caption{The difference $\D \ka=\ka-\ka_0$ as a function of the doping along the phase boundary line for four values of $\ka_0$ indicated at the logarithmic scale. The optimal doping for $\ka_0a^4\eqa 10^{-9}$ is indicated by $\D \tilde n_*$.
 \label{fig:KappaVsKappa0}}
\end{figure}

\subsection{Relation between $\ka$ and system parameters}
In the simulation we controlled the value of $\ka$ by the variation of the electron-photon coupling coefficient $g_\ell$ and measured $\ka$ from measurements of excitation densities [via Eq.~\eqref{eq:DefinitionofKa}]. The definition of $\ka$, given in Eq.~\eqref{eq:Balance_parameter}, suggests that it is also sensitive to other parameters of the system controlling the heating and cooling rates.
In particular, even though we fixed the value of $g_\ell$ throughout the phase diagram in Fig.~2a in the main text, the value of $\ka$ still may vary as a function of $U$ and $\D n$.
We chose the point $U=0$, $\D n=0$ in the phase diagram as a reference and denoted the value of $\ka$ at this point by $\ka_0$.

The value of $\ka$ is approximately equal to $\ka_0$ throughout most of the phase diagram in Fig.~2a (for fixed $g_\ell$). Yet, it differs from $\ka_0$ deep in the broken symmetry phase, as the single-particle Floquet band structure in this regime is significantly deformed.  Furthermore, the value of $\ka$ slightly varies (at low doping)  even in the symmetric phase, due to higher order doping-sensitive terms omitted from Eq.~\eqref{eq:ContinuityNe}.
Likewise, we expect slight changes in $\ka$ relative to $\ka_0$ at the phase boundary.
As we show in  Fig.~\ref{fig:KappaVsKappa0}, the relative change in $\ka$ at the phase boundary near the optimal doping (indicated by $\D \tilde n_*$ in the figure) is less than $5\%$. This finding justifies the assumptions made in our extended rate equation (outlined in Sec.~\ref{sec:ExtendedRateModel}), where we treated $\ka$ as a doping-independent quantity.


\begin{figure}
  \centering
  \includegraphics[width=8.6cm]{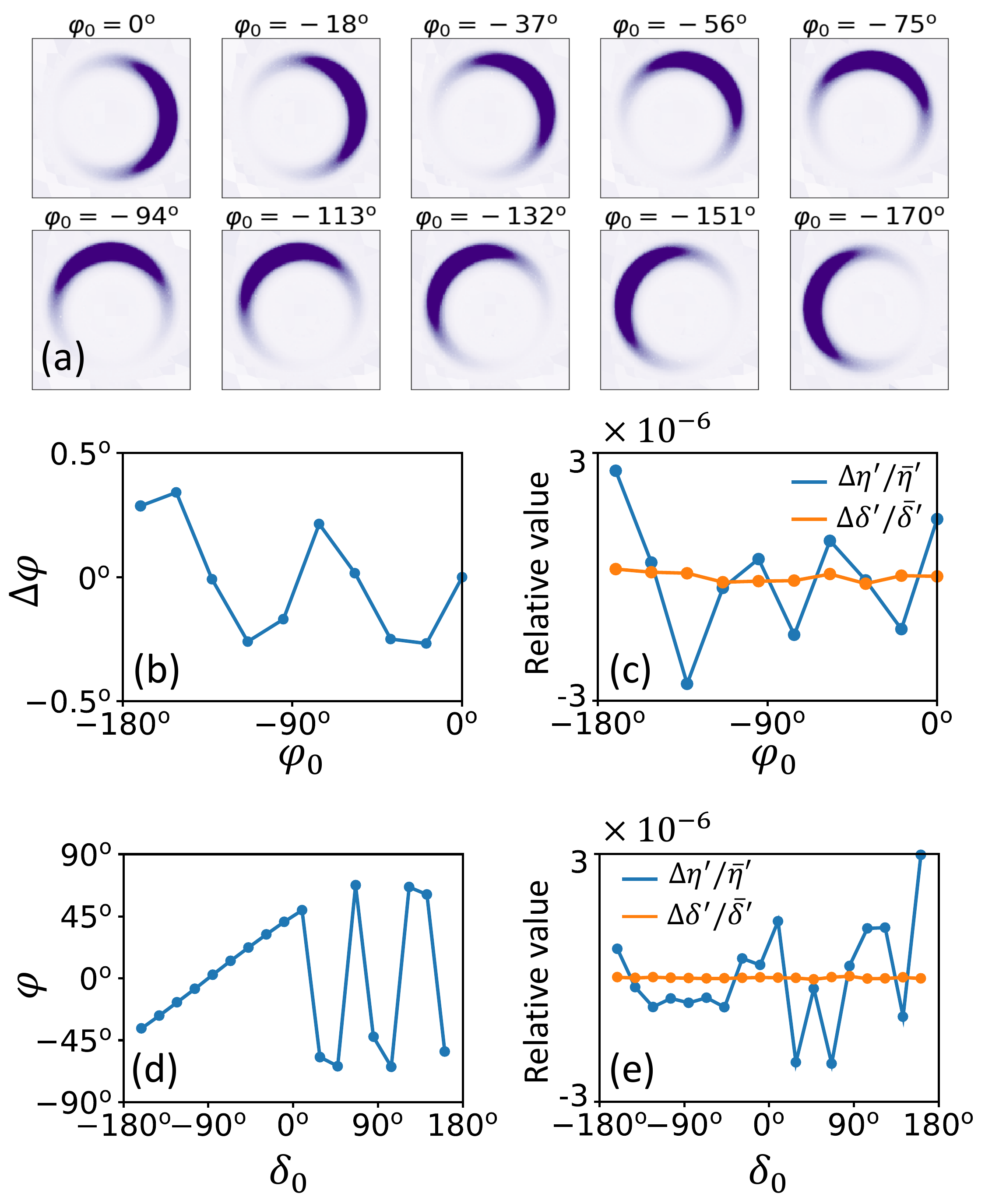}\\
  \caption{Sensitivity of the self-consistent solution to initial conditions. All the results are obtained for $\D \tilde n = 0.004$ and $\tilde U= 2.68$, indicated by the red square in Fig.2a in the main text. (a). Steady-state distribution of the electrons in the upper Floquet band, $f_{\vk+}$, as a function of $k_x$ and $k_y$, near the resonance ring for several values of the phase shift in the initial magnetization, $\vf_0$, see text. (b). The angle mismatch between  $\vf_0$ and the phase of the self-consistent field, $\vf$, for the data shown in panel a. (c). Fluctuations of the relative polarization amplitudes and angles, $\D \h'=\h'-\bar\h'$, $\D \dl'=\dl'-\bar\dl'$, where $\bar \h'\eqa 1.078$ and $\bar\dl'=-90^\circ$. (d). Phase shift of the self-consistent field $\vf$, as a function of the relative initial polarization angle $\dl_0$. (e). Fluctuations of the relative polarization amplitudes and angles as a function of $\dl_0$.
 \label{fig:Sensitivity}}
\end{figure}

\subsection{Sensitivity of the self-consistent solution to the initial conditions}
Here we examine the sensitivity of the self-consistent solution to the initial magnetization, $\vh^{(0)}(t)$. All the results are obtained for $\D \tilde n = 0.004$ and $\tilde U= 2.66$, indicated by the red square in Fig.2a in the main text.

First, we tested the response to different phase shifts of the initial magnetization, $\vh^{(0)}(t)=h_1^{(0)} e^{i\vf_0}(\nx-i\ny)e^{i\W t}+c.c.$. We swept $\vf_0$ in the range $[-180^{\circ},0^{\circ}]$, and measured the in-plane component of the first harmonics of the self-consistent mean-field solution, $\vh_1^{(xy)}=|h_x|e^{i\vf_x}\nx+|h_y|e^{i\vf_y}\ny$ [see definition of $\vh_1^{(xy)}$ in the text following Eq.~(8)]. Our simulation provides the complex-valued amplitudes of the components in the coordinate system spanned by $\nx$ and $\ny$. We define a new coordinate system (spanned by $\nx'$ and $\ny'$) in which the $\nx'$ component is real. The angle between $\nx$ and $\nx'$ is  given by
\Eq{
\tan(\vf)=-\frac{|h_x|\sin(\vf_x)}{|h_y|\sin(\vf_y)}.
\label{eq:RotationNewCoordinates}
}
In the new coordinate system, the in-plane component of the magnetization reads
\Eq{
\vh_1^{(xy)}=|h_{x'}|\nx'+|h_{y'}|e^{i\dl'}\ny'.
\label{eq:RotatedMagnetization}
}
In a system described by a perfectly rotational symmetric interacting Hamiltonian, we expect $\vf=\vf_0$. In turn, we expect the $\nx'$ and $\ny'$ components to be same as the magnetization obtained in the case of $\vf_0=0$, shown in Fig.~2b in the main text. Namely, the relative angle between $\nx'$ and $\ny'$ polarizations, $\dl'$, is expected to be equal $\sim -90^\circ$. And the ratio of the amplitudes,
\Eq{
\h'=h_{x'}/h_{y'}
\label{eq:AmplitudesRatio}
}
is expected to be equal $\sim 1.078$.

Fig.~\ref{fig:Sensitivity}a demonstrates the steady-state distributions of the electrons as a function of the phase shift, $\vf_0$. Clearly, the angle along which the system breaks the symmetry in the rotating frame, $\vf$, follows $\vf_0$. Fig.~\ref{fig:Sensitivity}b shows the deviation between the two angles, $\D\vf=\vf-\vf_0$ as a function of $\vf_0$. The deviation exhibits fluctuations around $0^\circ$ within the range of $\pm 0.5^\circ$ presumably due to the lattice effects, spoiling the perfect rotational symmetry of the Hamiltonian. Fig.~\ref{fig:Sensitivity}c shows the relative fluctuations of the angle $\dl'$ [see Eq.~(\ref{eq:RotatedMagnetization})], and the ratio of the amplitudes, $\h'$ [Eq.~\eqref{eq:AmplitudesRatio}] as a function of $\vf_0$. Both quantities are almost constant with relative fluctuations of the order of $\sim 10^{-6}$ around their averages over the data set, $\bar \dl'\eqa -90^\circ$ and $\bar\h'\eqa 1.078$.

Next, we checked the stability of the self-consistent solution to different elliptical polarizations of the initial conditions. This time, we set the initial mean-field to $\vh^{(0)}(t)=h_1^{(0)} (\nx+e^{i\dl_0}\ny)e^{i\W t}+c.c.$, and swept $\dl_0$ in the range $[-180^\circ,180^\circ]$. For each value of $\dl_0$, we found the self-consistent mean-field solution and extracted the first harmonic of an in-plane component of the magnetization. The result rotated to the new coordinate system by $\vf$ [Eq.~\eqref{eq:RotationNewCoordinates}] is of the form given in Eq.~\eqref{eq:RotatedMagnetization}. Fig.~\ref{fig:Sensitivity}e shows the fluctuations of $\dl'$ and $\h'$ [defined in Eq.~\eqref{eq:AmplitudesRatio}], relative to the averages $\bar \dl'\eqa -90^\circ$ and $\bar\h'\eqa 1.078$, as a function of $\dl_0$. It follows that the quantities $\dl'$ and $\h'$ are almost insensitive to $\dl_0$, with relative fluctuations of the order of $\sim 10^{-6}$.
Fig.~\ref{fig:Sensitivity}d shows the rotation angle, $\vf$, as a function of $\dl_0$. When $\dl_0<0$, the phase shift of the self-consistent solution approximately equals the average between the phases of the initial $x$ and $y$ polarizations, $\vf\eqa (\dl_0+90^\circ)/2$. This relation breaks when $\dl_0>0$, i.e., when the polarization has a strong right-handed circular polarization component. Exactly at $\dl_0=180^\circ$, the self-consistent algorithm does not find a solution with a nonzero magnetization.

\section{Analysis when the Floquet-Auger rate is significant}
\begin{figure}[t]
  \centering
  \includegraphics[width=8.6cm]{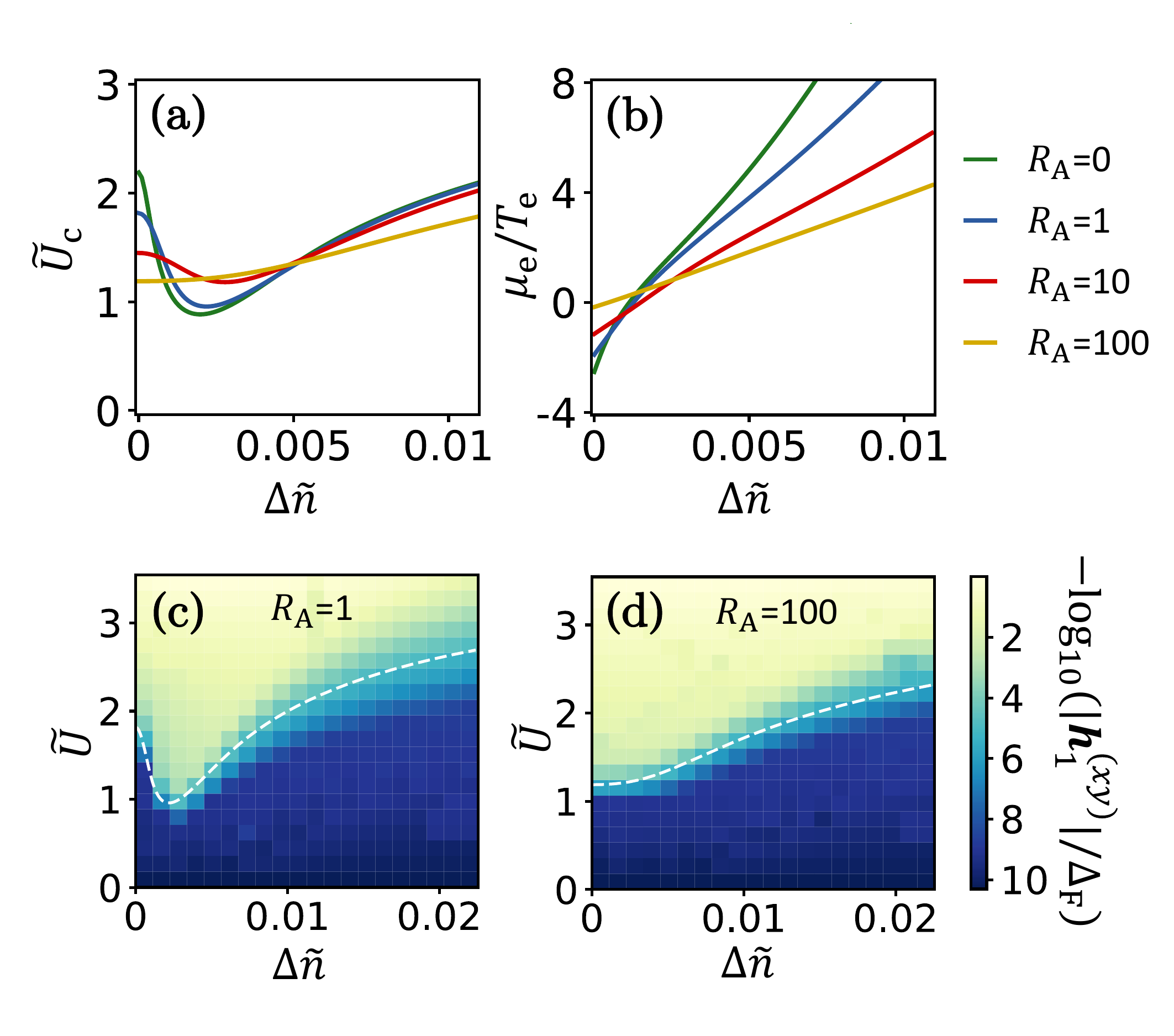}\\
  \caption{Phase diagram as a function of the strength of the Auger processes, $R_\yA=\ch_{\rm ee}/\ka_{\rm ph}$ for $\ka_{\rm ph}\eqa 10^{-9}$ and $v_\ys = 0.0086 \D_\yF/k_\yR$. (a). $\tilde U_\yc$ obtained by solving Eq.~(10) where $n_\ye$, $n_\yh$, $\m_\ye/T_\ye$, and $\m_\yh/T_\yh$ were taken from the analytical rate equation approach [Eqs.~\eqref{eq:ElectronHoleDensities}, \eqref{eq:FitMT} and \eqref{eq:FitMTholes}] and $\ka$ is given by Eq.~\eqref{eq:KappaAuger}. We used the same parameters for $\z$, $\tilde U_{\rm ex}$, and $\tilde U_{\rm fb}$ as in Fig.~3 in the main text. (b). The ratio $\m_\ye/T_\ye$ extracted from Eqs.~\eqref{eq:FitMT} and \eqref{eq:KappaAuger}, for the values of $\tilde U$ obtained from the data in panel a. (c,d). Spontaneous magnetization strength, $|\vh^{(xy)}_1|$ obtained from the self-consistent mean-field calculation, as a function of a normalized electron doping and normalized interaction strength, for two values of $R_\yA$. White dashed lines represent the analytical calculation, shown in panel a for the corresponding values of $R_\yA$.
 \label{fig:DominantAuger}}
\end{figure}

In this section we analyze the phase diagram when the excitation rate due to Floquet-Auger heating described in Sec.~\ref{sec:AugerHeating} is significant. The effect of Floquet-Auger processes is introduced in the model by an additional heating channel with a strength depending on $U^2$ (see Sec.~\ref{sec:NumericalDetails}). In this case, $\ka$ reads
\Eq{
\ka(\tilde U)=\ka_{\rm ph}+\ch_{\rm ee}\tilde U^2,
\label{eq:KappaAuger}
}
where $\ch_{\rm ee}=\ka_{\rm ee}/\tilde U^2$ and $\ka_{\rm ee}=\G_{\rm ee}/\Lm_{\rm inter}$ [cf. Eqs.~\eqref{eq:AugerRate} and \eqref{eq:Lambda_inter}]. Note that $\tilde U=\cA_{\yR} U/\dl E$ is the normalized interaction strength, such that $\ch_{\rm ee}$ and $\ka_{\rm ph}$ have the same dimensions.
To see how the phase boundary is affected by the Floquet-Auger processes, recall that $\tilde U_\yc$, given by Eq.~(10) in the main text, depends on $\ka$ through the dependence of the densities $n_\ye$ and $n_\yh$ on $\ka$ [see Eq.~\eqref{eq:ElectronHoleDensities}] and through the dependence of $\m_\ye/T_\ye$ and $\m_\yh/T_\yh$ on $\ka$, cf. Eq.~\eqref{eq:FitMT}. Therefore, Eq.~(10) becomes a transcendental equation for $\tilde U_\yc$ which can be solved numerically. The primary effect of a strong Floquet-Auger rate would be increasing the excitation density and heating the steady-state distribution. \newline\indent
We evaluate the relative effect of the Floquet-Auger heating by a dimensionless parameter $R_\yA=\ch_{\rm ee}/\ka_{\rm ph}$. Figs.~\ref{fig:DominantAuger}a and b show $\tilde U_\yc$ and $\m_\ye/T_\ye$  resulting from the numerical solution of the transcendental equation obtained from combining Eq.~(10) in the main text with Eqs.~\eqref{eq:ElectronHoleDensities}, \eqref{eq:FitMT}, \eqref{eq:FitMTholes} and \eqref{eq:KappaAuger}. We plot the results for four values of $R_\yA$. The effect of increasing $R_\yA$ is very similar to the effect of changing $\ka_0$ in Fig.~3c and d in the main text, as both control the effective heating. Panels c and d in Fig.~\ref{fig:DominantAuger} show the results of the self-consistent mean-field calculation of the in-plane magnetization component, $\vh_1^{(xy)}$, as a function of $\tilde U$ and $\D \tilde n$ for $R_\yA=1$ and $R_\yA=100$. The parameter $R_\yA$ is controlled in the simulation by tuning $r_\yA$ (see Sec.~\ref{sec:NumericalDetails}). White dashed lines represent the analytical solution to Eq.~(10) similar to the corresponding line in the panel a.

\bibliography{Bibliography.bib}